\documentclass[twocolumn,times]{aastex631}

\usepackage{comment}
\usepackage{multirow}
\usepackage[flushleft]{threeparttable}
\usepackage{graphicx}
\usepackage{amsmath}

\newcommand{\tilda}{$\sim$}

\newcommand{\lessthan}{$<$}
\newcommand{\greaterthan}{$>$}
\newcommand\boldred[1]{\textcolor{red}{\textbf{#1}}}
\newcommand{\jwst}{\textit{JWST}}
\newcommand{\hst}{\textit{HST}}
\newcommand{\sst}{\textit{Spitzer}}
\newcommand{\db}{\texttt{DENSE BASIS}}
\newcommand{\bagpipes}{\texttt{Bagpipes}}

\graphicspath{{./}{}}

\begin{document}

\title{CANUCS\footnote{The public release of the CANUCS dataset is dedicated to the memory of John Hutchings, whose decades of work enabled Canada's involvement in JWST and the creation of this dataset.}/Technicolor Data Release 1: Imaging, Photometry, Slit Spectroscopy, and Stellar Population Parameters}


\author[0000-0002-0786-7307]{Ghassan T. E. Sarrouh}
\altaffiliation{Equal contribution.}
\affiliation{Department of Physics and Astronomy, York University, 4700 Keele St., Toronto, Ontario, M3J 1P3, Canada}

\correspondingauthor{Ghassan T. E. Sarrouh (gsarrouh@yorku.ca) \& Yoshihisa Asada (asada@kusastro.kyoto-u.ac.jp)}

\author[0000-0003-3983-5438]{Yoshihisa Asada}
\altaffiliation{Equal contribution.}
\affiliation{Department of Astronomy and Physics and Institute for Computational Astrophysics, Saint Mary's University, 923 Robie Street, Halifax, Nova Scotia B3H 3C3, Canada}
\affiliation{Department of Astronomy, Kyoto University, Sakyo-ku, Kyoto 606-8502, Japan}
\affiliation{Waseda Research Institute for Science and Engineering, Faculty of Science and Engineering, Waseda University, 3-4-1 Okubo, Shinjuku, Tokyo 169-8555, Japan}

\author[0000-0003-3243-9969]{Nicholas S. Martis}
\affiliation{NRC Herzberg, 5071 West Saanich Rd, Victoria, BC V9E 2E7, Canada}
\affiliation{Department of Astronomy and Physics and Institute for Computational Astrophysics, Saint Mary's University, 923 Robie Street, Halifax, Nova Scotia B3H 3C3, Canada}
\affiliation{University of Ljubljana, Faculty of Mathematics and Physics, Jadranska ulica 19, SI-1000 Ljubljana, Slovenia}

\author[0000-0002-4201-7367]{Chris J. Willott}
\affiliation{NRC Herzberg, 5071 West Saanich Rd, Victoria, BC V9E 2E7, Canada}

\author[0000-0001-9298-3523]{Kartheik G. Iyer}
\affiliation{Columbia Astrophysics Laboratory, Columbia University, 550 West 120th Street, New York, NY 10027, USA}
\affiliation{Dunlap Institute for Astronomy and Astrophysics, 50 St. George Street, Toronto, Ontario M5S 3H4, Canada}

\author{Gaël Noirot}
\affiliation{Department of Astronomy and Physics and Institute for Computational Astrophysics, Saint Mary's University, 923 Robie Street, Halifax, Nova Scotia B3H 3C3, Canada}
\affiliation{Space Telescope Science Institute, 3700 San Martin Drive, Baltimore, MD 21218, USA}

\author[0000-0002-9330-9108
]{Adam Muzzin}
\affiliation{Department of Physics and Astronomy, York University, 4700 Keele St., Toronto, Ontario, M3J 1P3, Canada}

\author[0000-0002-7712-7857]{Marcin Sawicki}
\affiliation{Department of Astronomy and Physics and Institute for Computational Astrophysics, Saint Mary's University, 923 Robie Street, Halifax, Nova Scotia B3H 3C3, Canada}

\author[0000-0003-2680-005X]{Gabriel Brammer}
\affiliation{Cosmic Dawn Center (DAWN), Denmark}
\affiliation{Niels Bohr Institute, University of Copenhagen, Jagtvej 128, DK-2200 Copenhagen N, Denmark}

\author[0000-0001-8325-1742]{Guillaume Desprez}
\affiliation{Department of Astronomy and Physics and Institute for Computational Astrophysics, Saint Mary's University, 923 Robie Street, Halifax, Nova Scotia B3H 3C3, Canada}
\affiliation{Kapteyn Astronomical Institute, University of Groningen, P.O. Box 800, 9700AV Groningen, The Netherlands}

\author[0009-0009-4388-898X]{Gregor Rihtar\v{s}i\v{c}}
\affiliation{University of Ljubljana, Faculty of Mathematics and Physics, Jadranska ulica 19, SI-1000 Ljubljana, Slovenia}

\author{Johannes Zabl}
\affiliation{Department of Astronomy and Physics and Institute for Computational Astrophysics, Saint Mary's University, 923 Robie Street, Halifax, Nova Scotia B3H 3C3, Canada}

\author[0000-0002-4542-921X]{Roberto Abraham}
\affiliation{David A. Dunlap Department of Astronomy and Astrophysics, University of Toronto, 50 St. George Street, Toronto, Ontario, M5S 3H4, Canada}

\author[0000-0001-5984-0395]{Maru\v{s}a Brada{\v c}}
\affiliation{University of Ljubljana, Faculty of Mathematics and Physics, Jadranska ulica 19, SI-1000 Ljubljana, Slovenia}
\affiliation{Department of Physics and Astronomy, University of California Davis, 1 Shields Avenue, Davis, CA 95616, USA}

\author[0000-0001-5485-4675]{Ren\'{e} Doyon}
\affiliation{Département de Physique and Observatoire du Mont-Mégantic, Université de Montréal, C.P. 6128, Succ. Centre-ville, Montréal, H3C 3J7, Québec, Canada}
\affiliation{Institut de Recherche sur les exoplanètes, Université de Montréal, Québec, Canada}

\author[0000-0002-0243-6575]{Jacqueline Antwi-Danso}
\altaffiliation{Banting Postdoctoral Fellow}
\affiliation{David A. Dunlap Department of Astronomy and Astrophysics, University of Toronto, 50 St. George Street, Toronto, Ontario, M5S 3H4, Canada}

\author[0000-0001-7549-5560]{Samantha Berek}
\affiliation{David A. Dunlap Department of Astronomy and Astrophysics, University of Toronto, 50 St. George Street, Toronto, Ontario, M5S 3H4, Canada}
\affiliation{Dunlap Institute for Astronomy \& Astrophysics, University of Toronto, 50 St. George Street, Toronto, ON M5S 3H4, Canada}
\affiliation{Data Sciences Institute, University of Toronto, 17th Floor, Ontario Power Building, 700 University Avenue, Toronto, ON M5G 1Z5, Canada}

\author[0000-0002-6741-078X]{Westley Brown}
\affiliation{Department of Physics and Astronomy, York University, 4700 Keele St., Toronto, Ontario, M3J 1P3, Canada}

\author{Vince Estrada-Carpenter}
\affiliation{Department of Astronomy and Physics and Institute for Computational Astrophysics, Saint Mary's University, 923 Robie Street, Halifax, Nova Scotia B3H 3C3, Canada}

\author[0009-0005-6999-2073]{Jeremy Favaro}
\affiliation{Department of Astronomy and Physics and Institute for Computational Astrophysics, Saint Mary's University, 923 Robie Street, Halifax, Nova Scotia B3H 3C3, Canada}

\author[0009-0001-0778-9038]{Giordano Felicioni}
\affiliation{University of Ljubljana, Faculty of Mathematics and Physics, Jadranska ulica 19, SI-1000 Ljubljana, Slovenia}

\author[0000-0001-6003-0541]{Ben Forrest}
\affiliation{Department of Physics and Astronomy, University of California Davis, One Shields Avenue, Davis, CA, 95616, USA}

\author[0000-0001-9293-4449]{Gaia Gaspar}
\affiliation{Department of Astronomy and Physics and Institute for Computational Astrophysics, Saint Mary's University, 923 Robie Street, Halifax, Nova Scotia B3H 3C3, Canada}

\author[0000-0003-4196-5960]{Katriona M. L. Gould}
\affiliation{Cosmic Dawn Center (DAWN), Denmark}
\affiliation{Niels Bohr Institute, University of Copenhagen, Jagtvej 128, DK-2200 Copenhagen N, Denmark}

\author{Rachel Gledhill}
\affiliation{Cosmic Dawn Center (DAWN), Denmark}
\affiliation{Niels Bohr Institute, University of Copenhagen, Jagtvej 128, DK-2200 Copenhagen N, Denmark}

\author[0000-0001-9414-6382]{Anishya Harshan}
\affiliation{University of Ljubljana, Faculty of Mathematics and Physics, Jadranska ulica 19, SI-1000 Ljubljana, Slovenia}

\author[0009-0004-7069-1780]{Nusrath Jahan}
\affiliation{Department of Physics, Shahjalal University of Science and Technology, Sylhet 3114, Bangladesh}

\author[0009-0009-9848-3074]{Naadiyah Jagga}
\affiliation{Department of Physics and Astronomy, York University, 4700 Keele St., Toronto, Ontario, M3J 1P3, Canada}

\author[0009-0000-2101-1938]{Jon Jude\v{z}}
\affiliation{University of Ljubljana, Faculty of Mathematics and Physics, Jadranska ulica 19, SI-1000 Ljubljana, Slovenia}

\author[0000-0001-9002-3502]{Danilo Marchesini}
\affiliation{Department of Physics and Astronomy, Tufts University, 574 Boston Avenue, Suite 304, Medford, MA 02155, USA}

\author[0000-0002-5694-6124]{Vladan Markov}
\affiliation{University of Ljubljana, Faculty of Mathematics and Physics, Jadranska ulica 19, SI-1000 Ljubljana, Slovenia}

\author[0000-0002-7547-3385]{Jasleen Matharu}
\affiliation{Cosmic Dawn Center (DAWN), Denmark}
\affiliation{Niels Bohr Institute, University of Copenhagen, Jagtvej 128, DK-2200 Copenhagen N, Denmark}

\author{Shannon MacFarland}
\affiliation{Department of Astronomy and Physics and Institute for Computational Astrophysics, Saint Mary's University, 923 Robie Street, Halifax, Nova Scotia B3H 3C3, Canada}

\author[0009-0000-5385-8674]{Maya Merchant}
\affiliation{Niels Bohr Institute, University of Copenhagen, Jagtvej 128, DK-2200 Copenhagen N, Denmark}

\author[0000-0001-8115-5845]{Rosa M. M\'erida}
\affiliation{Department of Astronomy and Physics and Institute for Computational Astrophysics, Saint Mary's University, 923 Robie Street, Halifax, Nova Scotia B3H 3C3, Canada}

\author[0000-0002-8530-9765]{Lamiya Mowla}
\affiliation{Whitin Observatory, Department of Physics and Astronomy, Wellesley College, 106 Central Street, Wellesley, MA 02481, USA}

\author[0009-0009-2307-2350]{Katherine Myers}
\affiliation{Department of Physics and Astronomy, York University, 4700 Keele St., Toronto, Ontario, M3J 1P3, Canada}

\author[0000-0002-8432-6870]{Kiyoaki C. Omori}
\affiliation{Department of Astronomy and Physics and Institute for Computational Astrophysics, Saint Mary's University, 923 Robie Street, Halifax, Nova Scotia B3H 3C3, Canada}

\author[0000-0003-4196-0617]{Camilla Pacifici}
\affiliation{Space Telescope Science Institute, 3700 San Martin Drive, Baltimore, MD 21218, USA.}

\author[0000-0002-5269-6527]{Swara Ravindranath}
\affiliation{Space Telescope Science Institute (STScI), 3700 San Martin Drive, Baltimore, MD 21218, USA\\}

\author[0000-0002-6265-2675]{Luke Robbins}
\affiliation{Department of Physics and Astronomy, Tufts University, 574 Boston Avenue, Suite 304, Medford, MA 02155, USA}

\author[0000-0002-6338-7295]{Victoria Strait}
\affiliation{Cosmic Dawn Center (DAWN), Denmark}
\affiliation{Niels Bohr Institute, University of Copenhagen, Jagtvej 128, DK-2200 Copenhagen N, Denmark}

\author[0000-0003-0780-9526]{Visal Sok}
\affiliation{Department of Physics and Astronomy, York University, 4700 Keele St., Toronto, Ontario, M3J 1P3, Canada}

\author[0000-0002-3503-8899]{Vivian Yun Yan Tan}
\affiliation{Department of Physics and Astronomy, York University, 4700 Keele St., Toronto, Ontario, M3J 1P3, Canada}

\author[0000-0002-9909-3491]{Roberta Tripodi}
\affiliation{University of Ljubljana, Faculty of Mathematics and Physics, Jadranska ulica 19, SI-1000 Ljubljana, Slovenia}

\author[0000-0002-6572-7089]{Gillian Wilson}
\affiliation{Department of Physics, University of California Merced,
5200 North Lake Road, Merced, CA 95343, USA}

\author[0009-0000-8716-7695]{Sunna Withers}
\affiliation{Department of Physics and Astronomy, York University, 4700 Keele St., Toronto, Ontario, M3J 1P3, Canada}





\begin{abstract}

We present the first data release of the CAnadian NIRISS Unbiased Cluster Survey (CANUCS), a JWST Cycle 1 GTO program targeting 5 lensing clusters and flanking fields in parallel (Abell 370, MACS0416, MACS0417, MACS1149, MACS1423; survey area \tilda100 arcmin$^{2}$), with NIRCam imaging, NIRISS slitless spectroscopy, and NIRSpec prism multi-object spectroscopy. Fields centered on cluster cores include imaging in 8 bands from 0.9-4.4$\mu$m, alongside continuous NIRISS coverage from 1.15-2$\mu$m, while the NIRCam flanking fields provide 5 wide and 9 medium band filters for exceptional spectral sampling, all to \tilda29 mag$_{AB}$.
We also present JWST in Technicolor, a Cycle 2 follow-up GO program targeting 3 CANUCS clusters (Abell 370, MACS0416, MACS1149). The Technicolor program adds NIRISS slitless spectroscopy in F090W to the cluster fields while adding 8 wide, medium, and narrow band filters to the flanking fields. This provides NIRCam imaging in all wide and medium band filters over \tilda30 arcmin$^{2}$. 
This paper describes our data reduction and photometry methodology. We release NIRCam, NIRISS, and HST imaging, PSFs, PSF-matched imaging, photometric catalogs, and photometric and spectroscopic redshifts. We provide lens models and stellar population parameters in up to 19 filters for \tilda53,000 galaxies in the cluster fields, and \tilda44,000 galaxies in up to 29 filters in the flanking fields. We further present 733 NIRSpec spectra and redshift measurements up to $z=10.8$. Comparing against our photometric redshifts, we find catastrophic outlier rates of only 4-7\% and scatter  of $\sigma_{\rm NMAD}$ of 0.01-0.03.



\end{abstract}



\section{Introduction}\label{sec:intro}



\begin{figure*}[t!]
\plotone{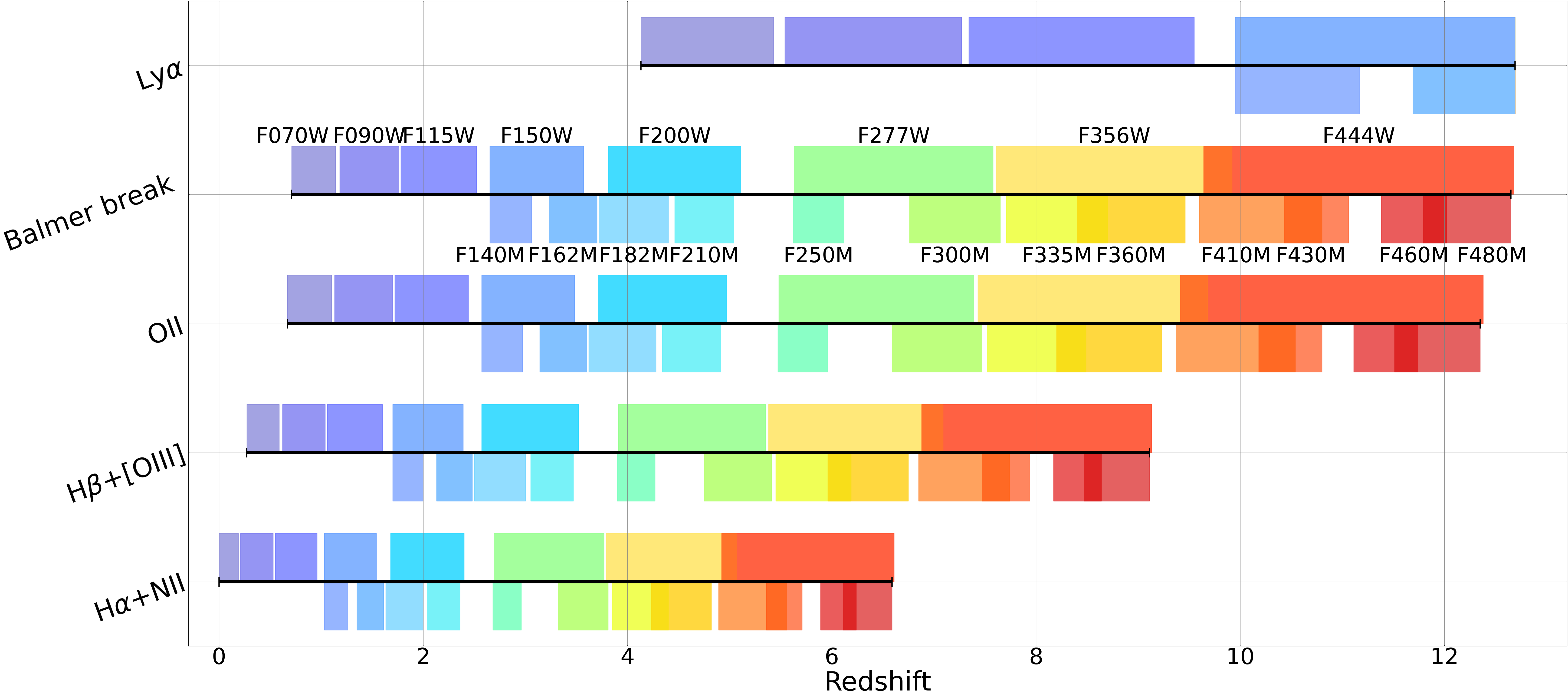}
\caption{Spectral features as a function of redshift. Colored bars represent NIRCam photometric filters, and span the range of redshifts where the observed wavelength of the spectral feature falls within the bandwidth of the filter.}
\label{fig:spectral_props_by_filt}
\end{figure*}


Extragalactic deep fields offer the furthest glimpse into the past that astronomical observations can achieve. Limited by the impossibility of observing cosmic evolution directly on human timescales, unbiased, wide field surveys are necessary to provide statistical snapshots of galaxy populations at different cosmological epochs. Infrared observations in particular are essential to probe the rest frame optical spectrum of galaxies in the high redshift universe, allowing their properties to be constrained and their natures understood.

The first two years of operations of the James Webb Space Telescope (JWST) have afforded the clearest picture of the distant universe to date. Cycle 1 \& 2 programs have pursued both area (e.g. CEERS, \citealt{Finkelstein2023}; COSMOS-Web, \citealt{Casey2023}; FRESCO, \citealt{Oesch2023}, PRIMER, GO 1837, PI: J. Dunlop) and depth (e.g. JADES, \citealt{Eisenstein2023}; GLASS, \citealt{Treu2022}; UNCOVER, \citealt{Bezanson2022}), revealing new populations at high redshifts (e.g. \citealt{Nelson2023,Matthee2024,Greene2024,Robertson2024,Tripodi2024,Witstok2024,Willott2025,Merida2025}), distant quiescent galaxies (e.g. \citealt{Carnall2023a,Strait2023a,Looser2024,Glazebrook2024}), and bursty star forming and extreme emission line galaxies during the Epoch of Reionization (e.g. \citealt{Asada2023,Withers2023,Asada2024a,Harshan2024,Endsley2024}). 

Some programs have targeted strong lensing clusters (e.g. \citealt{Treu2022,Bezanson2022,Windhorst2023,Kokorev2024c}), seeking to exploit the natural magnification afforded by foreground galaxy clusters to peer even deeper into the unknown, revealing some of the faintest and most distant galaxies (e.g. \citealt{Mascia2023,Hsiao2024,Fujimoto2024}). While observing galaxy clusters offers obvious advantages, they are not without caveats. Foreground cluster galaxies and intracluster light (ICL) may contaminate targets of scientific interest behind the cluster (e.g. \citealt{Shipley2018,Bhatawdekar2019,Martis2024,Weaver2024}), and lensing models may be poorly constrained and highly uncertain \citep{Strait2018}. Furthermore, the source-plane area behind the lensing cluster is smaller, and so probes smaller volumes of space. Nevertheless, lensing clusters present a unique opportunity to detect novel phenomena that would otherwise be out of reach.

The characterization of galaxy physical properties however relies heavily on the results of spectral energy distribution (SED) modeling. While photometric data provide an efficient way to characterize a large number of galaxies, insufficient sampling of the SED leads to known degeneracies (see \citealt{Iyer2025} and references therein). These include confusing spectral features such as the Lyman and Balmer breaks (\citealt{Naidu2022,ArrabalHaro2023,Donnan2023,Zavala2023}), leading to dramatically different redshifts; and an inability to distinguish emission line from continuum flux at higher redshift, leading to underestimated line strengths and inflated stellar masses (\citealt{Yabe2009,Stark2013,Smit2014,Laporte2023,Desprez2024,Sarrouh2024}). Ground-based medium band surveys have addressed these problems at lower redshift (\citealt{Wolf2003,Whitaker2011, Perez-Gonzalez2013,Straatman2016,Antwi-Danso25}), however the presence of the atmosphere limits their ability to sample the SED beyond observed wavelengths of $\gtrsim$ 2$\mu$m. 

While multi-band photometry is efficient at sampling the SEDs of all sources within the instrumental field of view, it lacks the precision and spectral resolution of spectroscopy. JWST's NIRSpec multi-object spectroscopy offers exceptional sensitivity and wavelength coverage from 0.6-5.2$\mu$m that has already lead to discoveries of confirmed galaxies at z \greaterthan \ 10 (e.g. \citealt{Carniani2024,CurtisLake2023}), early massive quiescent galaxies out to z \tilda 5 (e.g. \citealt{Carnall2023a,Glazebrook2024}), and the first ``napping" or ``mini-quenched" galaxies (e.g. \citealt{Strait2023a,Looser2024}). 


Most early JWST imaging surveys focused on wide band photometry, with limited exceptions (JADES ORIGINS, \citealt{Eisenstein2023a}; JEMS, \citealt{Williams2023}; MegaScience, \citealt{Suess2024}). Early JWST results based on wide band photometry alone revealed many massive, high redshift galaxies candidates (e.g. \citealt{Labbe2023,Atek2023,Trussler2023}), testing the limits of the $\Lambda$CDM paradigm. However, follow-up studies utilizing NIRSpec prism spectroscopy and NIRCam's medium bands, which provide vastly superior SED sampling (see Figure \ref{fig:spectral_props_by_filt}), have found such sources to be at lower mass than when analyzed with wide bands alone, attributing the difference to strong line emission in high redshift SEDs (\citealt{Zavala2023,Desprez2024,Sarrouh2024}). 

\begin{figure*}[t!]
\includegraphics[width=\textwidth]{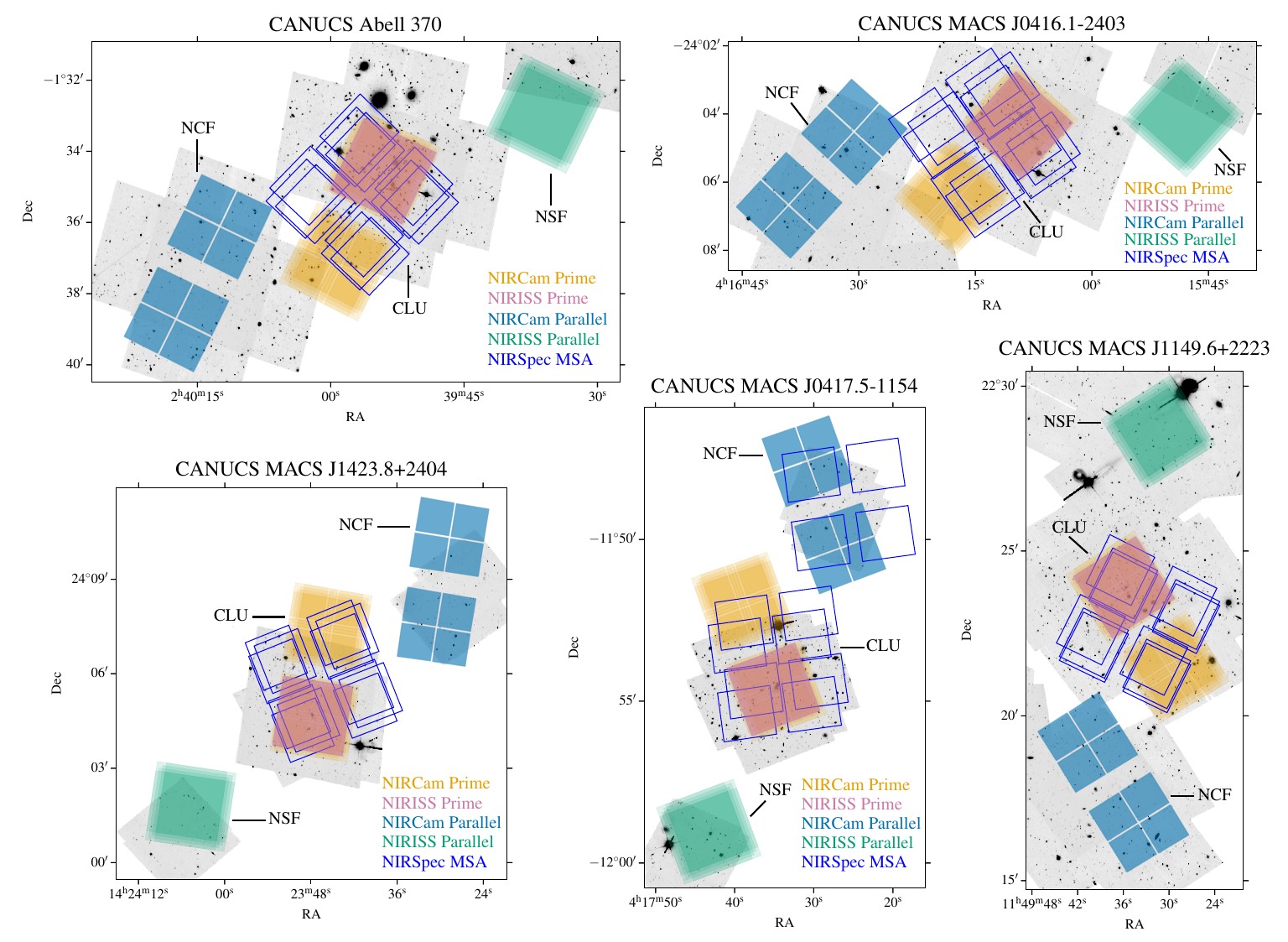}
\caption{Layouts for the CANUCS observations in the five cluster target fields. For each cluster, the background greyscale images are HST optical F606W (F606W and F606WU combined for MACS0417 and MACS1423). NIRCam and NIRISS coverage when prime are shaded different colors from when the parallel instrument. This shows how the set of two coordinated parallel observations lead to three fields per cluster; the central cluster field (CLU) being covered by NIRCam (with NIRISS overlapping module B), the NIRCam flanking field (NCF) only with NIRCam and the NIRISS flanking field (NSF) only with NIRISS. The three NIRSpec follow-up pointings per field are shown with blue outlines for each MSA quadrant.}
\label{fig:FOVs}
\end{figure*}


The CAnadian NIRISS Unbiased Cluster Survey (CANUCS, GTO: 1208, PI: C. Willott, \citealt{Willott2022}) used \tilda200 hours of observatory time in Cycle 1 with NIRCam \citep{Rieke2023a}, NIRISS (\citealt{Willott2022,Doyon2023}), and NIRSpec \citep{Jakobsen2022} to conduct a coordinated parallel survey of 5 massive lensing clusters with both NIRISS and NIRCam, with flanking fields to either side of the cluster field. Follow-up NIRspec MOS prism spectroscopy was taken for all 5 cluster fields as well as the MACS0417 NIRCam flanking field. The \emph{JWST in Technicolor} program in Cycle 2 (GO: 3362, PI: A. Muzzin) revisited 3 of the clusters, extending NIRISS wide field slitless spectroscopy (WFSS) coverage in the cluster with F090W in both GR150C and GR150R grisms, and completing the full set of NIRCam wide and medium bands in the flanking fields.

This paper presents CANUCS/Technicolor Data Release 1 (DR1) including space-based imaging and PSF-matched photometric catalogs, NIRSpec spectroscopy, photometric and spectroscopic redshifts and stellar population synthesis (SPS) parameters for 10 CANUCS fields; 5 centered on the lensing clusters and 5 in the adjacent NIRCam flanking fields. Imaging, catalogs, and other data products including lens models are made available online\footnote{https://niriss.github.io/data.html}. We also release all NIRSpec spectra obtained as part of the Cycle 1 follow-up program. NIRISS imaging is included, however all other NIRISS WFSS data products will be released separately and are not included in this first data release.

In Section \ref{sec:data} we describe the survey design and data available in each field. An overview of our data processing, including modelling and subtraction of foreground cluster galaxies and ICL, is given in Section \ref{sec:data_reduction}. In Section \ref{sec:photometry} we discuss source detection, photometry and the construction of the photometric catalogs, as well as our methodology for empirically measuring and homogenizing the point spread function (PSFs) and photometric and spectroscopic redshifts. Catalog properties including number counts and depths are presented in Section \ref{sec:cat_properties}. SPS parameters and a discussion of the fitting techniques used is given in Section \ref{sec:SPS}. Gravitational lensing models and related data products for the central CLU fields only are presented in Section \ref{sec:lensing}. Section \ref{sec:product_summary} discusses specific naming conventions and details regarding the various types of individual files included in this release. Finally, we summarize our work in Section \ref{sec:summary}. 

All magnitudes are expressed in the AB system \citep{Oke1983}. Flux density $f_{\nu}$ is measured in nJy ($10^{-28}$ erg cm$^{-2}$ s$^{-1}$ Hz$^{-1}$), where AB$_{\nu}$ = -2.5 log$_{10}$($f_{\nu}$/nJy) + 31.4. We adopt a flat $\Lambda$CDM cosmology with $H_{0}$ = 70 km s$^{-1}$Mpc$^{-1}$, $\Omega_{M,0}$ = 0.3, and $\Omega_{\Lambda,0}$ = 0.7.


%


%

\section{Survey Design, Data Set Description \& Data Products Included in Release} \label{sec:data}




\begin{deluxetable*}{lcccccl}
    \label{tab:cluster_info}
    \tablecaption{CANUCS target clusters
  	}
    \tablewidth{0pt}
    \tablehead{
    \colhead{Cluster} & \colhead{R.A.} & \colhead{Dec.} & \colhead{Redshift}& \colhead{Galactic extinction\tablenotemark{a}}  & \colhead{JWST V3PA\tablenotemark{b}} & \colhead{HST Provenance}\\
    \colhead{} & \colhead{(J2000)} & \colhead{(J2000)} & \colhead{} & \colhead{$A_V$ [mag]} & \colhead{(degrees)} & \colhead{}
    }
    \startdata
    Abell\,370         & 02:39:54.1 & -01:34:34 & 0.375 & 0.0843 (0.0814) & 63.96 & Frontier Fields \citep{Lotz2017}\\
    MACS\,J0416.1-2403 & 04:16:09.4 & -24:04:21 & 0.395 & 0.1090 (0.1111) & 48.56 & Frontier Fields \citep{Lotz2017}\\
    MACS\,J0417.5-1154 & 04:17:35.1 & -11:54:38 & 0.443 & 0.0992 (0.1091) & 289.96 & CLASH \citep{Postman2012}\\
    MACS\,J1149.6+2223 & 11:49:36.7 & +22:23:53 & 0.543 & 0.0613 (0.0632) & 121.96 & Frontier Fields \citep{Lotz2017}\\
    MACS\,J1423.8+2404 & 14:23:47.8 & +24:04:40 & 0.545 & 0.0711 (0.0603) & 260.90 & RELICS \citep{Salmon2020}\\
    \enddata
    \tablenotetext{a}{Galactic dust extinction is measured from the \citet{Schlafly2011} dust map (see Sec.~ \ref{subsec:fluxes_and_errors}). Extinction values in NCF fields are provided in parentheses.} 
    \tablenotetext{b}{Position angle of telescope V3 axis for the NIRCam and NIRISS observations of the cluster.}
\end{deluxetable*}


\begin{figure*}
\plotone{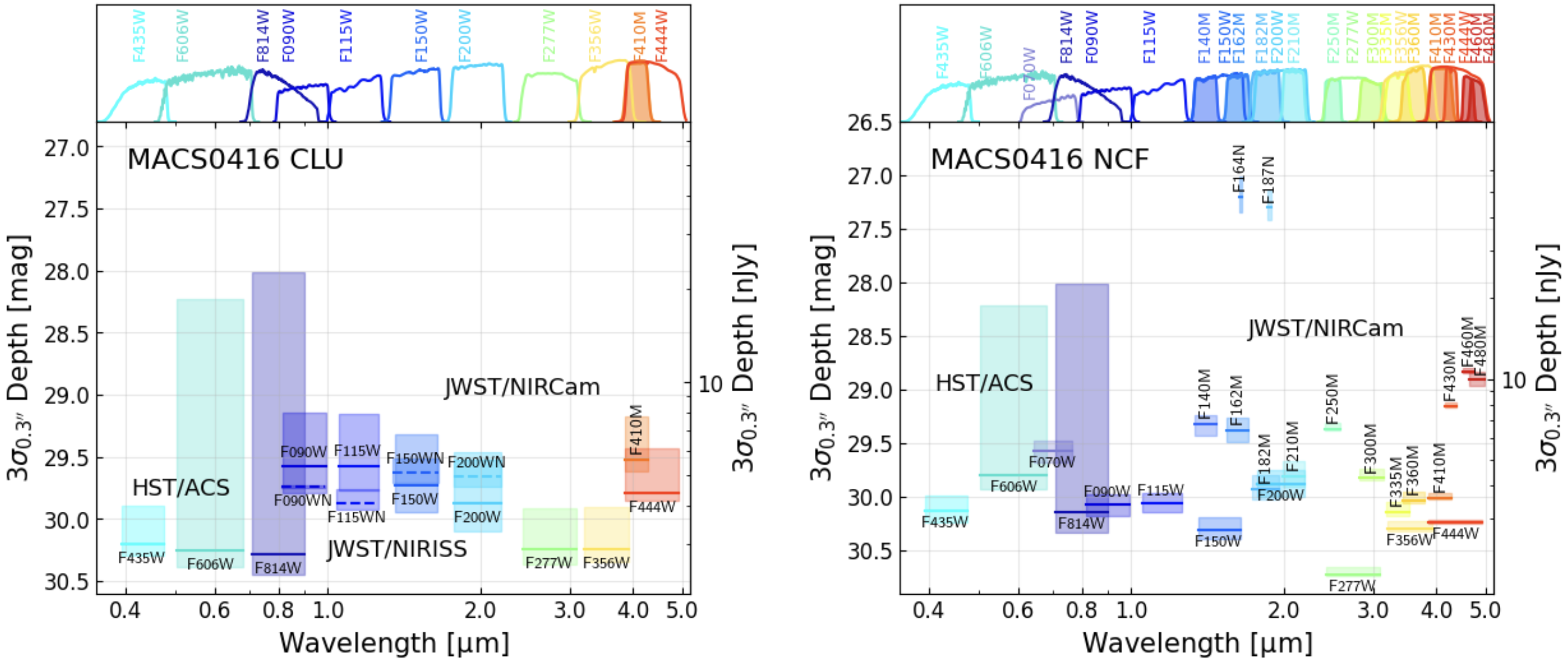}
\caption{Image depths and available filter coverage of the CANUCS observations supplemented with archival \hst\ images. The MACS0416 CLU and NCF fields are shown as the example, but depths and filter sets are almost identical across the five pointings for CLU fields, but some filters in this plot are missing in some NCF fields (see Table \ref{tbl:filts_and_fields} for available filters in each field). 3-$\sigma$ depths for a point source are quoted in $0.^{\prime\prime}3$-diameter aperture photometry. Thick solid lines (or dashed lines for NIRISS) present the median 3-sigma flux uncertainties in the catalog, with the shaded regions showing the range of 10th to 90th percentiles of them. \hst/WFC3 IR filters are not shown for clarity.}
\label{fig:Depth_FTC}
\end{figure*}




The CANUCS survey consists of coordinated parallel observations of \emph{JWST} Near-Infrared Imager and Slitless Spectrograph (NIRISS) and Near-Infrared Camera (NIRCam), with both instruments pointing at the primary “cluster” field in turn, and NIRISS/NIRCam parallel fields to either side of the main cluster field (see Figure \ref{fig:FOVs}). Observations of a given field preserve a fixed position angle between instruments and are observed close together in time, yielding a uniform point spread function (PSF) between cluster and flanking fields. We restrict our discussion to the cluster and NIRCam flanking fields (CLU and NCF, respectively), as the NIRISS flanking field (NSF), which has similar NIRISS data as the CLU field, is not included in the current data release. 

Five massive lensing clusters were selected as targets to leverage the gravitational lensing provided by the foreground cluster, allowing deeper observations of faint background galaxies. Three well-known clusters were selected from the Hubble Frontier Fields program (Abell\,370, MACS\,J0416.1-2403, MACS\,J1149.6+2223; \citealt{Lotz2017}), and one each from the CLASH (MACS\,J1423.8+2404; \citealt{Postman2012}) and RELICS (MACS\,J0417.5-1154; \citealt{Coe2019}) surveys. Figure \ref{fig:FOVs} shows the design layouts with the footprints of NIRISS and NIRCam imaging in the cluster and flanking fields overlaid on existing HST data for each target (greyscale). The NIRSpec MSA quadrant footprints are shown with blue outlines. The design, motivation, and strategy of the overall CANUCS program is detailed in \citet{Willott2022}.



\begin{figure*}[t]
\plotone{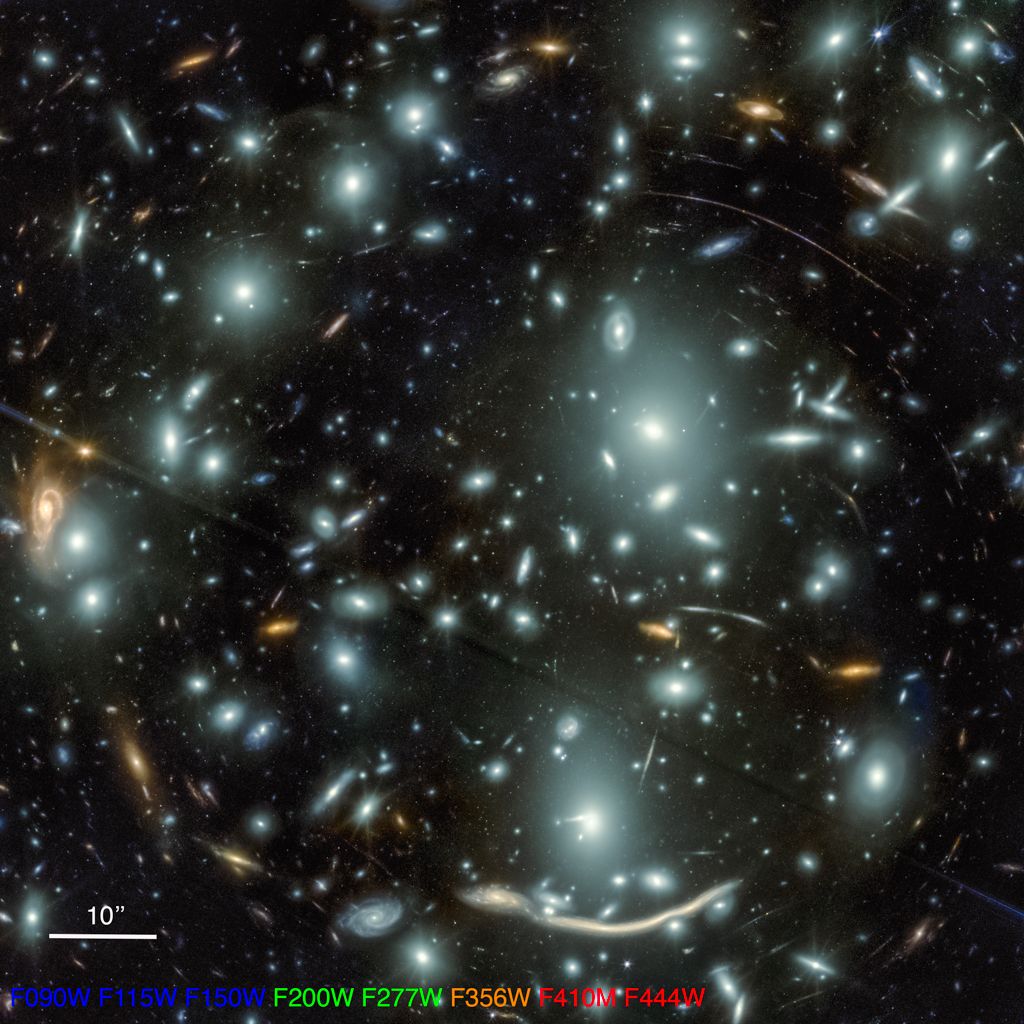}
\caption{RGB image of the center of Abell 370 CLU module B using NIRCam F090W, F115W, and F150W for the blue channel, F200W, F277W, and F356W for the green one, and F356W, F410M, and F444W for the red. North is up and East is toward the left. }
\label{fig:RGB_clu}
\end{figure*}




\begin{figure*}[t]
\includegraphics[width=\textwidth]{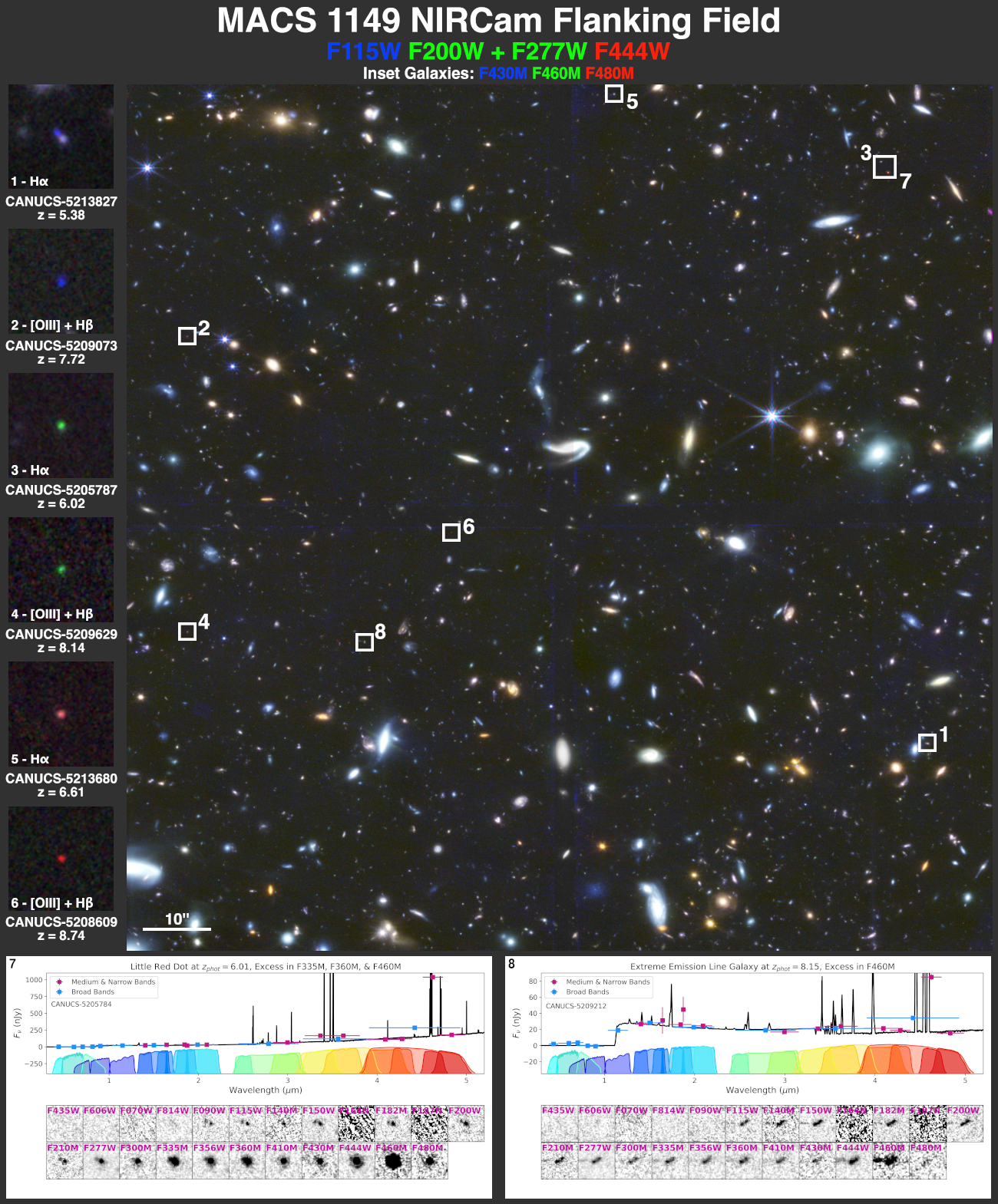}
\caption{RGB image of MACS1149 NCF in NIRCam F115W, F200W+F277W, and F444W. Example RGB cutouts highlighting the medium-band excesses (left, F430M-F460M-F480M) and SEDs (below) are shown for select emission line galaxies at 5 \lessthan \ z \lessthan \ 9. SEDs for two sources show individual cutouts and filter transmission curves, highlighting the spectral resolution of the medium bands.}
\label{fig:RGB_ncf}
\end{figure*}


\subsection{Cluster Field}\label{subsubsec:data_clu}

The primary cluster field contains overlapping observations of the central cluster with both NIRISS and NIRCam on the galaxy cluster's central region, with the NIRISS footprint overlapping one of the two NIRCam modules (see Fig. \ref{fig:FOVs}). To distinguish NIRISS filters from NIRCam filters at the same wavelength, we append an ``N" to each NIRISS filter designation.

CANUCS NIRISS imaging observations consist of 2.28ks of exposure in each of F115WN, F150WN, and F200WN, providing continuous NIRISS coverage between \tilda1-2$\mu$m. F090WN is also added for the three Technicolor fields (Abell370, MACS0416, MACS1149) with deeper 3.86ks of exposure.

CANUCS NIRCam imaging is taken in F090W, F115W, F150W, F200W, F277W, F356W, F410M, \& F444W with 6.4ks of exposure. We used the INTRAMODULEX 6-point dither pattern to fill in the gaps between the short-wave (SW) detectors. Table \ref{tbl:filts_and_fields} shows all photometric filters (\emph{JWST} and \emph{HST}) available in each field and lists 3-$\sigma$ depths for extended sources, taken in 0\farcs3 diameter apertures. Filter depths are also shown for MACS0416 for the cluster field (left panel) in Figure \ref{fig:Depth_FTC}, along with filter transmission curves. Figure \ref{fig:RGB_clu} shows an RGB color image of the Abell370 CLU field, constructed using F090W+F115W+F150W filters for the blue channel, F200W+F277W for the green, and F356W+F410M+F444W for the red. Color images of the remaining CLU fields are shown in Appendix \ref{apx:clu_RGBs}.

Several other JWST programs have observed some of the CANUCS target fields in various combinations of NIRCam Wide and Medium band filters. We do not include these other observations in our dataset, due to different pointing centers and orientations providing inhomogeneous overlap. Additionally, the PSFs for combined observations become more complex and field-dependent, due to outlier rejection of rotated diffraction spikes.


\begin{figure*}[t]
\includegraphics[width=\textwidth]{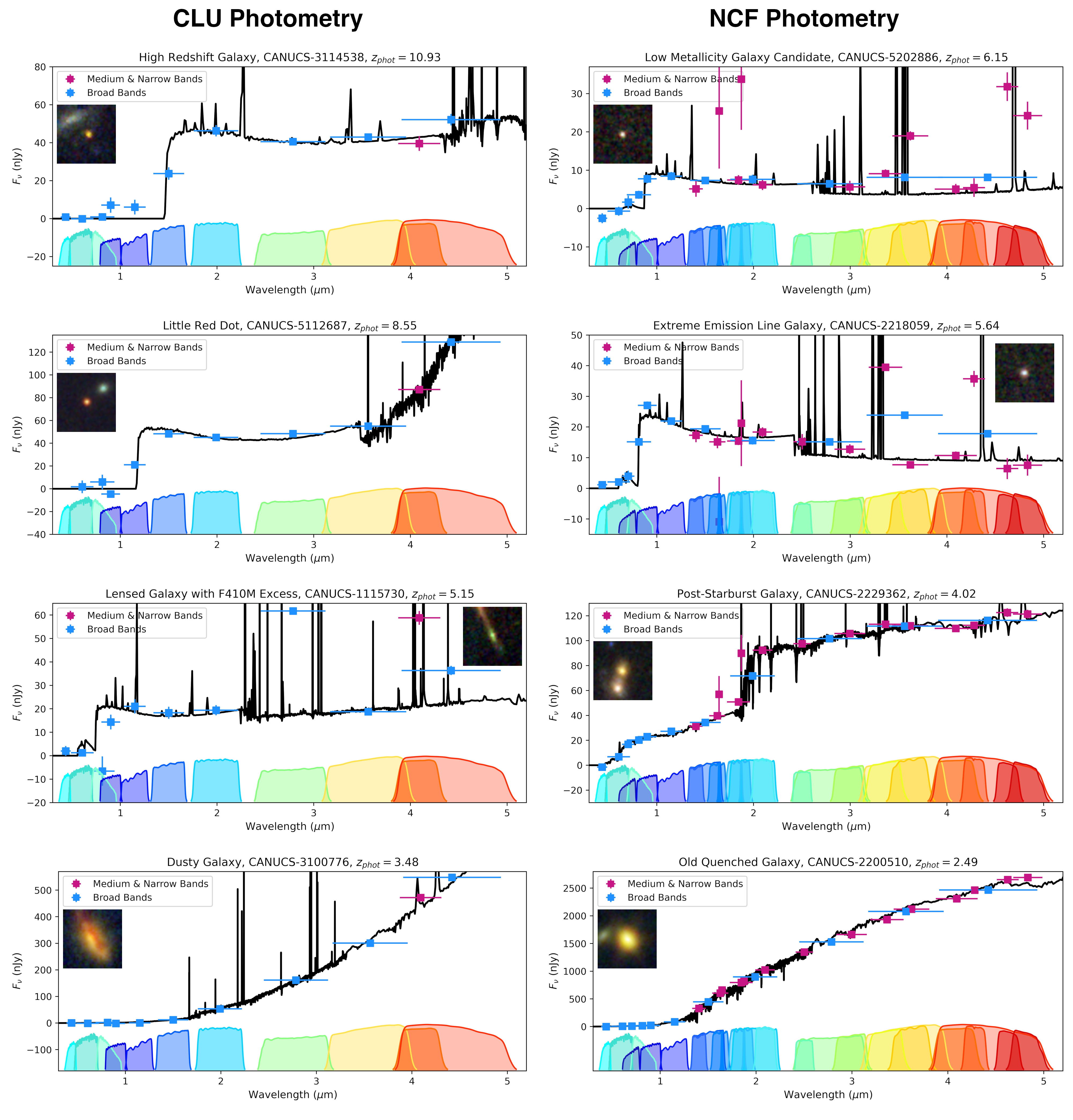}
\caption{Example SEDs in the CANUCS fields in up to 29 filters from \hst\ and \jwst. Horizontal errorbars denote filter width, with transmission curves displayed below each SED. The four galaxies on the left are in CLU fields and the four on the right are in NCF/Technicolor fields.}
\label{fig:SEDs}
\end{figure*}


\subsubsection{CLU Supporting Data}

We supplement our observations with existing \emph{HST} data available in the CANUCS fields. 

The three Frontier Fields targets (Abell370, MACS0416, MACS1149) have ultra-deep imaging in the optical with \emph{HST} Advanced Camera for Surveys (ACS) in F435W, F606W, \& F814W, and in the infrared with \emph{HST} Wide Field Camera 3 (WFC3/IR) in F105W, F125W, F140W, F160W \citep{Lotz2017}. Coverage is not continuous across the entire CANUCS observational footprints owing to the different instrumental geometries of \hst\ and \jwst\ and smaller footprint of the F435W and WFC3/IR detectors (see Figure \ref{fig:FOVs}). We also include WFC3/IR F110W \citep{Postman2012} which is slightly shallower than the Frontier Fields WFC3/IR filters (see Table \ref{tbl:filts_and_fields}).

We further supplement the Frontier Fields data with imaging from the BUFFALO program \citep{Steinhardt2020}, which expands the footprint around the original prime and parallel pointings, but with substantially reduced depths (about \tilda 2 mags shallower). Additionally, we add F475W and F625W in Abell370 CLU (PID: 11507, PI: K.Noll).

MACS0417 and MACS1423 are supported by optical data using \emph{HST/ACS} F435W, F606W, \& F814W, observed in HST-GO-16667 (PI: M.Bradac), as well as shallower data from the RELICS \citep{Salmon2020} and CLASH \citep{Postman2012} surveys, respectively.



\begin{table*}[]
    \centering
    \begin{tabular}{|c|c||cc|cc|cc|cc|cc|}
    \multicolumn{12}{c}{CANUCS Data \& Depths by Field} \\
    \hline
        Instrument & Filter & \multicolumn{2}{c|}{Abell370} & \multicolumn{2}{c|}{MACS0416} & \multicolumn{2}{c|}{MACS0417} & \multicolumn{2}{c|}{MACS1149} & \multicolumn{2}{c|}{MACS1423} \\
            & &  CLU & NCF &  CLU & NCF &  CLU & NCF &  CLU & NCF &  CLU & NCF \\
            \hline
    \hline
\multirow{7}{*}{\hst/ACS WFC}& F435W & 29.94 & 30.08 & 30.20 & 30.13 & 29.31 & -- & 29.82 & 30.11 & 29.64 & -- \\
& F475W & 29.24 & -- & -- & -- & -- & -- & -- & -- & -- & -- \\
& F606W & 29.97 & 30.26 & 30.35 & 29.87 & 29.71 & -- & 30.10 & 30.20 & 29.82 & -- \\
& F606W* & (28.17) & (28.14) & (28.38) & (28.36) & -- & -- & (28.51) & (28.24) & -- & -- \\
& F625W & 28.14 & -- & -- & -- & -- & -- & -- & -- & -- & -- \\
& F814W & 30.25 & 30.18 & 30.42 & 30.30 & 28.23 & -- & 30.26 & 30.23 & 29.02 & -- \\
& F814W* & (27.98) & (28.01) & (28.16) & (28.16) & -- & -- & (28.24) & (28.15) & -- & -- \\
\hline
\multirow{2}{*}{\hst/WFC3 UVIS}& F438WU & -- & -- & -- & -- & -- & 28.56 & -- & -- & -- & 28.45 \\
& F606WU & -- & -- & -- & -- & -- & 29.27 & -- & -- & -- & 29.20 \\
\hline
\multirow{8}{*}{\hst/WFC3 IR}& F105W & 29.91 & 30.10 & 30.06 & 30.38 & 27.95 & -- & 30.14 & 30.17 & 28.72 & -- \\
& F105W* & (27.75) & (27.74) & (27.86) & (27.99) & -- & -- & (27.95) & (27.87) & -- & -- \\
& F110W & 28.90 & -- & 28.45 & -- & -- & -- & 29.12 & -- & 28.51 & -- \\
& F125W & 29.50 & 29.67 & 29.64 & 29.86 & 27.28 & -- & 29.96 & 29.64 & 28.55 & 28.61 \\
& F125W* & (27.78) & (27.75) & (27.89) & (27.99) & -- & -- & (27.78) & (27.88) & -- & -- \\
& F140W & 29.34 & 29.60 & 29.65 & 29.91 & 27.34 & -- & 29.39 & 29.63 & 28.51 & -- \\
& F160W & 29.50 & 29.69 & 29.48 & 29.89 & 27.68 & -- & 29.71 & 29.61 & 28.39 & 28.41 \\
& F160W* & (27.42) & (27.37) & (27.42) & (27.58) & -- & -- & (27.52) & (27.51) & -- & -- \\
\hline
\multirow{4}{*}{\jwst/NIRISS} & F090WN & 29.56 & -- & 29.74 & -- & -- & -- & 29.74 & -- & -- & -- \\
& F115WN & 29.64 & -- & 29.87 & -- & 29.68 & -- & 29.68 & -- & 29.81 & -- \\
& F150WN & 29.50 & -- & 29.63 & -- & 29.50 & -- & 29.53 & -- & 29.57 & -- \\
& F200WN & 29.48 & -- & 29.66 & -- & 29.61 & -- & 29.53 & -- & 29.65 & -- \\
\hline
\multirow{22}{*}{\jwst/NIRCam}& F070W & -- & 29.48 & -- & 29.57 & -- & -- & -- & 29.53 & -- & -- \\
& F090W & 29.47 & 30.18 & 29.58 & 30.07 & 29.39 & 29.99 & 29.51 & 29.94 & 29.51 & 30.04 \\
& F115W & 29.47 & 29.95 & 29.57 & 30.06 & 29.38 & 29.99 & 29.48 & 30.17 & 29.58 & 29.99 \\
& F140M & -- & 29.22 & -- & 29.32 & -- & 29.25 & -- & 29.23 & -- & 29.28 \\
& F150W & 29.56 & 30.13 & 29.74 & 30.31 & 29.56 & 30.14 & 29.61 & 30.38 & 29.73 & 30.22 \\
& F162M & -- & 29.31 & -- & 29.38 & -- & 29.26 & -- & --- & -- & 29.35 \\
& F164N & -- & 27.14 & -- & 27.20 & -- & -- & -- & 27.21 & -- & -- \\
& F182M & -- & 29.81 & -- & 29.93 & -- & 29.88 & -- & 29.87 & -- & 29.94 \\
& F187N & -- & 27.27 & -- & 27.30 & -- & -- & -- & 27.24 & -- & -- \\
& F200W & 29.81 & 29.80 & 29.87 & 29.89 & 29.74 & -- & 29.83 & 29.88 & 29.86 & -- \\
& F210M & -- & 29.64 & -- & 29.81 & -- & 29.69 & -- & 29.76 & -- & 29.77 \\
& F250M & -- & 29.25 & -- & 29.37 & -- & 29.10 & -- & --- & -- & 29.19 \\
& F277W & 30.08 & 30.53 & 30.24 & 30.72 & 30.12 & 30.53 & 30.21 & 30.77 & 30.12 & 30.64 \\
& F300M & -- & 29.71 & -- & 29.83 & -- & 29.58 & -- & 29.69 & -- & 29.59 \\
& F335M & -- & 30.05 & -- & 30.14 & -- & 29.90 & -- & 30.09 & -- & 29.93 \\
& F356W & 30.09 & 30.21 & 30.25 & 30.30 & 30.15 & -- & 30.19 & 30.22 & 30.21 & -- \\
& F360M & -- & 29.96 & -- & 30.03 & -- & 29.89 & -- & 30.06 & -- & 29.94 \\
& F410M & 29.46 & 29.87 & 29.53 & 30.01 & 29.47 & 29.88 & 29.51 & 29.88 & 29.50 & 29.80 \\
& F430M & -- & 29.21 & -- & 29.15 & -- & -- & -- & 29.17 & -- & -- \\
& F444W & 29.67 & 30.05 & 29.79 & 30.24 & 29.84 & 30.20 & 29.78 & 30.19 & 29.84 & 30.25 \\
& F460M & -- & 28.83 & -- & 28.83 & -- & -- & -- & 28.80 & -- & -- \\
& F480M & -- & 28.86 & -- & 28.91 & -- & -- & -- & 28.81 & -- & -- \\
\hline
    \end{tabular}
    \caption{Effective 3$\sigma$ catalog depths (AB mag) in 0$.\!\!^{\prime\prime}$3 diameter apertures. For \hst\ filters where we supplement with the BUFFALO program (filters with asterisk), depths are measured in the HFF deep footprint and the shallower BUFFALO region separately. Values in parentheses are the depths in BUFFALO-only regions. Depths are determined by measuring the distribution of fluxes in apertures in empty sky regions on the F444W PSF-convolved images. Depths are provided for all available filters in each field.}
    \label{tbl:filts_and_fields}
\end{table*}


\subsection{NIRCam Flanking Field}\label{subsubsec:data_ncf}

The NIRCam flanking fields (NCF) utilize NIRCam observations in conjunction with ancillary \emph{HST} data, and employ a combination of 14 wide and medium band filters in Cycle 1. The NCF does not have NIRISS imaging or WFSS, nor extensive NIRSpec prism follow-up. In each NCF field, as part of the CANUCS observations, the following filters are available: wide band photometry in F090W, F115W, F150W, F277W, \& F444W, and medium band photometry in F140M, F162M, F210M, F250M, F300M, F335M, F360M, \& F410M. Most filters have 10.3ks exposure times, excepting F140M, F162M, F250M, \& F300M which have 5.7ks exposures. In MACS1149, a program definition error resulted in F162M \& F250M not being observed, with deeper exposures being obtained for F150W \& F277W in this particular field. 

Note that since Cycle 1 NCF observations were conducted in coordinated parallel with NIRISS CLU observations, available dithers in NCF were set by the NIRISS WFSS dither pattern on CLU, resulting in a smaller dither pattern which does not fill the cross-shaped detector gaps of the NIRCam SW channel. As a result these detector gaps are present in the SW mosaics in the NCF field. 

The Cycle 2 "JWST in Technicolor" program (ID: 3362, PI: A.Muzzin) adds 8 wide, medium, and narrow band filters to three NCF fields (Abell370, MACS0416, MACS1149): F070W, F164N, F187N, F200W, F356W, F430M, F460M, \& F480M. Most filters have \tilda10ks exposures. Together with Cycle 1, these programs yield uniquely rich fields containing all NIRCam wide and medium band imaging from 0.7-4.4$\mu$m (excepting 2 medium band filters in MACS1149, as noted above).

Figure \ref{fig:RGB_ncf} shows an RGB color image of the MACS1149 NCF field, constructed with F150W, F200W+F277W, and F444W wide bands. Inset thumbnails in the F430M, F460M, and F480M highlight examples of galaxies with strong [OIII]+Hbeta or [NII]+Halpha lines captured in the medium bands. The photometric filters and depths available by field are listed in Table \ref{tbl:filts_and_fields}. Filter depths are also shown for MACS0416 NCF (right panel) in Figure \ref{fig:Depth_FTC}, along with filter transmission curves. Some example SEDs are shown in Figure \ref{fig:SEDs} to demonstrate the exquisite sampling in the CANUCS NCF fields. Horizontal errorbars denote filter width, with bandpasses displayed underneath.

\subsubsection{NCF Supporting Data}

As in the CLU field, we supplement our observations with existing \emph{HST} data. 

The three Frontier Fields targets (Abell370, MACS0416, MACS1149) have the same imaging in the prime and parallel pointings (roughly overlapping the CLU/NCF fields, respectively), however F110W is not available in the flanking field. We use all 7 bands in the flanking field: F435W, F606W, F814W, F105W, F125W, F140W, F160W.

MACS0417 and MACS1423 are supplemented with \emph{HST}/WFC3 UVIS imaging in F438W \& F606W (HST-GO-16667; PI: M.Bradac). Note that all UVIS filters end with the letter designation ``U" (e.g. F438WU).

\subsection{NIRSpec Spectroscopy}\label{subsec:nirspec}

As part of the CANUCS program (ID: 1208), follow-up multi-object spectroscopy using the Micro-Shutter Assembly (MSA, \citealt{Ferruit2022}) of NIRSpec was carried out. These observations were planned using target selection, astrometry and target acquisition reference objects from the NIRCam and NIRISS imaging. Due to field observability and the need for time after imaging acquisition for MSA configuration planning, only two of these five `Cycle 1' observations were executed during Cycle 1 and three during Cycle 2. The low-resolution ($R\sim100$) prism was used to capture the full near-infrared spectrum from 0.6 to 5.5 $\mu$m and provide a high level of multiplex. 

Due to the high galactic latitudes of these fields there are relatively few stars bright enough to be used for MSA target acquisition. Therefore we opted to include compact galaxies that did not show wavelength-dependent structure in the set of acceptable MSA target acquisition (MSATA) reference objects. The positions of all MSATA objects were remeasured on the NIRCam F200W image (c.f. the imaging catalogs use the position in the $\chi$-mean detection image, see Section \ref{subsec:source_detection}). Using the NIRCam positions for MSATA objects allows us to fix the astrometry to the \jwst\ imaging. This is particularly important for stars with proper motion offsets between the \jwst\ and \hst\ epochs, where the detection image incorporates data from significantly different epochs.

For most clusters, three MSA configurations were observed in the CLU field to enable the large gaps between the four MSA quadrants to be dithered across. For MACS0417 we used two configurations in the CLU field and one in the NCF field to enable spectroscopy of targets selected from the set of 14 NIRCam Wide and Medium filters. Each configuration was observed for 2.9\,ks with the targets nodded along a 3-shutter slitlet. A small set of high-priority targets were observed in multiple MSA configurations, providing total exposure time up to 8.7\,ks.  

Target selection was performed using a diverse set of interesting targets selected from the NIRCam, NIRISS and \hst\ imaging. The MSA Planning Tool was used to set up the configurations with targets allocated `weights' depending on their relative priorities. The priority scheme was adapted over the course of the year during which the different fields were observed, as lessons learnt from prior fields were incorporated. In approximate order of priority the main target selection classes were: (1) $z>7.5$ galaxies; (2) galaxies at $5<z<7.5$ showing emission line excesses in F410M or F444W photometry; (3) high-redshift quiescent or dusty galaxies (including {\it Little Red Dots}); (4) Lyman break galaxies at $5<z<7.5$; (5) strongly-lensed galaxies with magnification $\mu>10$ or multiply-imaged; (6) galaxies with particularly red or blue F277W-F356W colors indicating emission line excesses; (7) galaxies with photometric redshifts $z>3$; (8) galaxies in the lensing clusters.








%


%

\subsection{Data Products} \label{subsec:data_products}


This first CANUCS data release includes the following data products from Cycle 1 \& 2 observing programs:

\begin{itemize}
    \item Background-subtracted images in the CLU and NCF fields, with bright cluster galaxies modeled out in the CLU field (see \S \ref{sec:data_reduction} \& \ref{subsec:bcgs_icl} in particular for bCG/ICL modeling and subtraction);
    \item PSFs, convolution kernels, and F444W PSF-matched images (see \S \ref{subsec:psf_matching});
    \item Photometric catalogs including fluxes in a variety of apertures measured on PSF-matched images, photometric and spectroscopic redshifts and detection, segmentation and RMS maps for all filters (see \S \ref{subsec:source_detection}, \S \ref{subsec:fluxes_and_errors} \& \S \ref{sec:cat_properties});
    \item NIRSpec prism spectra taken in all 5 cluster fields and one NCF field (see \S \ref{subsec:nirspecproc});
    \item Stellar population properties for photometric catalogs (see \S \ref{sec:SPS});
    \item Lensing maps using the latest models derived using constraints from CANUCS (see \S \ref{sec:lensing});
    \item Demonstration notebooks  for working with catalogs, visualizing cutouts/SEDs\footnote{https://niriss.github.io/data.html}
\end{itemize} 

NIRISS wide-field slitless spectra will be released in a future CANUCS data release. 

The remainder of this paper will discuss in detail the preparation and processes used in creating these data products.

%


%

\section{Data Reduction} \label{sec:data_reduction}

\subsection{Imaging Pipeline Processing} 
\label{subsec:imagepipe}

Processing of imaging data from \hst\ uses the \texttt{grizli} code \citep{Brammer2019} and follows the same steps that are well-documented in the literature (e.g. \citealt{Kokorev2022}). In this section we focus on the processing of \jwst\ imaging data.

The processing of imaging data from the two instruments NIRCam and NIRISS follows mostly similar procedures. The up-the-ramp exposures are converted into count rate images using the STScI \texttt{jwst} stage 1 pipeline. First, the data quality arrays are initialized. The bad pixel masks used in this step are the best available at the time, either from the STScI Calibration Reference Data System (CRDS) or custom versions we generated ourselves incorporating bad pixels from the most up-to-date dark calibration observations. We observed a small increase in the number of hot pixels between commissioning and Cycle 2, making the use of recent dark calibration observations important. After identifying bad pixels the next step is saturation flagging. For NIRISS, we use custom saturation reference files with a lower saturation threshold to mitigate the effects of charge migration \citep{Goudfrooij2024}. 

Next, the CRDS superbias reference frame is subtracted from the ramps and a linearity correction is applied. The pipeline persistence correction step is skipped since the robustness of this correction has not been demonstrated. The best available dark reference cube (either CRDS or our own) is subtracted off each group of the ramps. At this point, we exit the pipeline for NIRISS and apply the custom \texttt{columnjump}\footnote{https://github.com/chriswillott/jwst} step that corrects for random DC offsets in some columns (in the original detector coordinate format; these are rows in the STScI coordinate format) in each group of the ramp. The pipeline jump step is performed next to correct for cosmic ray impacts. We use a jump rejection threshold of $5\sigma$ and flag the four neighboring pixels for jumps of more than 10 counts. We do not include the \texttt{expand\_large\_events} option to flag snowballs, since at the time of our processing we found that the \texttt{snowblind}\footnote{https://github.com/mpi-astronomy/snowblind} package provides superior results. We use a custom modification of \texttt{snowblind} that allows the correction of negative snowball residuals on the final count rate images. Next, we perform a custom persistence flagging step that flags pixels as bad if they reach a level close to saturation in previous exposures. A variable time constant is adopted, with groups within 900\,s of the end of the offending exposure flagged as bad for NIRISS and most NIRCam detectors. The NIRCam detectors A3, B2, B3, and B4 show worse persistence behaviour and for these the time constant is increased to 1800\,s. Finally, the ramp slope fitting step is performed to derive count rate images. 

Image processing at the stage 2 level of the STScI pipeline is accomplished using a combination of standard and custom steps. First the world coordinate system is assigned for each integration based on the offset from the known guide star position. The relative positions of different dithers are then cross-matched and aligned. Absolute astrometry is achieved by cross-matching to Gaia DR3 stars proper-motion-corrected to the date of observation. We build a reference catalog using the NIRCam F444W image and ensure that all \jwst\ and \hst\ filters are consistently aligned. 1/{\it f} noise is removed using our own custom code in \texttt{grizli}, taking care to avoid any bias in cluster fields that have highly-structured backgrounds. The individual integrations are flat-fielded and a constant background is subtracted to account for varying DC levels in the detectors. For NIRCam, `wisps' are removed by fitting and subtracting templates built from a large set of sky frames. For NIRISS, the 'light saber' structure is modeled as in \cite{Doyon2023} and removed assuming it is dominated by zodiacal light. Photometric calibration is performed using the latest values available from CRDS. For NIRISS, ghosts are masked using the NIRISS ghost detection code\footnote{https://github.com/spacetelescope/niriss\_ghost}.

Mosaic images are built by drizzling each exposure per filter onto a common pixel grid with the \texttt{grizli} code \citep{Brammer2019}. We utilize a pixel scale of 40 milliarcsec per pixel for all \jwst\ and \hst\ images. For the NIRCam SW filters that have a native pixel scale of 33 milliarcsec we produce an extra data product of a mosaic image on a 20 milliarcsec per pixel scale. These images are intended for use in morphological analysis where the finer sampling improves the results of model fitting. During the mosaicing process, bad pixels and outliers are masked. Whilst this process generally works well, a small number of outliers are visible in the final mosaics, mostly in the corner and edge regions where very few dithers hamper the automated detection of outliers. Our final images are scaled to a flux value per pixel of 1\,nJy, which establishes an AB magnitude zeropoint of 31.4. Each science mosaic has an accompanying weight image providing the relative weights of each pixel, accounting for readnoise and Poisson noise from the sky and sources.


%

\subsection{Bright Cluster Galaxies \& Intra-Cluster Light Modelling \& Subtraction} \label{subsec:bcgs_icl}

The choice of rich galaxy clusters as the primary targets for CANUCS enables us to take advantage of the magnification effect of gravitational lensing in order to study intrinsically fainter background sources than equivalent exposure times in blank fields allow. One of the primary challenges associated with this choice is the presence of bright cluster member galaxies and ICL which significantly contaminate the photometry of background sources if not accounted for. To address this issue, we construct models for bright cluster and foreground galaxies as well as the diffuse ICL which we subtract from the reduced images, resulting in clean residual images with a well-behaved background on which we perform photometry. This process significantly improves the effective area of the images in which we can perform reliable photometry, particularly close to the cluster centers. Our strategy is based on that introduced by \citet{Shipley2018} and explained in detail in Section 3 of \citet{Martis2024}. The general strategy is to construct isophote models for individual cluster and foreground galaxies and remove remaining diffuse ICL by applying median filtering of the background. Following \citet{Shipley2018}, we refer to galaxies within our model as ``bCGs" (bright Cluster Galaxies) as distinct from the single, central BCG (Brightest Cluster Galaxy), noting again that in a few cases this includes foreground galaxies as well. In this section, we outline the most important steps in this process and show the results for this data release.

\begin{figure*}[t]
\plotone{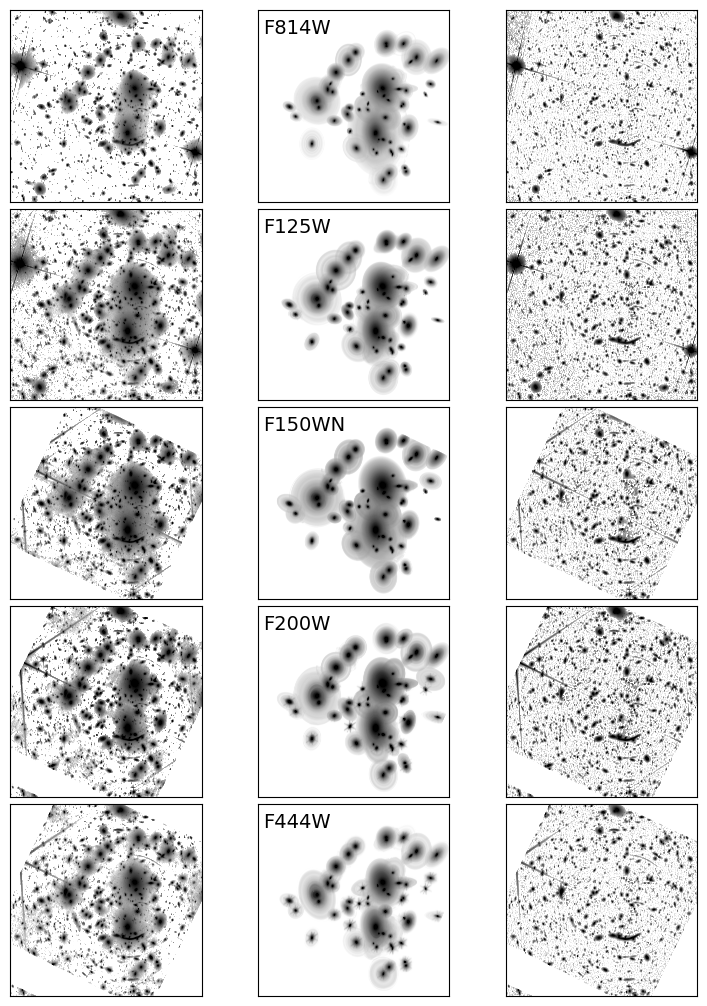}
\caption{Original, drizzled science image (left), bCGs model (center), and residual image (right) for each indicated filter for the Abell 370 cluster. The filters shown highlight a range of instruments: F814W (HST/\emph{ACS}), F125W (HST/\emph{WFC3/IR}), F150WN (NIRISS), F200W (NIRCam SW), and F444W (NIRCam LW).}
\label{fig:bcg_gallery}
\end{figure*}

\begin{figure*}[t]
\plotone{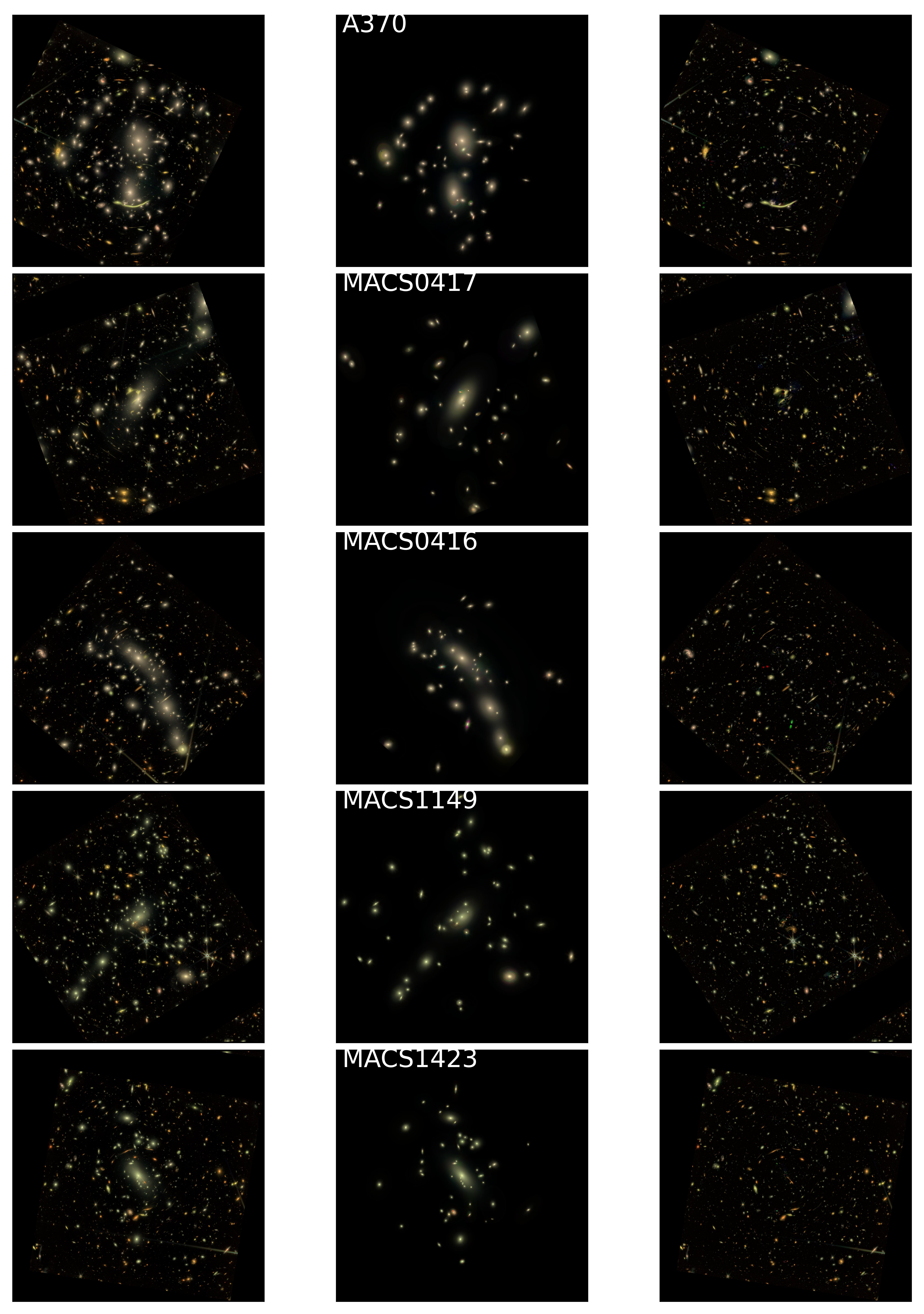}
\caption{Color images constructed from F277W, F356W, and F444W for the central region of each cluster. As in Figure \ref{fig:bcg_gallery}, left column shows the original image, the center the bCGs model, and the right the residual image}
\label{fig:bcg_color}
\end{figure*}

\subsubsection{bCG Selection} \label{subsubsec:bcgs_select}

The first step in constructing our bCG model is to select which galaxies to include. For the three CANUCS clusters which were previously observed by the Hubble Frontier Fields program (Abell370, MACS0416, and MACS1149), we adopt the same list of cluster galaxies as \citet{Shipley2018} since we are performing a similar process. For the remaining two clusters (MACS0417, MACS1423) we construct our own list of cluster and foreground galaxies to include based on similar principles. Namely, we prioritize galaxies which are bright, contaminate large areas, and fall within the footprint of the NIRCam module covering the cluster center. Due to the limitations of the adopted isophote modeling process, some spiral galaxies are rejected because they leave behind significant residuals. We made use of RGB images and photometric redshift catalogs provided by CLASH \citep{Molino2017} and RELICS \citep{Coe2019} to aid in this process which was started before JWST data were in hand. The number of galaxies modeled in each field ranges from 42 to 77.

\subsubsection{Preliminary Background Subtraction} \label{subsubsec:prelim_background}

We note that before beginning any further processing of the drizzled science images, we bin the images by a factor of two to a coarser pixel grid of $0.08''$ per pixel. This significantly speeds up the isophote modeling step and has no effect on the quality of final models when they are interpolated back to the original resolution. 

The presence of bright ICL features in the images causes the default pipeline background subtraction to overestimate the background level, leading to negative sky levels on the outskirts of the images and significant ICL remaining in the central regions. Before performing modeling of individual bCGs, it is necessary to bring the background level of the images closer to a flat level. In the central regions, failure to do so results in ICL being incorporated in the isophote models as they extend to radii well beyond the extent of the galaxy in an attempt to capture all of the light in the image. Conversely, in the outer regions, the isophote models attempt to add negative light to the models in order to bring the over-subtracted background up to zero in the residual. To avert these issues we use the spatially varying \textsc{photutils} Background2D function with a box size of 20x20 pixels and filter size of 11 pixels (0\farcs08 pixels) as a first approximation to remove large-scale ICL and correct the over-subtracted background in the outer regions of the image. 

\subsubsection{bCG Modelling} \label{subsubsec:bcgs_modelling}

To act as a base for the modeling process, we perform preliminary source detection and segmentation using \textsc{photutils}. The segmentation image is used as an initial mask for neighboring sources when performing the modeling. Additionally, the morphological parameters from the source detection are used as initial conditions for the isophote models. Frequently these need to be adjusted manually in order to achieve optimal fits, but given the large number of galaxies included when modeling five clusters, some automation is required. The galaxy size from the segmentation image is also used as a guide to determine the possible spatial extent of the model (though we note we allot significantly larger sizes than the source detection to allow the low surface brightness outskirts to be captured).

We utilize the \textsc{photutils} elliptical isophote fitting framework in order to perform our fits. Beginning with the brightest galaxy according to the preliminary source detection, we extract a cutout centered on the galaxy to be modeled. This image is fit using an isophote model and the segmentation image to mask out neighboring sources. Due to the crowded nature of the cluster environment and the large spatial extent of the cluster galaxies, numerous sources often overlap with the model, particularly in the case of the BCG. In order to create a more refined mask to apply to the modeling, we perform a second round of source detection on the residual cutout image. This enables the detection of sources close to the center of the model and more reliable deblending of neighboring bright sources. This new segmentation image is used as a new mask and the isophote modeling is repeated, usually resulting in a significantly improved model. We refer to this resulting model as the ``iteration 0" model. This model is subtracted from the original full image before moving on to the next brightest bCG, such that as we progress to fainter galaxies in our model, there is less contamination from the brightest galaxies interfering with the modeling. This step is completed when we have an iteration 0 model for every bCG. In practice, we find that even with this careful process, the masks need to be manually adjusted for optimal fitting. To keep the modeling process from becoming untenable, we choose to create one mask for each instrument: F814W for \textit{Hubble} ACS, F160W for WFC3, F200W for NIRCam, and F200W for NIRISS. When combined with sigma clipping during the isophote modeling process, these masks sufficiently account for enough of the differences between filters (including diffraction spikes from bright stars) to achieve optimal models.

The resulting iteration 0 residual approximates the bCGs' light profiles well, but several improvements can be made. We observe two issues relating to the behavior of the point spread function (PSF) in the residual images. First, the central regions of the models exhibit strong alternating positive and negative residuals extending $2-3''$ in the shape of the PSF likely due to the presence of a point source (possibly either a nuclear star cluster or AGN). Additionally, diffraction spikes from this PSF feature extend to even larger distances, particularly in the WFC3 and NIRCam long-wave filters, contaminating the photometry of nearby objects. Both of these features are visible in the bCG-subtracted images shown in \citet{Shipley2018} and \citet{Weaver2024}. Furthermore, the iteration 0 models, especially for the BCG, unavoidably suffer from contamination by fainter bCGs even when applying the refined masks due to source overlap.

We overcome both of these issues by iterating our modeling process with an additional prescription to account for PSF effects. Again beginning with the BCG, we insert only the model for this single galaxy back into the residual image. We make a new cutout around this galaxy which now contains no light from other bCGs and deconvolve it using the \textsc{skimage} restoration package \citep{VanDerWalt2014}. We perform the isophote modeling on the deconvolved image, then reconvolve the model with the PSF. The resulting model now includes the PSF features described above and results in significantly better residuals in terms of both spatial extent and residual flux \citep[see Figure 1 of][]{Martis2024}. This enables reliable photometry of background sources very close to the bCG centers. As in the iteration 0 step, we progress through our list of bCGs in order of decreasing brightness, adding a single galaxy back into the residual and remodeling it in the absence of contaminating light from other bCGs. When all bCGs have been thus remodeled, we have an iteration 1 model for the full cluster. 

This entire process is iterated a total of ten times in order to achieve convergence in the models. The final model is the median of the iterated models (excluding iteration 0). We use the median because although later iterations generally converge, small changes to neighboring sources sometimes cause instabilities in the fitting algorithm leading to failed fits. The final model and residual image are visually inspected and any unsatisfactory models are manually adjusted and individually re-run through the iteration process. Once a satisfactory total bCG model is obtained, the model and residual image are interpolated back to the original $0.04''$ pixel grid (or $0.02''$ pixel grid in the case of the higher resolution NIRCam SW images).

\subsubsection{Final Image Adjustments} \label{subsubsec:final_adjustments}

After the light from bCGs has been accurately modeled, we can better account for the remaining diffuse ICL. We again use the \textsc{photutils} Background2D median background function, but this time in an iterative fashion. We begin with the residual image obtained after subtracting the bCG model only (the preliminary background subtraction is not included). We first perform source detection on the residual image with a low detection threshold ($2.5\sigma$), mask detected sources, and calculate a background using a box size of 12 pixels and filter size of 5 pixels. This captures the large scale ICL structure and corrects the over-subtraction by the default pipeline in outer regions. However, some smaller regions of the ICL are significantly brighter and are picked up by source detection. In order to account for these features we repeat source detection after subtracting the first background with a higher source detection threshold dependent on the filter (since ICL brightness varies with wavelength). We dilate the segmentation map to ensure that the outer regions of galaxies are still included with the higher detection threshold, then repeat the background subtraction which now removes brighter ICL features without interfering with the outskirts of other galaxies. The final residual has a well-behaved, flat background across the entire image.

Even after this careful process, some unavoidable residuals caused artifacts when generating our images convolved to a common PSF (see Sec~\ref{subsec:psf_matching}). We removed this final issue by manually setting the residual image to zero in a small region (radius 15 pix) at the center of each bCG, similar to the way saturated stars are treated. The final result is our 'bgsub-sci' image. The same process was applied to the $0.02''$ per pixel NIRCam short-wave images with corresponding changes to parameters quoted in pixel units. In this data release we provide the original drizzled science images, the bcgmodel-sci, and the bgsub-sci image for each filter. The user may calculate the total removed background by taking the difference of the original and bcgmodel-sci + bgsub-sci images. Here we show two visual indications of the bCG subtraction quality. Figure \ref{fig:bcg_gallery} shows from left to right, an original image, bcgmodel-sci image, and bgsub-sci image for five different instrument/filter combinations for the Abell370CLU. Figure \ref{fig:bcg_color} shows a color version of original, bCG model, and bcgs\_out images for each of the five cluster fields. The images demonstrate that the models properly match the bCG colors, indicating accurate modeling in all filters.

\subsubsection{NCF Background Subtraction}

Some of our flanking field observations include bright cluster or foreground galaxies, but we have elected to perform the modeling process only on the cluster fields. We have however performed our secondary background subtraction on the flanking field images in order to maintain consistency in our photometry. This has the benefit of removing some image defects including persistence. Even though we perform no bCG modeling, these background-subtracted images are labeled as bgsub-sci in the data release to preserve a consistent file naming scheme.  

\begin{figure}
\includegraphics[width=0.5\textwidth]{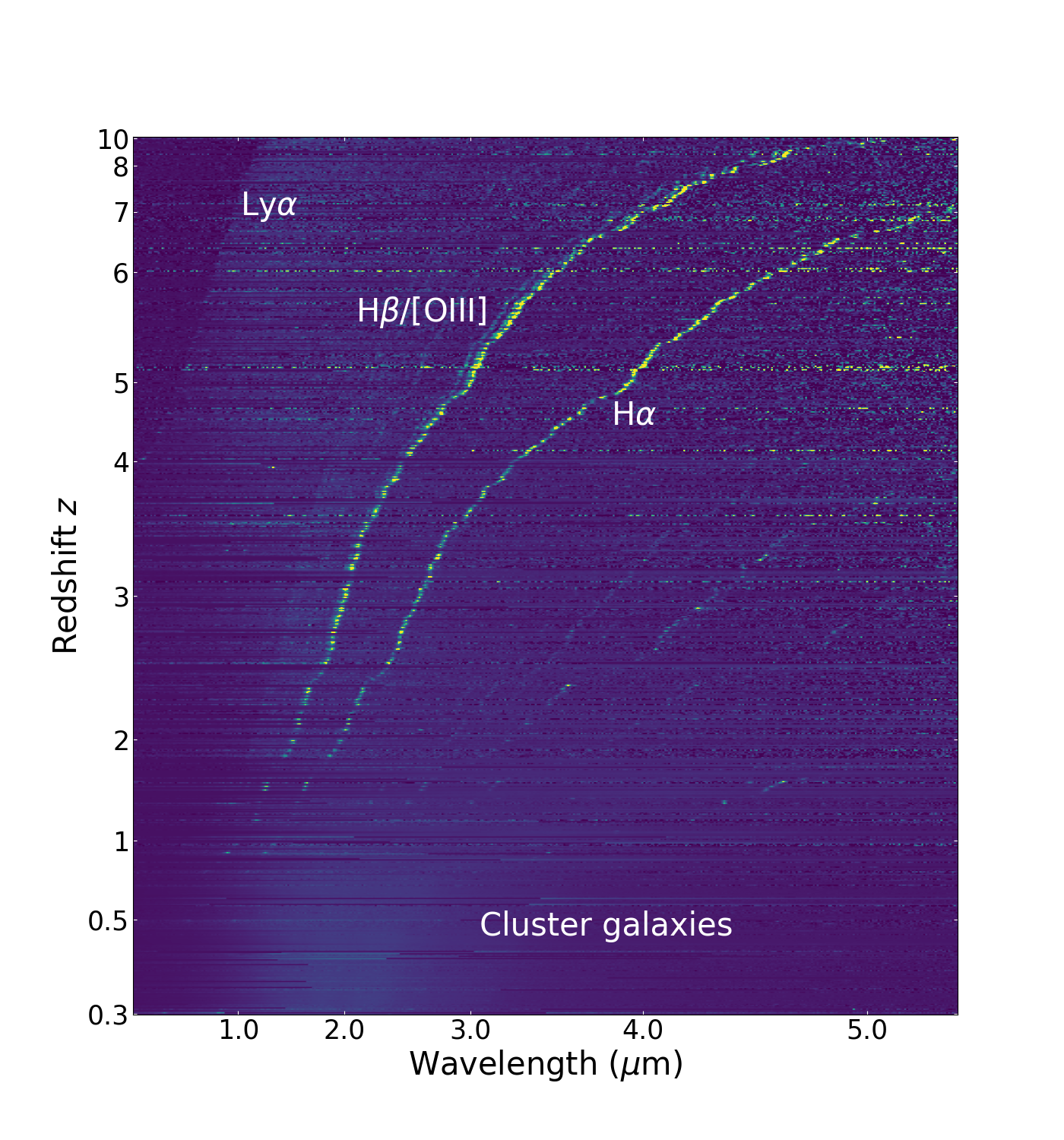}
\caption{Vertical stack of 733 CANUCS-NIRSpec prism spectra ordered by redshift increasing from bottom to top. The most prominent emission lines are labeled. Also labeled is the dark wedge caused by the Lyman-$\alpha$ break at high-redshift and the evolved, continuum-dominated galaxies at low redshift, predominantly in the lensing clusters.}
\label{fig:nirspec2D}
\end{figure}

\subsection{NIRSpec Spectroscopy Processing}\label{subsec:nirspecproc}
Details of the NIRSpec processing are given in \cite{Desprez2024} and also largely follow the procedures described in \cite{Heintz2025}. Initial processing uses the STScI \texttt{jwst} stage 1 pipeline with custom snowball and 1/{\it f} noise correction. The read-noise uncertainty array of each exposure is adjusted to match the noise measured in un-illuminated pixels. The \texttt{jwst} stage 2 pipeline is run up to the photometric calibration step. Further processing is performed using the \texttt{grizli} \citep{Brammer2019} and \texttt{msaexp}  \citep{Brammer2022msaexp} packages. Wavelength calibration uses a correction for the known intra-shutter offset along the dispersion direction. The spectral background is removed using the standard nodded background subtraction. One-dimensional spectra are extracted using a wavelength-dependent optimal extraction that accounts for the increase in PSF FWHM with wavelength.

Redshifts are determined via automated fitting using \texttt{msaexp}. This is followed by visual inspection and interactive fitting for those galaxies that required it. In most cases, redshifts are determined from the observed wavelengths of two or more emission lines (quality grade \texttt{Z\_Q\_REF}=1). In some galaxies only one emission line is detected, but a confident redshift can be assigned based on combining the line wavelength with the photometric redshift and/or a measured spectral break (\texttt{Z\_Q\_REF}=2). Our final class of confident NIRSpec redshifts (\texttt{Z\_Q\_REF}=3) are from strong breaks (Lyman, Balmer, D4000) and/or continuum features (blended atomic/molecular absorption lines) in the spectra that match to galaxy templates available in \texttt{msaexp}. Redshifts are not reported when the fitting resulted in ambiguous or tentative redshifts (\texttt{Z\_Q\_REF}$>3$). Therefore, only high-quality NIRSpec redshifts are included in this data release. 

In total, we obtain 747 NIRSpec spectroscopic redshift measurements for sources in the photometric catalogs covering the NIRCam footprints. Among these 747 NIRSpec redshifts, 555 are based on multiple emission lines (\texttt{Z\_Q\_REF}=1), 25 are from one single emission line but determined with the aid of photometry (\texttt{Z\_Q\_REF}=2) and 167 are from breaks or continuum shapes (\texttt{Z\_Q\_REF}=3).
Fourteen of the spectroscopic redshifts are stars at $z=0$. A 2D plot of the 733 $z>0$ spectra is shown in Figure \ref{fig:nirspec2D}. This illustrates the high occurrence rate of $H\beta$, [OIII] and H$\alpha$ emission lines in $z>2$ galaxies.


%

\section{Photometry and catalog construction} \label{sec:photometry}

In this section we discuss the construction of specific data products and procedures, including: source detection and image segmentation, PSF measurement and image convolution, aperture photometry, point-source identification, and catalog construction, organization, and suggested usage of flags. 

%


%

\subsection{Source Detection \& Segmentation} \label{subsec:source_detection}
A deep $\chi_{mean}$ detection image is created by co-adding background-subtracted images for all available JWST and HST-optical filters in each field following the approach described in \citet{Drlica-Wagner2018a}, a modified version of \citet{Szalay1999}'s chi-squared technique. Note that \hst/WFC3 IR filter images are excluded from the detection image owing to the broad instrumental PSF and larger native pixel scales. In the CLU field the images are both bCG-subtracted and background-subtracted. A non-linear $\chi$-mean co-added image allows for faint sources detected by emission lines in even a single band to be propagated through to the detection image at high signal-to-noise ratios (S/Ns), while minimizing discontinuities in regions where the number of available dithers differs (e.g. at the edge of the field of view for each instrument). The $\chi_{mean}$ detection image is produced according to:
\begin{equation}
    \chi_{\rm mean} = \frac{\sqrt{\sum_{i}^{n} f_{i}^{2}w_{i}} - \mu_{\chi}}{\sqrt{n - (\mu_{\chi})^{2}}},
\end{equation}
where $f_{i}$ and $w_{i}$ are the flux and weight in filter \emph{i}, \emph{n} is the total number of filters, and $\mu_{\chi}$ is given by
\begin{equation}
    \mu_{\chi} = \sqrt{2}\frac{\Gamma((n-1)/2)}{\Gamma(n/2)}.
\end{equation}
The weights $w_i$ are given by the inverse of the root mean square (RMS) map for each filter $i$. The RMS map is given by the RMS noise level in each pixel in the filter, and the RMS map is constructed based on the weight image by rescaling it so that the sky background of RMS-normalized science image follows the standard normal distribution.

Individual filter images are inspected for artifacts with masks made for the affected regions. Masks on individual filters are applied at the detection image co-adding stage to avoid generating spurious detections in the segmentation map, while still performing source detection in the affected regions. This is particularly important in certain fields (e.g. MACS0417NCF, MACS1149NCF) where multiple filter images in the NIRCam short wavelength ("SW") channel are affected by high persistence from the prior visit or other artifacts such as scattered light (e.g., dragon's breath, claws\footnote{\url{https://jwst-docs.stsci.edu/known-issues-with-jwst-data/nircam-known-issues/nircam-scattered-light-artifacts}}). Also in the CLU fields, areas within 1.2 arcsec radii from bCG centers are masked from the detection image and excluded from source detection. 

Source detection and segmentation are performed using the \texttt{photutils} \citep{bradley_2024_10967176} implementation of \texttt{SourceExtractor}'s \citep{BERTINE.ARNOUTS1996} source detection and watershed segmentation algorithm. In source detection and the downstream processes, we further mask out bright stars and high residual areas of the bCG subtraction in Section \ref{subsec:bcgs_icl} within the NIRCam field of view (hereafter, we refer to this exclusive mask as ``\texttt{mask\_exclude}''). The \texttt{mask\_exclude} mask is determined by visually inspecting the $\chi$-mean detection image.

We use \texttt{photutils.segmentation.Background2D} to compute the global RMS value of background region in the $\chi$-mean detection image, which is used as the noise level of the detection image.
The RMS value of a $\chi$-mean image should be unity ideally by definition, but the actual RMS values of our detection images are slightly offset from unity (typically $\sim5\ \%$).
Based on the noise level estimation of the detection image, we compute the S/Ns of each pixel in the detection image.

We adopt the two-step detection approach commonly known as the ``cold+hot'' mode strategy \citep[e.g.,][]{Rix_2004,Guo_2013,Galametz_2013}.
This approach consists of two runs of source detection, where we first perform the ``cold'' mode detection that is optimized for bright, extended sources, followed by ``hot'' mode with a configuration optimized for faint, compact sources.
The source detection performance is characterized with four independent parameters; \texttt{DETECT\_THRESH}, \texttt{DETECT\_MINAREA}, \texttt{DEBLEND\_NTHRESH}, and \texttt{DEBLEND\_MINCONT}. 
When the number of connected pixels that are above the S/N threshold (\texttt{DETECT\_THRESH}) is equal to \texttt{DETECT\_MINAREA} or larger, the connected pixels are identified as a source.
Then each source is deblended with the watershed segmentation algorithm with the two parameters (\texttt{DEBLEND\_NTHRESH} and \texttt{DEBLEND\_MINCONT}).

The two ``cold'' and ``hot'' mode detections are combined to derive the final list of detected sources following the strategy of \texttt{GALAPAGOS} \citep{Barden_2012}.
All cold mode sources are kept in the final detection catalog, while the position of each hot mode source is checked if it is close enough to any of the cold mode sources.
When a hot mode source is located within the (scaled) Kron ellipse of a cold mode source, the hot mode source is regarded as a substructure of the brighter nearby source and does not enter the final detection catalog.
The scale factor in this combining process is a parameter, and setting this parameter to e.g., 1.2 means the Kron ellipse is enlarged by 20\%\ in this process.
The Kron ellipse scaling factor in the merging process is set to 1.2 and 1.3 in CLU and NCF fields, respectively.

As the bright source light profiles in the CLU and NCF fields are significantly different, we slightly vary the two parameters regarding deblending in the ``cold'' mode detection and the scaling factor in the merging process from field to field (Table \ref{tab:detection}).
In the cold mode detection, we require at least 12 adjacent pixels above S/N$=2.7$, the number of deblending thresholds (\texttt{DEBLEND\_NTHRESH}) is set in the range of 16 to 32, and the deblending contrast threshold (\texttt{DEBLEND\_MINCONT}) is in the range of 0.002 to 0.008 in the cold mode detection.
While in the hot mode detection, we require at least 4 connected pixels above S/N$=3$, and the deblending parameters are set as \texttt{DEBLEND\_NTHRESH} $=16$ and \texttt{DEBLEND\_MINCONT} $=0.01$ in all fields.
We chose these parameters to avoid any discontinuity between the cold and hot mode photometry, particularly at the faint end of cold mode detection where the two mode detections are merged \citep[c.f.,][]{Galametz_2013}.

In this data release, we include photometric sources that are observed with at least one NIRCam filter with the smallest aperture we use ($0.\!\!^{\prime\prime}3$-diameter circular aperture).
Since we co-add all NIRCam, NIRISS, and \hst\ optical filter images to build the detection image, the detection $\chi$-mean image contains a number of \emph{HST}-only sources in the footprint.
However in the light of our purpose of releasing \jwst\ data products, we do not include the \emph{HST}-only sources in the released photometric catalog.
A total of 121261 photometric sources are included across all CLU and NCF fields combined.

\begin{deluxetable*}{cc|cccccccccc}[t!]
\tablecaption{Source detection parameters}
\tablewidth{0pt}
\label{tab:detection}
\tablehead{
\colhead{} & \colhead{} & \multicolumn{2}{c}{A370} & \multicolumn{2}{c}{M0416} & \multicolumn{2}{c}{M0417} & \multicolumn{2}{c}{M1149} & \multicolumn{2}{c}{M1423}  \\
\colhead{Mode} & \colhead{Parameter} & \colhead{CLU} & \colhead{NCF} & \colhead{CLU} & \colhead{NCF} & \colhead{CLU} & \colhead{NCF} & \colhead{CLU} & \colhead{NCF} & \colhead{CLU} & \colhead{NCF}
}
\startdata
\multirow{4}{*}{COLD}& \texttt{Detect\_thresh} & 2.7 & 2.7 & 2.7 & 2.7 & 2.7 & 2.7 & 2.7 & 2.7 & 2.7 & 2.7 \\
& \texttt{Detect\_minarea} & 15 & 15 & 15 & 15 & 12 & 15 & 15 & 15 & 15 & 15 \\
& \texttt{Deblend\_nthresh} & 16 & 32 & 16 & 32 & 32 & 24 & 16 & 32 & 16 & 24 \\
& \texttt{Deblend\_mincont} & 0.005 & 0.005 & 0.005 & 0.005 & 0.008 & 0.002 & 0.002 & 0.001 & 0.005 & 0.002 \\
\hline
\multirow{4}{*}{HOT}& \texttt{Detect\_thresh} & 3.0 & 3.0 & 3.0 & 3.0 & 3.0 & 3.0 & 3.0 & 3.0 & 3.0 & 3.0 \\
& \texttt{Detect\_minarea} & 4 & 4 & 4 & 4 & 4 & 4 & 4 & 4 & 4 & 4 \\
& \texttt{Deblend\_nthresh} & 16 & 16 & 16 & 16 & 16 & 16 & 16 & 16 & 16 & 16 \\
& \texttt{Deblend\_mincont} & 0.01 & 0.01 & 0.01 & 0.01 & 0.01 & 0.01 & 0.01 & 0.01 & 0.01 & 0.01 \\
\hline
Merging & \texttt{Kron\_increase} & 1.2 & 1.3 & 1.2 & 1.3 & 1.2 & 1.3 & 1.2 & 1.3 & 1.2 & 1.3 
\enddata
\end{deluxetable*}

\subsection{PSF Matching \& Image Convolution} \label{subsec:psf_matching}


PSFs are extracted empirically from each image following the methodology described in \citet{Skelton2014} and summarized in \citet{Sarrouh2024}. For NIRCam or NIRISS filters common to both the cluster and flanking field pointings, a single PSF is constructed using stars from both fields as visits were scheduled in quick succession and the observations have identical position angles. For Technicolor filters taken in Cycle 2 at identical PAs, curves of growth were measured and analyzed to ensure consistency of the PSF across a 1-year timespan, before combining the PSF stars with those from Cycle 1. Our PSF-matching strategy is to PSF-match to the largest FWHM filter which is available in all fields. We choose F444W as the target filter for all fields.

\begin{figure*}
\plotone{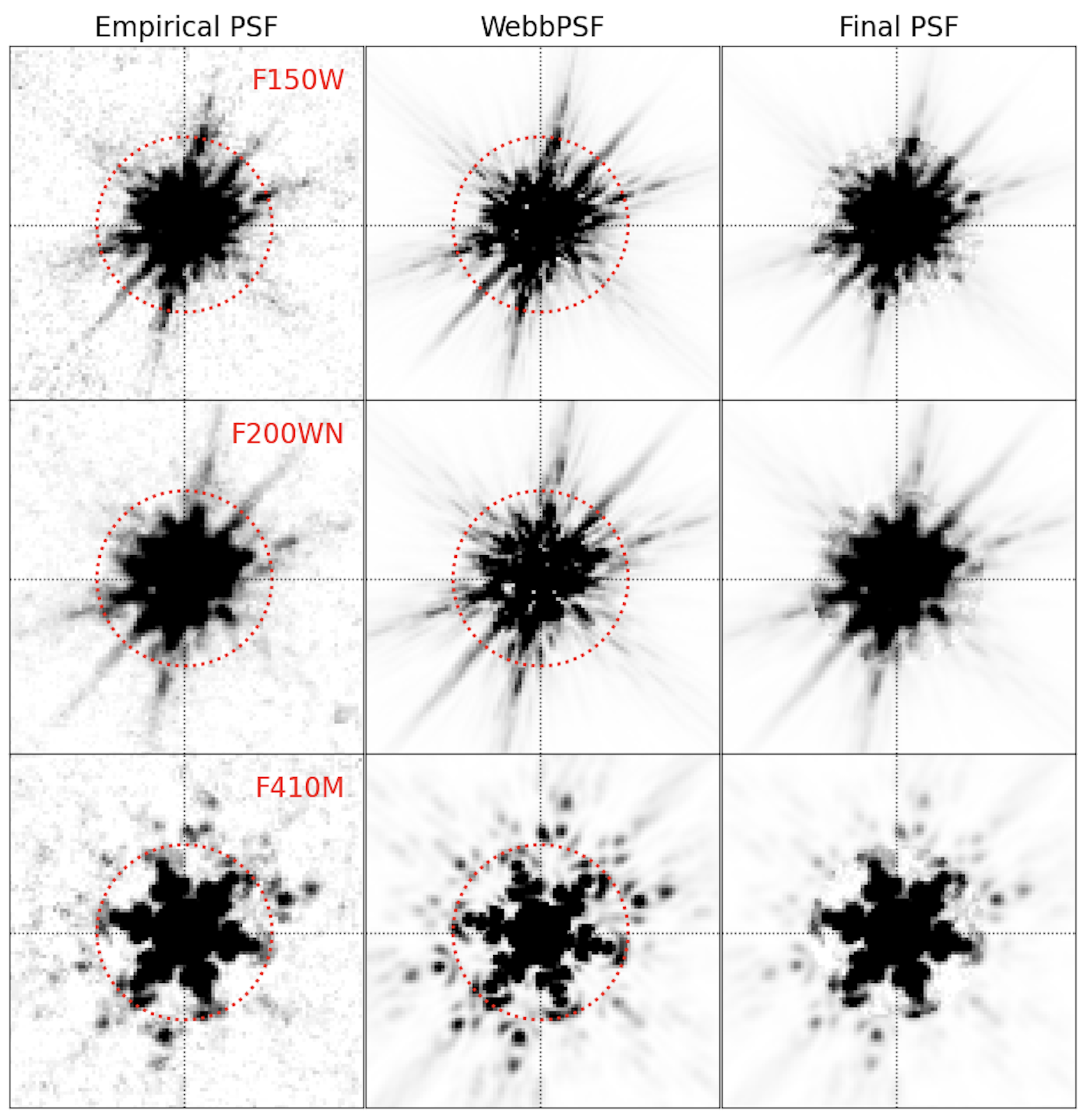}
\caption{A demonstration of the PSF construction method. Empirical PSFs are measured in each filter and then combined with the WebbPSF model.;   \emph{Left:} Empirical PSF constructed from median stacking bright stars. The central 1\arcsec of the empirical PSF is used in the final PSF (inside the dotted red circle); \emph{Middle:} WebbPSF model. The peripheral regions outside the central 1\arcsec \ disc are used in the final PSF (outside the dotted red circle). \emph{Right:} Final PSFs consist of the empirical PSF in the central region, and the WebbPSF in the outer regions.}
\label{fig:psf_construction}
\end{figure*}

\begin{figure*}
\plotone{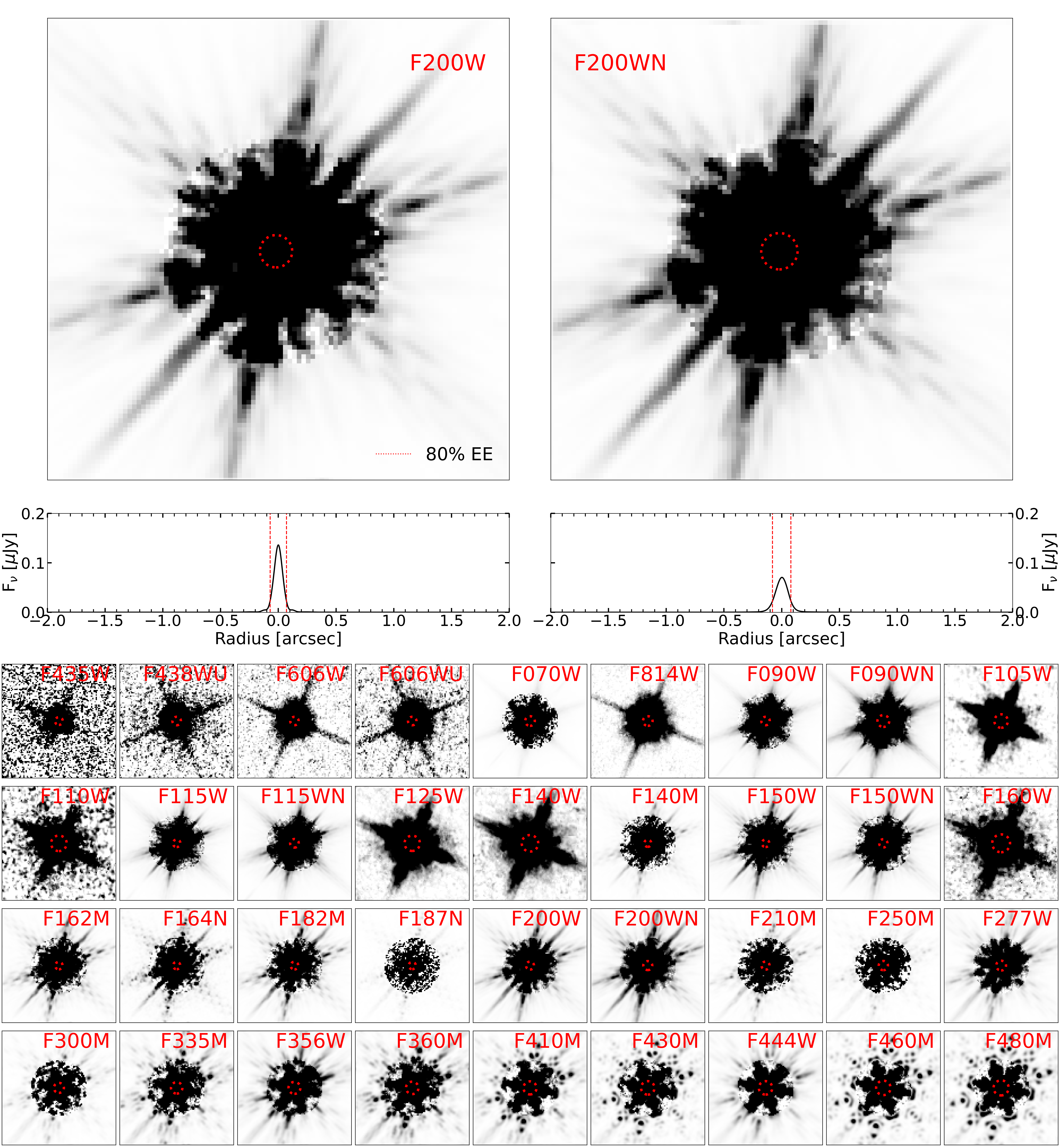}
\caption{4\arcsec x4\arcsec PSF cutouts for F200W NIRCam (\emph{top left}) and NIRISS (\emph{top right}). The aperture which encloses 80\% of the total energy is marked with a red dashed circle, and a 1D slice of the PSF profile along the horizontal axis is shown below. 4\arcsec x4\arcsec PSF cutouts for all filters available in the CLU field are shown below. All PSFs are from the MACS0416 CLU/NCF fields, excepting the \emph{HST}/WFC3 UVIS filters F438WU \& F606WU which are in MACS0417 NCF.}
\label{fig:psfs_all}
\end{figure*}

Point sources are identified using the surface brightness to magnitude relation, where point sources occupy a tight locus in this space. We visually inspect all sources with 20 $\leq$ \ mag$_{AB}$(\emph{filt}) $\leq$ \ 23, rejecting non-isolated or saturated sources, and masking nearby sources within a 4\arcsec x4\arcsec cutout as necessary. All star cutouts are then re-centered using \texttt{photutils.centroids.centroid\_2dg}, and their fluxes normalized within a radius of 1\arcsec. An empirical PSF in each filter is constructed by stacking all stars and taking the median value of unmasked pixels at each position. The empirical PSF is then normalized to unity within a radius of 1\arcsec. Encircled energies of the stars and empirical PSF are then measured for each filter. Stars whose profiles are outliers relative to the PSF are removed iteratively until a homogeneous set of stars are found which vary by no more than $\pm1\%$ from the PSF profile. This removes both stars with low signal-to-noise as well as bright stars which, while not quite saturated, exhibit strong non-linearities in their flux profiles due to nearly-full potential wells on the detector. The final PSF used in generating convolution kernels is constructed by combining the central 1\arcsec \ disc, or "core", of the empirical PSF with the outer regions of the WebbPSF model \citep{Perrin2014} where the SNR of empirical PSF is low and the diffuse light contained in the PSF diffraction spikes is poorly measured, as shown in Figure \ref{fig:psf_construction}. This is achieved by re-normalizing the entire empirical PSF cutout to unity and conserving the encircled energy contained within the empirical core and the outer regions. The WebbPSF normalization is scaled such that the total flux in the outer region matches that of the outer region of the empirical PSF, which it is replacing. The final PSF is then the core of the empirical PSF with the scaled outer regions of the WebbPSF model. This reduces noise in the convolution kernels and residuals in the convolved data, and is of particular importance for the bCG and ICL modelling discussed in \S \ref{subsec:bcgs_icl}. Utilizing the WebbPSF model in the outer regions also forces the convolution kernel to go to zero along the periphery, which eliminates edging effects and cutout-shaped residuals (i.e. "boxing") around bright sources upon convolution of the 4\arcsec x 4\arcsec \ kernel with the much larger data images. It can however create an apparent discontinuity between the empirical and WebbPSF components, in particular for the shorter wavelength filters with sharper PSFs (see Figure \ref{fig:psfs_all}). This is an inevitable consequence of the discrepancy between the empirical NIRCam PSFs and the WebbPSF models, and the need to conserve flux in each component when combining the two. The effect is negligible on the PSF curves of growth, and the flux enclosed within the empirical 1\arcsec \ disc is $\geq$ 98\% of the total flux of the PSF for all filters.

Figure \ref{fig:psfs_all} shows the difference in PSF structure between corresponding NIRCam and NIRISS filters at 2$\mu$m. As shown in the upper panel NIRISS has a marginally broader PSF owing to the finer sampling of the NIRCam SW detector. PSFs for all filters available in the CLU field are also shown in order of ascending wavelength.

\begin{figure*}[t]
\plotone{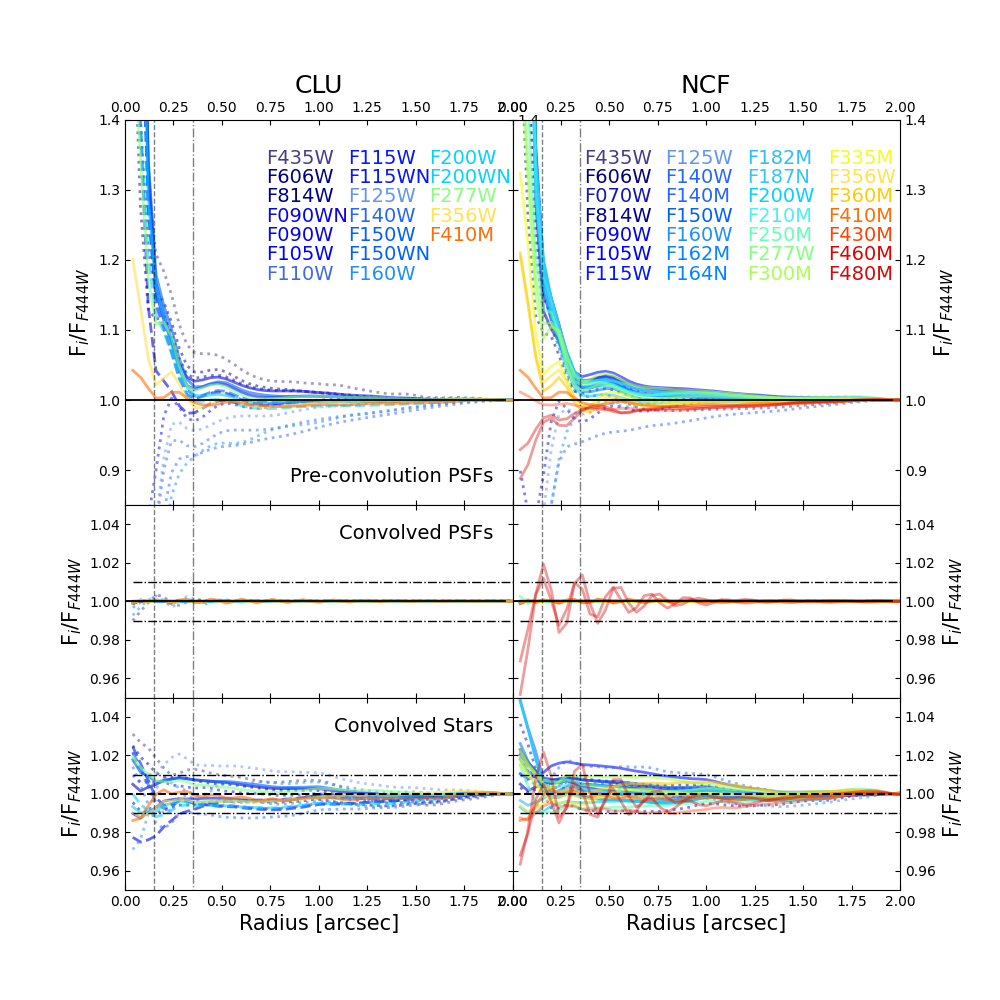}
\caption{\emph{Top:} PSF encircled energies relative to the F444W PSF for MACS0416 CLU (left) and NCF (right), pre-convolution; \emph{Middle:} Convolved PSFs relative to the target F444W PSF; \emph{Bottom:} Convolved empirical stars, measured relative to empirical stars in F444W. In all panels the 0\farcs15 and 0\farcs35 radial apertures use to compute total color fluxes are shown as dashed and dot-dashed lines, respectively. JWST filters are shown as solid curves; HST filters as dotted curves.}
\label{fig:psf_matching}
\end{figure*}

\begin{sloppypar}
Convolution kernels are constructed with \texttt{create\_matching\_kernel()} from \texttt{photutils.psf.matching}, which uses the ratio of Fourier transforms to create a kernel to match PSFs. We apply a  \texttt{SplitCosineBellWindow()} windowing function, which is necessary to remove high-frequency noise which results from floating-point imprecision when taking the ratio of Fourier transforms. We optimize the window functions by varying the parameters which control the function's shape. All stars used in constructing the source PSF (i.e. the shorter wavelength PSF) are then convolved and compared against the stars used to construct the target PSF (i.e. the target PSF - F444W). We select the kernel which minimizes the median residual of convolved stars from the source filter as compared against all stars across the target filter. For poorly-sampled PSFs (e.g. from supporting HST/\emph{WFC3/IR}), the kernel RMS is taken into account as well to minimize convolution residuals and boundary effects around bright sources due to higher noise in the outer regions of the kernel. 
\end{sloppypar}
We conduct this empirical test as it is a more informative diagnostic of how well convolution will perform with the constructed PSFs on empirical sources within real data, as compared to testing convolutions on the PSFs themselves, which in general will yield good results as the kernels effectively convolve the source PSF to the target PSF by construction. Figure \ref{fig:psf_matching} shows the encircled energies diagnostic pre-convolution (top row) and post-convolution for PSFs and empirical stars (middle and bottom rows, respectively) for the CLU and NCF fields. The 0\farcs3 and 0\farcs7 diameter apertures used for photometry, photo-z and SED fitting (see \S \ref{subsec:fluxes_and_errors} \& \S \ref{sec:SPS}) are shown as dashed and dot-dashed green lines. In general we see excellent results at the \tilda0.1\% level for convolved PSFs, the exception being the two reddest medium bands, F460M \& F480M. As they have broader PSFs than the target F444W filter, these are deconvolutions which exhibit larger residuals at the smallest apertures, though they are limited to the \tilda1\% level at the photometry apertures. However, this uniform result convolving idealized PSFs can obscure actual convolution performance on empirical sources, shown in the lower panels of Fig. \ref{fig:psf_matching}. Empirical stars exhibit larger median residuals as well as a larger scatter. Our optimization of convolution kernels based on empirical criteria in general yields convolution residuals contained to within $\leq5$\% at the central pixel, and residuals at the \tilda1\% level within the photometry apertures.

%


%

\subsection{Measurement of Photometric Fluxes \& Errors} \label{subsec:fluxes_and_errors}
We use \texttt{photutils} to perform fixed aperture photometry in circular apertures of 0$.\!\!^{\prime\prime}$3, 0$.\!\!^{\prime\prime}$5, 0$.\!\!^{\prime\prime}$7, 1$.\!\!^{\prime\prime}$5, \& 3$.\!\!^{\prime\prime}$0 in diameter on the PSF-convolved images described above. We also perform photometry in a Kron-like elliptical aperture, scaling the Kron radius by a factor of 2.5, which should enclose \greaterthan96\% of the source total flux \citep{Kron1980}. Note that this Kron aperture is determined based on the light profile of each source in the $\chi$-mean detection image, and a single aperture is fixed to each source and not varied filter by filter.
A local background correction is performed during the photometric flux measurements. The local background level (per pixel) for each source is estimated by 3-sigma-clipped mode within the rectangular annulus aperture with $\sim2^{\prime\prime}$ width around the source segmentation. We provide three total flux measurements: ``KRON'', ``COLOR03'', and ``COLOR07''.

The "KRON" total flux measures flux in elliptical apertures scaled by 2.5 $\times$ KRON\_RADIUS \citep{Kron1980} directly in each band with an aperture correction (as described later). The ``COLOR03'' total flux is based on the 0$.\!\!^{\prime\prime}$3-diameter fixed aperture photometry but scaling fluxes of all filters by the ratio of 0$.\!\!^{\prime\prime}$3-to-Kron aperture fluxes in the NIRCam F277W filter. The ``COLOR07'' is a similar measurement but uses 0$.\!\!^{\prime\prime}$7-diameter photometry instead.
Generally, total COLOR fluxes (both ``COLOR03'' and ``COLOR07'') yield a higher signal-to-noise flux than ``KRON'' total flux measurements, as it assumes no color gradient between filters. In both cases a minimum circularized Kron radius of 0$.\!\!^{\prime\prime}$35 is imposed for total fluxes; sources below this limit have total fluxes measured in circular apertures of 0$.\!\!^{\prime\prime}$35 radius. Total fluxes have aperture corrections applied based on F444W PSF encircled energies, as the images have all been homogenized to the F444W PSF.
The aperture correction factor is determined based on the size of the Kron aperture of each source, and it is $\sim1.1$ for the smallest Kron aperture (0\farcs7-diameter aperture).
Finally, we correct all photometry for Galactic dust attenuation along the line-of-sight, with E(B-V) values obtained from the \citet{Schlafly2011} dust map and adopting a \citet{Fitzpatrick1999} attenuation law with selective extinction of $R_V$=3.1 (Table \ref{tab:cluster_info}). 
In all photometry, we replace the measured flux with NaN when the aperture contains one or more bad pixels identified by the weight map or any pixel outside the FoV of the filter of interest.


\begin{figure*}[t]
\plotone{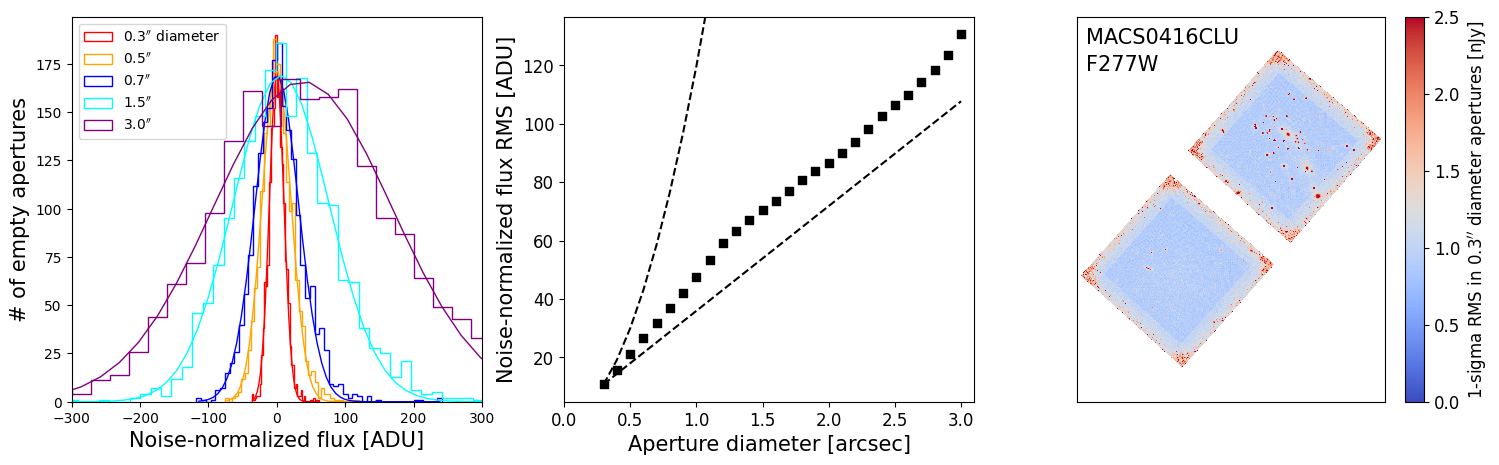}
\plotone{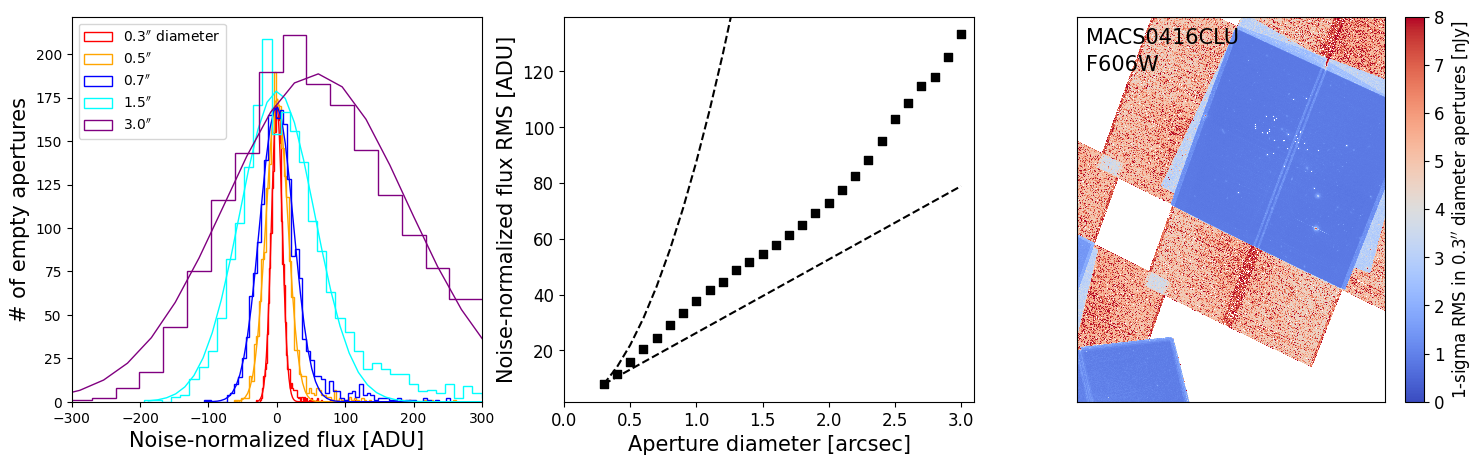}
\caption{Flux error measurements in the MACS0416 CLU field for the JWST/NIRCam F277W image (top) and HST/ACS F606W image (bottom).
We perform 2000 random empty aperture photometry measurements in the noise-normalized image with each fixed aperture size, and measure the RMS of the empty aperture flux distribution by fitting a Gaussian function. Left panels show the flux distribution color-coded by aperture sizes.
Middle panels present measured RMSs as a function of the aperture size. The dashed curves denote the linear ($\propto N$) and $N^2$ scalings, which correspond to correlation-free and full correlation between pixels, respectively.
The right panels show the flux error maps for 0.3$^{\prime\prime}$-diameter aperture photometry, based on the empty aperture flux RMS. 
To reconsider: F160W might be better to show instead of F606W.
}
\label{fig:fluxerr}
\end{figure*}


To estimate photometric uncertainty, we follow the method described in \citet{Skelton2014}. We measure the background noise level in an empirical way, as drizzling is known to create pixel correlation owing to charge being moved around at the sub-pixel scale \citep{Casertano2000}. First we measure flux in 2000 apertures placed in blank regions of sky for circular apertures from 0$.\!\!^{\prime\prime}$3 to 3$.\!\!^{\prime\prime}$0. Due to the varying depth, we measure empty aperture fluxes on RMS-normalized images. Left panels in the Figure \ref{fig:fluxerr} show examples for empty aperture flux distributions in MACS0416CLU for F606W \& F277W. Each distribution is well described by a Gaussian, with distribution width increasing with aperture size. The growth of noise level as a function of the aperture size is shown in the middle panels in the figure. The dashed lines in the middle panels denote the growth curves corresponding to correlation-free and perfect correlations between pixels, and our measurements show that the background error is pixel correlated to some extent.
The width is described as a function of aperture size by fitting a power law of the form
\begin{equation}
    \sigma = \alpha N^{\beta}, \\
\end{equation}
where $\alpha$ is the normalization, $\beta$ is the power law slope, and $N = \sqrt{S}$ where S is the area of the aperture. This is then multiplied by the RMS level at the position of each source to yield the final error reported in the catalog.
Right panels in the figure show the resulting flux uncertainty maps for 0$.\!\!^{\prime\prime}$3-diameter aperture photometry.


%

\subsection{Bright Cluster Galaxies in the catalog}\label{subsec:bcg_photometry}
We append the bCGs that are removed from the final science image (Section \ref{subsec:bcgs_icl}) in our photometric catalog, for the sake of catalog completeness.
For bCGs, we include their positions, morphological parameters, and the total flux estimations, which are simply the total fluxes of the models, but not the fixed circular aperture photometry.
We assume 1 \% flux errors on the bCG total fluxes, and use the same value for all three total flux measurements (``KRON'', ``COLOR03'', and ``COLOR07'').
All bCGs in the catalog can be identified via the ``\texttt{FLAG\_BCG}'' flag.
The bCGs are not included in the segmentation map but are in the photometric catalog, which results in an inconsistency in the number of sources between the catalog and segmentation map.

The definitions of morphological parameters are not completely homogeneous between bCGs and other normal photometric sources.
For normal photometric sources, the lengths along the semi-major/minor axis (\texttt{A} and \texttt{B}) are estimated based on the 1-sigma standard deviation of the 2D Gaussian that has the same second-order moments as the source in the detection image.
While for bCGs, the semi-major axis length is the half-light radius, and the semi-minor axis length is based on the ellipticity at the half-light radius.
These values are almost consistent if a source has an ideal 2D Gaussian profile, though, the actual light profiles of the bcgs are more centrally concentrated, and thus semi-major/minor axis lengths for bcgs are estimated systematically smaller than for normal photometric sources.

\subsection{Useful flags} \label{subsec:flags}
In this subsection, we briefly introduce several useful flags in the photometric catalog.
A full description of the entries of our catalog is given in Appendix \ref{apx:catalog_entries}.

\subsubsection{Point source flag} \label{subsubsec:flagging_stars}
\begin{figure}[t]
\plotone{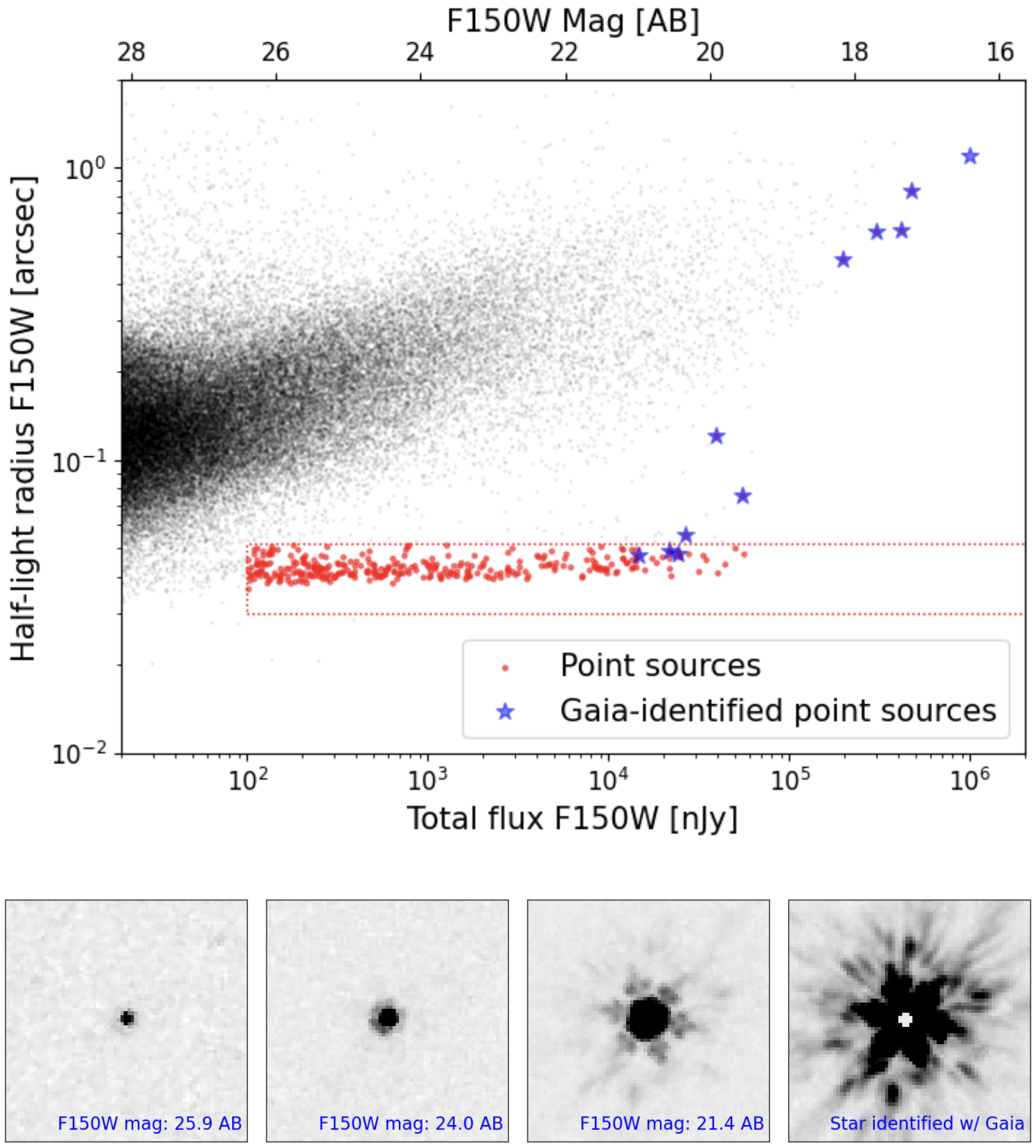}
\caption{Point source identification in the F150W size-magnitude diagram. Black points represent all photometric sources in our catalogs, and red points are the point sources identified with the criteria described in Section \ref{subsubsec:flagging_stars} (red dashed lines). Blue stars are those identified by Gaia observations. The bottom panels display four examples of point sources identified in Section \ref{subsubsec:flagging_stars}, at a wide range of F150W magnitudes. The cutouts are 1\farcs4 on the side.
Not all Gaia-identified stars are plotted in this figure, since a large number of them have completely saturated pixels at the center and their total Kron fluxes and sizes in the F150W image are not measurable.}
\label{fig:star_gal_sep}
\end{figure}

One of the main uses of the photometric catalog is to build a sample of extragalactic objects, and this requires flagging possible foreground stars in our Galaxy that could contaminate the galaxy sample.
Compact unresolved sources can be selected in a size-magnitude diagram, as they have constant small sizes regardless of their magnitudes.
We measure the circularized half-light radius of each source in the original, non-PSF-matched NIRCam F150W 20mas image, and Figure \ref{fig:star_gal_sep} shows the size-magnitude diagram in the NIRCam F150W image.
We select point sources in the photometric catalog having a half-light radius between 1.5 to 2.6 pixel and brighter than 100 nJy in the NIRCam F150W 20mas image (red points in Figure \ref{fig:star_gal_sep}).

However, as noted in the literature \citep[e.g.,][]{Weaver2024}, foreground stars in our Galaxy can easily saturate in deep \jwst\ images, and these saturated stars can deviate from the point-source locus in the size-magnitude diagram.
To flag them, we cross-matched our photometric catalog with the Gaia DR3 catalog (\citealt{Gaia2016,Gaia2023}) and flag a source if it exists in the Gaia catalog and its \texttt{classprob\_dsc\_combmod\_galaxy} is less than 0.1.
Blue stars in Figure \ref{fig:star_gal_sep} present locations of the Gaia-selected point sources in the size-magnitude diagram, and the rightmost subpanel at the bottom of the figure shows the F150W image of an example of Gaia-identified stars.
The figure demonstrates that flagging Gaia sources indeed selects saturated stars while the faint-end of Gaia sources includes unsaturated stars in the diagram, which suggests cross-matching with the Gaia catalog effectively selects potential stars that are missed with the size-based selection.
A total of 363 sources are flagged as point sources, and 49 of those are from Gaia catalogs.
All point sources identified in this subsection can be found via ``\texttt{FLAG\_POINTSRC}'' flag in the catalog.

%


%
\subsubsection{High bCG contamination flag} \label{subsubsec:flag_highbcg}
Although our bCG subtraction makes it possible to search for astronomical objects in the vicinity of bCGs, it also generates residuals which are identified as a source in some cases.
Particularly, faint detections around the bCGs are highly likely residuals, and it is useful to flag these possible fake detections.
We set a flag (``\texttt{FLAG\_HIGH\_BCGCONTAMI}'') for this, and this flag is true when the NIRCam F444W bCG model flux is larger than the science image flux within the segmented area of the source.

\subsubsection{Use phot flag} \label{subsubsec:use_flag}
A ``use'' flag can be employed when users want to work with the photometric catalog.
For this purpose, we introduce ``\texttt{USE\_PHOT}'' flag in the catalog, and the flag is True when the following criteria are all met:
\begin{enumerate}
    \setlength{\itemsep}{-0.em}
    \item the source is not a bCG and thus removed from the final science image,
    \item the model bCGs do not significantly contaminate the science flux in the F444W image (i.e., \texttt{FLAG\_HIGH\_BCGCONTAMI} is False),
    \item the Kron aperture does not overlap with the exclusive mask (``\texttt{mask\_exclude}''),
    \item the Kron aperture is within FoVs of 80 \% of NIRCam filters available in the field,
    \item the Kron aperture does not contain any bad pixel in 80 \% of NIRCam filters available in the field.
\end{enumerate}
This flag ensures that the total flux measurements of the source are reliable in almost all NIRCam filters at $\lambda_{\rm obs}\sim0.9-4.4\ \mu$m, and is useful for rejecting possibly problematic sources.
However, we do not impose any S/N cut or photo-$z$ fit quality cut in this flag, because such criteria should be determined depending on the science case.
In most science cases, additional cuts should be necessary, particularly based on the S/N in some filters or on the quality of template fitting.

We also have the ``\texttt{USE\_PHOT\_APER03}'' flag for a similar purpose. The definition of the ``\texttt{USE\_PHOT\_APER03}'' flag is the same as ``\texttt{USE\_PHOT}'' but using 0$.\!\!^{\prime\prime}$3-diameter aperture instead of the Kron aperture. 
The flag selects sources of which the smallest aperture photometry is reliable in all NIRCam filters, so it is useful when users may want to secure photometric colors and/or photo-$z$s,  but do not need the total flux measurements.
Since the aperture size used for the criteria is smaller, ``\texttt{USE\_PHOT\_APER03}'' flag selects slightly more sources than the ``\texttt{USE\_PHOT}'' flag.
A total of 97080 sources is selected with ``\texttt{USE\_PHOT}'' in all 10 fields (80.1 \% of the photometric sources), while 99824 sources are selected with ``\texttt{USE\_PHOT\_APER03}''  (82.3 \%).

We here highlight that we do not remove point sources from ``\texttt{USE\_PHOT}'' selection, considering that point sources identified in Section \ref{subsubsec:flagging_stars} are not necessarily stars in our Galaxy.
Extragalactic point sources, such as ``Little Red Dots'' \citep[e.g.,][]{Labbe2023a}, can also be flagged via the \texttt{FLAG\_POINTSRC} flag, and including \texttt{FLAG\_POINTSRC} in the ``use'' flag criteria can lead to missing this population from the sample unknowingly.
Users thus may impose an additional cut using \texttt{FLAG\_POINTSRC} to remove any potential Galactic stars with a risk of missing a certain population of extragalactic objects.


\begin{figure*}[t]
\includegraphics[width=0.99\textwidth]{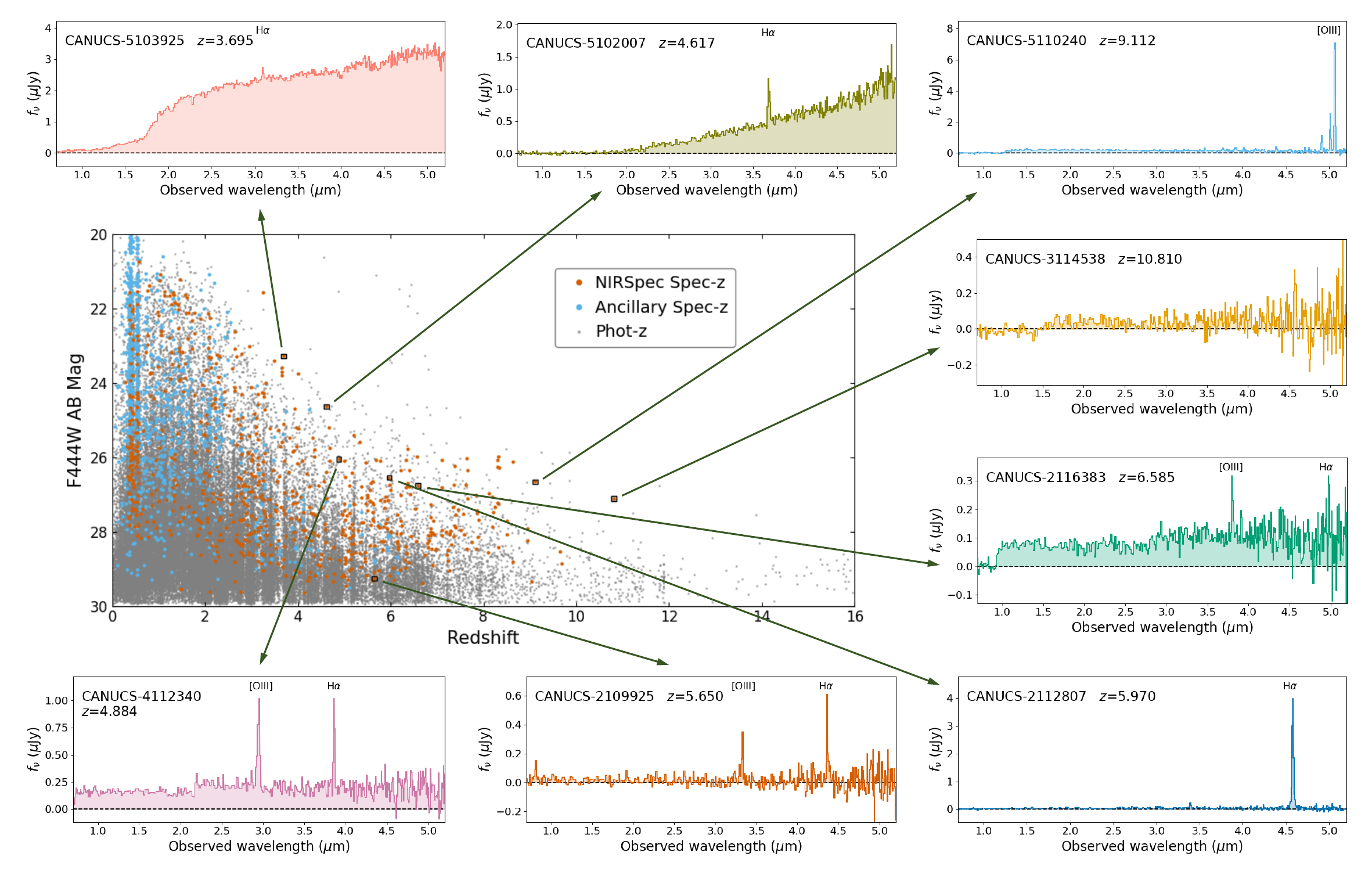}
\caption{Centre-left panel: F444W AB magnitude (0$.\!\!^{\prime\prime}$3 aperture) versus redshift for all sources in the CANUCS catalog with F444W S/N$>4$. Sources with spectroscopic redshifts are shown with colored circles: orange for CANUCS-NIRSpec redshifts and blue for ancillary. Sources with only photometric redshifts are shown in gray. Select CANUCS-NIRSpec sources are outlined with a black square and have green arrows connecting to their spectra, highlighting the diversity of the spectra including old quiescent, dusty star-forming, faint EELGs, high-$z$ LBGs, etc.}
\label{fig:nirspec}
\end{figure*}



\begin{figure*}[t]
\plotone{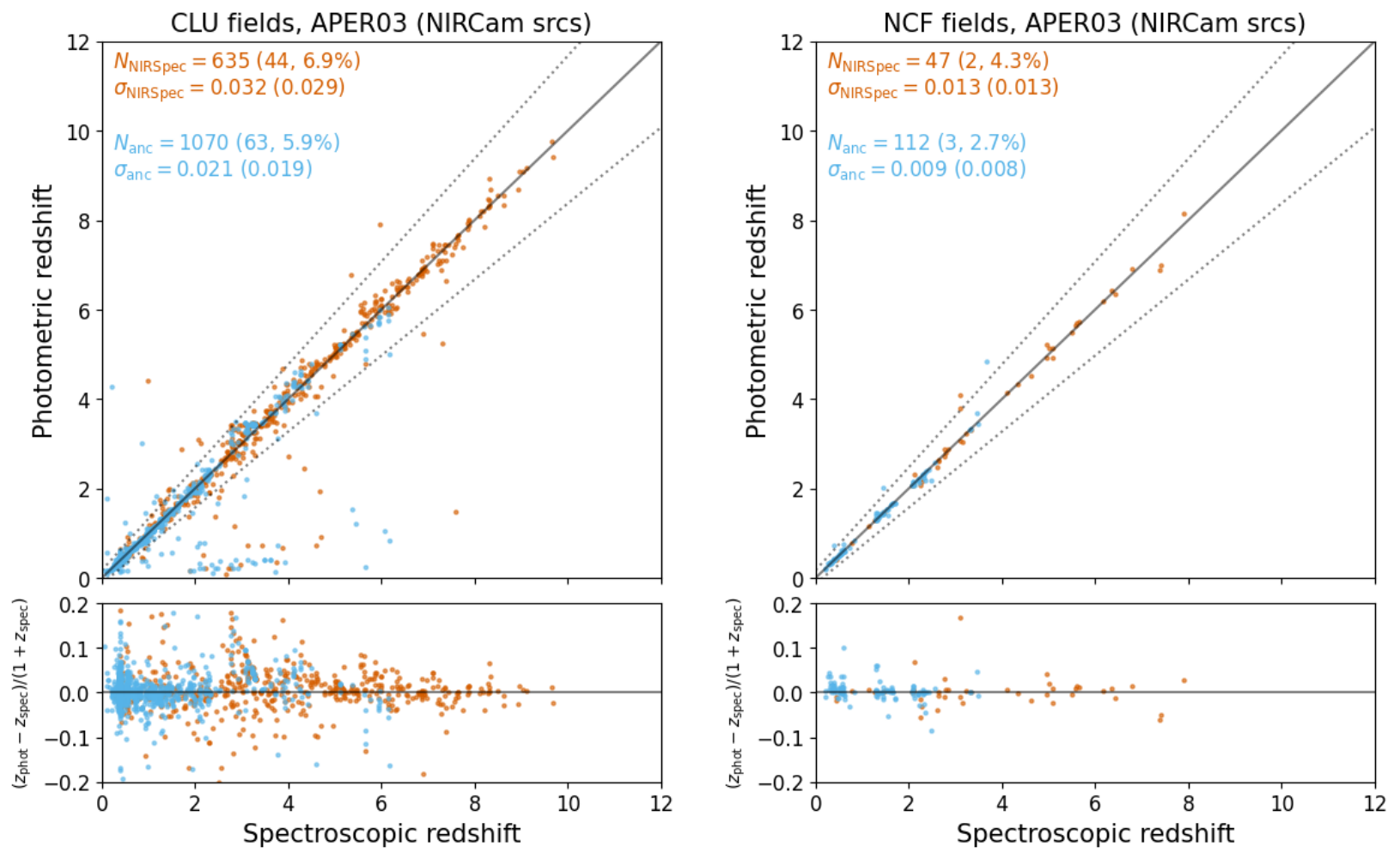}
\caption{Comparisons of photometric redshifts and $z_{\rm spec}$ in CLU fields (left) and NCF fields (right). Orange points represent the $z_{\rm spec}$ measurements based on our NIRSpec observations, while blue points denote those from literature. Note that our NIRSpec observation targets only one of five NCF fields, and NCF fields have typically less ancillary data. In the legend, we show the total number of sources in this plot, the outlier number and fractions (in parentheses), and the typical scatter of $\Delta z$. The $\sigma_{\rm NMAD}$ values in the parentheses are the scatters computed excluding the outliers.
The dashed lines correspond to 15 \% offset, $|\Delta z|=0.15$, which are used to define the outliers.}
\label{fig:zphot_zspec}
\end{figure*}



\begin{figure}[t]
\plotone{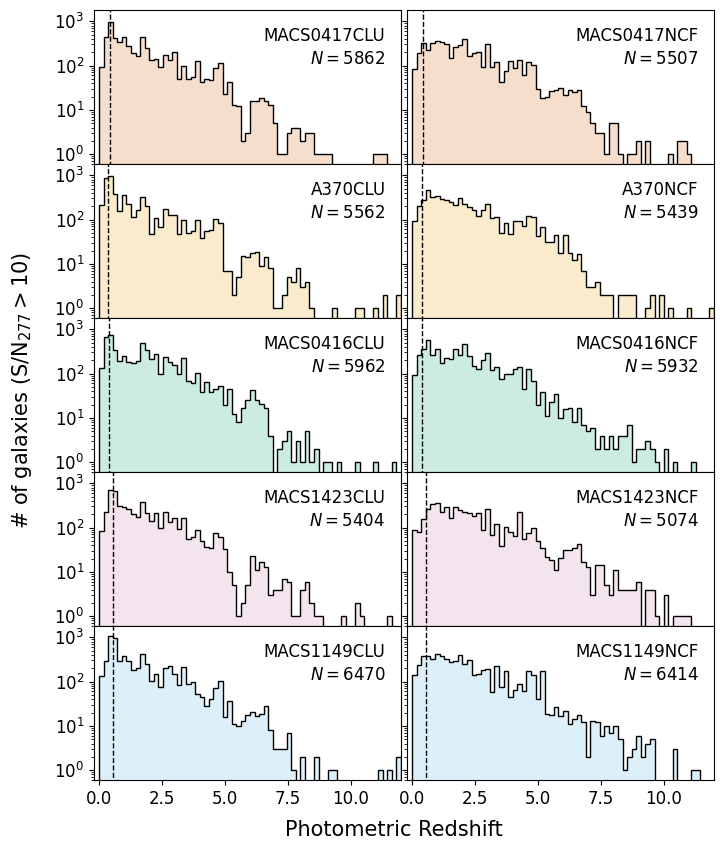}
\caption{Photometric redshift distributions in all 10 fields. Only S/N $>10$ sources in the NIRCam F277W filter are used. The vertical dashed lines represent the cluster redshifts in each galaxy cluster field.}
\label{fig:photz_dist}
\end{figure}


\subsection{Photometric redshifts} \label{subsec:photz}
We measured the photometric redshifts for the whole photometric catalog using the template-fitting code \texttt{EAzY-py} \citep[][]{Brammer2008}.
\texttt{EAzY-py} generates mock galaxy spectra by linearly combining a set of user-defined galaxy template spectra and searches for the best photometric redshift by fitting to the observed SEDs.
For the fit, we develop a galaxy template set starting with the ``set 3'' template by \citet{Larson2023}.


\begin{figure*}[t]
\plotone{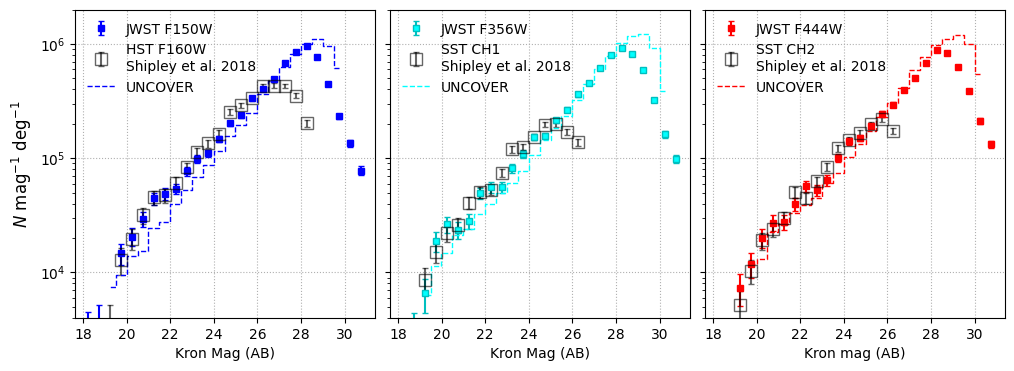}
\caption{The galaxy number counts in the MACS1149 CLU field as a function of the KRON total magnitudes in NIRCam F150W, F356W, and F444W filters. Error bars correspond to 1$\sigma$ Poisson uncertainties. The black open squares are the number count measurements in the same field from a \hst+\sst\ catalog \citep{Shipley2018}. Colored dashed lines are number counts in the same \jwst\ filters but in a different gravitational lensing field observed with longer NIRCam exposure time \citep{Weaver2023}.}
\label{fig:Numcounts}
\end{figure*}


\citet{Larson2023} presented template sets supplementing the standard \texttt{EAzY} template of \texttt{tweak\_fsps\_QSF\_12\_v3} with bluer, intensively star-forming galaxy spectra, and showed their template better reproduces high-$z$ ($z>6$) galaxy SEDs.
After the launch of \jwst, it has been shown that a large number of high-$z$ galaxies show extremely strong emission lines in the rest optical reaching EW$_{\rm rest}$([{\sc Oiii}]4959,5007) $>3000$ \AA\ or EW$_{\rm rest}$(H$\alpha$) $>2000$ \AA\ \citep[e.g.,][]{Withers2023,Boyett2024MN}.
The [{\sc Oiii}]4959,5007 lines are particularly strong, and the typical [{\sc Oiii}]5007/H$\beta$ and [{\sc Oiii}]5007/[{\sc Oii}]3727 line ratios at $z>4$ are $\sim3-10$ and $\sim3-30$, respectively \citep[e.g.,][]{Cameron2023a,Nakajima2023ApJS}.
However in the original template by \citet{Larson2023}, the [{\sc Oiii}]4959,5007 lines are not strong enough to reproduce many galaxies found with \jwst\ due to the very low gas-phase metallicity (5\% $Z_\odot$) and moderate ionizing parameter (log $U=-2$) assumptions.  The maximum [{\sc Oiii}]5007/H$\beta$ ratio among the template spectra is $2.7$ and the maximum EW$_{\rm rest}$([{\sc Oiii}]4959,5007) is $1900$ \AA.
The weak [{\sc Oiii}]4959,5007 lines in the template compared to typically observed SEDs can create (fake) photo-$z$ peaks at specific redshifts where the photometric filters do not completely cover the complete wavelength range and there are gaps between filter coverage.
This occurs because the observed SED can show a significantly higher flux excess from [{\sc Oiii}]4959,5007 than H$\alpha$, which the Larson templates cannot replicate. For such extreme [{\sc Oiii}] emitters, as a result, the photo-$z$ is erroneously offset by placing the H$\alpha$ line at the filter transmission edge, so as to mimic a high [{\sc Oiii}]4959,5007/H$\alpha$ flux excess ratio.

\begin{sloppypar}
To mitigate this issue, we generate a modified template spectrum based on the {\tt binc100z001age6\_cloudy\_LyaReduced} by \citet{Larson2023} by boosting the [{\sc Oiii}]4959,5007 lines with a factor of 3, so that the template spectrum can reproduce the SEDs of most extreme [{\sc Oiii}] emitters at $z>4$.
We also increase the sampling of the wavelength grid around the strong emission lines and assign reasonable emission line width of ${\rm FWHM}=200\ {\rm km\ s^{-1}}$, to alleviate the effect of filter edge in the template fitting.
Additionally, we generate another modified template spectrum from {\tt binc100z001age6\_cloudy\_LyaReduced} by removing all emission lines, so that the linear combination of them has the capability to reproduce SEDs with a wide range of emission line equivalent widths and [{\sc Oiii}]5007/H$\beta$ ratios.
In the end, we use a template set composed of 12 galaxy spectra from the standard templates {\tt tweak\_fsps\_QSF\_12\_v3} by {\tt EAzY} and three custom spectra based on {\tt binc100z001age6\_cloudy\_LyaReduced} (the original template spectrum, the one without emission lines, and [{\sc Oiii}] boosted one).
\end{sloppypar}

We use  $0.\!\!^{\prime\prime}3$-diameter aperture photometry in all available  \hst\ and \jwst\ filters for the fit, and add 5\ \% of the flux in each filter to the error budget in quadrature as the systematic uncertainty.
We turn off the zero-point offset corrections considering the uncertainty in the template spectra, but apply the magnitude prior in order to suppress the probability of unphysical solutions. 
We follow the \citet{Asada2025ApJ} prescription for the intergalactic medium (IGM) and circumgalactic medium attenuation correction, which takes into account the increased neutral hydrogen absorption at $z>7$ to avoid the overestimations of photometric redshifts at $z>7$ as is often reported in literature using a classical IGM attenuation curve \citep[e.g.,][]{Willott2024}.
We search for the best photometric redshift  over $z=0$-$20$, and for each source in the catalog, we report the best photometric redshift (\texttt{Z\_ML}) and 2.5, 16, 50, 84, and 97.5 percentiles of the posterior probability distribution (\texttt{Z\_025}, \texttt{Z\_160}, \texttt{Z\_500}, \texttt{Z\_840}, \texttt{Z\_975}).


\subsection{Spectroscopic redshifts} \label{subsec:specz}
We collect spectroscopic redshift ($z_{\rm spec}$, hereafter) measurements in our survey footprints from the literature, in addition to those from our NIRSpec observations.
The literature $z_{\rm spec}$ are from various papers over the last decade using ground-based instruments (e.g., {\it VLT}/MUSE) or \hst/grism observations \citep[][Rosati et al. in prep.]{Schmidt2014ApJ,Jauzac2014,Treu2015ApJ,Balestra2016ApJS,Grillo2016ApJ,Hoag2016ApJ,Treu2016ApJ,Caminha2017AA,Lagattuta2017MNRAS,Molino2017MNRAS,Shipley2018,Jauzac2019MNRAS,Richard2021AA}.
We also include $z_{\rm spec}$ measurements from a Keck/MOSFIRE observation (program ID U250 and U120, PI: G. Wilson), which targets star-forming galaxies in the NCF fields selected from our CANUCS catalog (Sok et al. in prep.).
We cross-matched our photometric sources with the $z_{\rm spec}$ sources in the literature (after correcting for the astrometry difference) with a search radius of $0.^{\prime\prime}2$, and keep only the most secure $z_{\rm spec}$ measurements.
When multiple $z_{\rm spec}$ measurements are matched to a single photometric source in our catalog, we keep the one with highest redshift quality.

The final catalog includes 1960 $z_{\rm spec}$ measurements. 1769 are located in the CLU fields and 191 are in the NCF fields. 747 of these are from our NIRSpec follow-up observations, as described in Section \ref{subsec:nirspecproc}.
Figure \ref{fig:nirspec} shows the F444W magnitudes (in 0$.\!\!^{\prime\prime}$3 aperture) against redshifts, while spec-$z$s are color-coded by its origin. Our spec-$z$ sample reaches down to $m_{444}\sim29.5$ mag, and contains sources at a wide range of brightness. As shown in the figure, our CANUCS-NIRSpec $z_{\rm spec}$ targets involve a variety of spectral types such as old quiescent, dusty star-forming, faint EELGs, high-$z$ LBGs, etc.
It demonstrates how powerful NIRSpec is, particularly for fainter ($m_{444}\gtrsim26$) and/or high-$z$ ($z\gtrsim3$) sources.
The $z_{\rm spec}$ measurements are stored in the \texttt{Z\_SPEC} column of the catalog.



\begin{figure*}[t]
\plotone{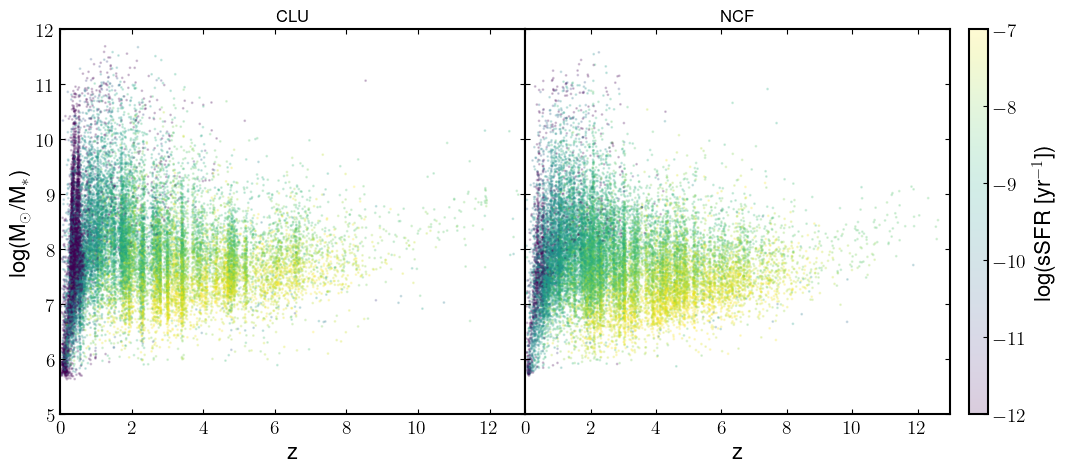}
\plotone{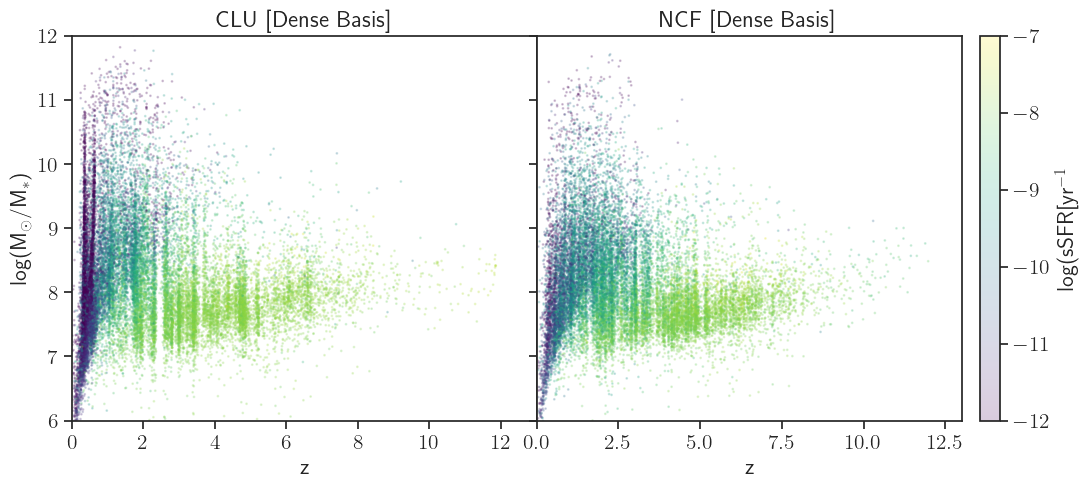}
\caption{Top: Stellar mass versus redshift for the full CANUCS sample using the \texttt{BAGPIPES} SED fits. Galaxies in the CLU (left) and NCF (right) fields are shown separately. The color bar indicates the specific SFR (sSFR). Bottom: As above using the \texttt{DENSEBASIS} SED fits.}
\label{fig:mass_z}
\end{figure*}

%


%

\section{Catalog Properties \& Diagnostics} \label{sec:cat_properties}
In this section, we provide several diagnostics to assess the quality of our photometric catalogs and/or to estimate typical uncertainties of our measurements in the catalogs.

\subsection{Photo-$z$ and spec-$z$ comparison} \label{subsec:zph_zsp}

We first compare the photometric redshifts and $z_{\rm spec}$, where available, to evaluate the quality of colors (or SEDs) of our photometry.
In the comparison, we only use sources that are observed with all NIRCam filters available in each field, and compare CLU and NCF fields separately considering the significant difference in the survey design between CLUs and NCFs.
We also compute the outlier fractions, which are defined as the number fraction of objects with $|\Delta z| = |z_{\rm spec} - z_{\rm phot}|/(1+z_{\rm spec})>0.15$, and the normalized median absolute distributions ($\sigma_{\rm NMAD}$) of $\Delta z$.
In this comparison, we remove our NIRSpec $z_{\rm spec}$ that are based on single emission line with the aid of photometry (see Sec.~\ref{subsec:nirspecproc}) to avoid circular reasoning.

Figure \ref{fig:zphot_zspec} shows a comparison between $z_{\rm spec}$ and $z_{\rm phot}$, color-coded by the source of spectroscopy (our NIRSpec observations or ancillary $z_{\rm spec}$ measurements).
Overall in CLU fields, the outlier fraction is $\sim7\ \%$ and the typical scatter of photo-$z$ estimations is $\sigma_{\rm NMAD}\sim0.03$.
The quality is even better in the NCF fields, with an outlier fraction of $\sim4 \%$ and typical scatter of $\sigma_{\rm NMAD}\sim0.01$. The lower outlier fraction and smaller scatter in NCF fields compared to CLU fields indicate the superb performance of NIRCam medium band filters in more accurately locating spectral signatures of high-$z$ galaxies such as emission line excesses and Balmer/Lyman breaks. The outlier fractions and typical scatters for CLU and NCF fields are comparable to other photometric catalogs based on \jwst\ observations \citep[e.g.,][]{Weaver2023, Rieke2023}.
The largest group of outliers in the CLU field is at $2<z_{\rm spec}<4$ and $z_{\rm phot}\lesssim0.5$, and these are mostly 1.6 $\mu$m bump vs strong emission line confusion. They are confused particularly when the source shows both Balmer break and strong emission line, because the SED appears as if it is a very red continuum with a bump at 1.6 $\mu$m.
We note that the results presented here could be somewhat biased by the target selection of spectroscopic observations, particularly the prioritization in some cases of galaxies with strong emission line excesses in photometry.

Another useful diagnostic is the photo-$z$ distribution, as it is free from any spectroscopic target selection bias.
Figure \ref{fig:photz_dist} shows the photo-$z$ distributions in all 10 fields with S/N $>10$ in the NIRCam F277W filter.
The distributions are reasonably smooth and there are no photo-$z$ spikes at the same redshift across the different fields, and those in CLU fields have photo-$z$ peaks right at the cluster redshifts (shown by vertical dashed lines).

\subsection{Galaxy number counts} \label{subsec:num_counts}
We next derive the galaxy number counts as a function of the total flux, and compare them with those measured in the same cluster fields in literature as measured by other surveys.
Since correcting the number counts for the gravitational lens effect is nontrivial and lens-model dependent, we derive the number counts without correcting for magnification.
We compute the number counts in NIRCam F150W, F356W, and F444W filters based on the ``Kron'' total flux measurements in one of the HFF cluster fields (MACS1149 CLU), and compare them to the (lensed) number counts in \hst\ F160W, \sst\ CH1, and CH2 filters measured in the same field \citep[][]{Shipley2018} assuming that F150W/F160W, F356W/CH1, and F444W/CH2 filter transmissions are similar enough.
The science area in each filter in our survey is computed based on the footprint of the filter and the excluse mask (``\texttt{mask\_exclude}'') applied in the source detection, and bCGs and highly contaminated sources are not used in the computation (i.e., use only sources whose \texttt{FLAG\_BCG} and \texttt{FLAG\_HIGH\_BCGCONTAMI} are both False).

As shown in Figure \ref{fig:Numcounts}, our galaxy number counts agree well with the literature \citep[][black open squares in the figure]{Shipley2018} within the Poisson uncertainties in general for bright sources ($m<24$ AB) in these filters.
There may be some offset at the intermediate magnitude ($m\sim24$-$26$ AB) in the F150W/F160W comparison, and the offset could be due to the difference in the matched PSF (we homogenize to F444W PSF while \citealt{Shipley2018} does to F160W PSF), in the detection image used to determine Kron parameters, or potential zero point offsets in the images.
At the faint end, our catalogs based on the deep \jwst\ $\chi$-mean detection image include significantly fainter sources than previously existing catalogs based on the \hst+\sst\ images.
Our number counts extend to $\sim 28$ AB or fainter, which is $\sim 2$ mag deeper than \hst\ and $\sim 4$ mag deeper than \sst.

We also compare our galaxy number counts with those in the same NIRCam filters from another lensed field \citep[UNCOVER;][targeting ABELL 2744; dashed lines in the figure]{Weaver2023} to visualize the depths of our images compared to other \jwst\ surveys.
We stress that both galaxy number counts from our catalog and UNCOVER shown in Figure \ref{fig:Numcounts} are not lens corrected and they target different lensing fields, so they cannot be directly compared due to gravitational lensing and the cosmic variance, but the locations of the histogram peak roughly tells the depth of the images.
In general, our photometric catalog seems slightly shallower than UNCOVER by $\sim0.5$ ABmag in the SW filter (F150W) and by $\sim1$ ABmag in LW filters. This is expected as UNCOVER observed deeper and in a smaller area than CANUCS.
We go slightly deeper in the SW filters (relative to UNCOVER) potentially because we detect sources in $\chi$-mean images while UNCOVER select sources in co-added NIRCam LW filter image.
We also note that the CANUCS imaging used small dither patterns so has very homogeneous depth across each field (see the top-right panel in Figure \ref{fig:fluxerr}).

\subsection{Depths} \label{subsec:depths}
In Table \ref{tbl:filts_and_fields} we provide the effective depths in all filters across the 10 fields to summarize our catalog properties.
Figure \ref{fig:Depth_FTC} shows the 3-sigma flux uncertainties in $0.^{\prime\prime}3$-diameter aperture in each filter in the MACS0416 CLU (left) and NCF (right) fields: the thick solid lines (or dashed line for NIRISS filters) are the median 3-sigma flux uncertainties of all photometric sources in the catalog, and the shaded region present the range of 10th to 90th percentiles of them.
We can see that the nominal median values are representative of the image depth except for the \hst\ filters where we combine the deep HFF iamge and shallower BUFFALO images.
We therefore take the median of 3-sigma flux uncertainties of all photometric sources in each filter/field to compute the effective depths except for the HFF+BUFFALO \hst\ filters (F606W, F814W, F105W, F125W, and F160W).
For the HFF+BUFFALO \hst\ filters, we compute the median and report the effective depths separately in HFF region and BUFFALO-only region.
The image depths in NIRCam and NIRISS filters are similar in each CLU and NCF fields, and we provide a set of very homogeneous photometric catalogs in ten fields.





%



\begin{figure*}[t]
\plotone{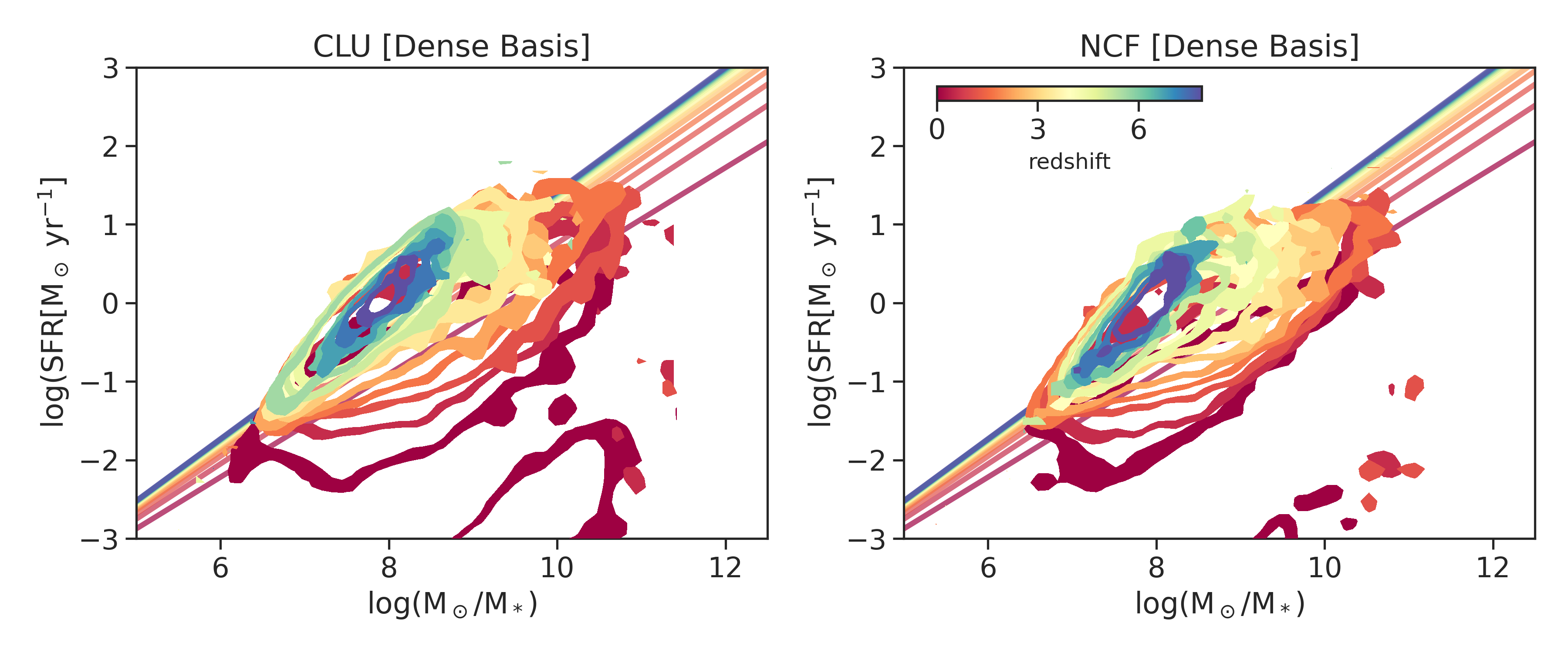}
\plotone{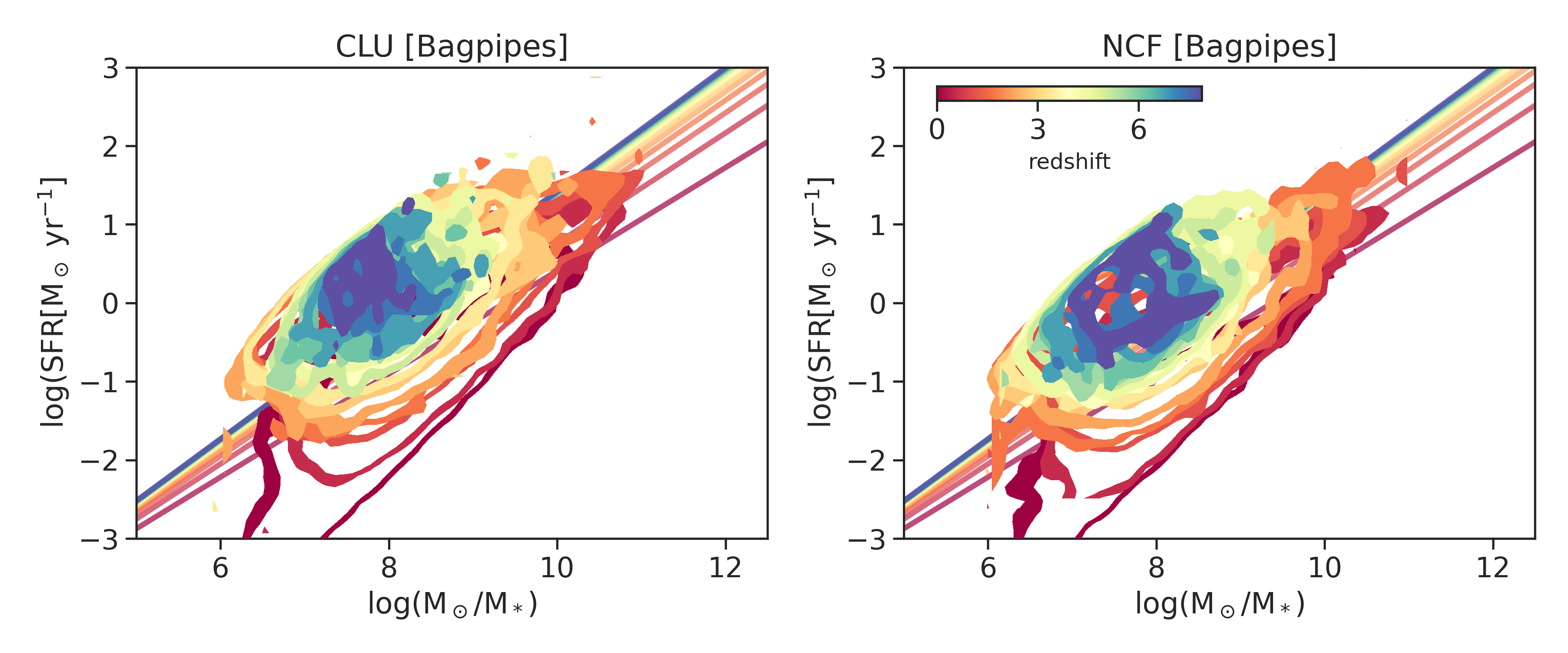}
\caption{Top: Star formation rate versus stellar mass relation for the full CANUCS sample using the \texttt{DENSEBASIS} SED fits. Galaxies in the CLU (left) and NCF (right) fields are shown separately. The color bar indicates redshift. The star-forming main sequence derived in \citet{Iyer2018} is shown for different redshifts as labeled. Bottom: As above using the \texttt{BAGPIPES} SED fits.}
\label{fig:sfr10_main_sequence}
\end{figure*}


%

\section{Photometric Redshifts \& Stellar Population Parameters} \label{sec:SPS}

Along with photometric catalogs and redshifts, we provide catalogs of derived physical properties from SED fitting of the photometry. Sources are fit with two standard SED codes, \texttt{DENSEBASIS} \citep{Iyer2017, Iyer2019} and \texttt{BAGPIPES} \citep{Carnall2018}. We provide results from two codes for a few reasons. First, different codes and modeling assumptions are well-known to systematically affect derived properties \citep[e.g.,][]{Pacifici2023}. Comparing the results from both codes for specific selections of galaxies allows the user to estimate the effect of these uncertainties on the chosen science case. Second, the general agreement between the two codes indicates results are driven primarily by the observations, rather than the fitting methodology. Particularly, identifying galaxies with physical properties in agreement between the two codes allows the user to construct samples which are robust against fitting assumptions/methods. For both codes, we fit observed colors in two apertures with total fluxes scaled by the F277W KRON flux (the 0\farcs3 and 0\farcs7 aperture 'COLOR' fluxes described in Section \ref{sec:photometry}). Redshifts are set to the \texttt{EAzY} photometric redshifts with a Gaussian prior with width 0.05 (\texttt{BAGPIPES}) and 0.1 (\texttt{DENSEBASIS}). When available, redshifts are fixed to spectroscopic redshifts. There are a small number of objects which have failed \texttt{EAzY} fits. For these objects, we allow the SED fitting to fit for redshift along with the stellar population parameters. Details of the fitting process for each code follow.

\begin{deluxetable}{ccc}[h]
\tablecaption{\db\, and \bagpipes\  Model Parameters and  Prior Ranges \label{tab:bagpipes}}
\tablewidth{0pt}
\label{tab:sedparam}
\tablehead{
\colhead{Parameter} & \colhead{Range} & \colhead{Prior} }
\startdata
& \db & \\
 log M$_*$ & (7, 12) & flat \\
 log sSFR &(-14,-7) & flat\\
 $\alpha_{SFH}$ & 3 & Dirichlet \\
 log Z/Z$_\odot$ & (-1.5, 0.25) & flat \\
 z & (0.005,12) & flat$^\dagger$ \\
 A$_V$ & (0, 4) & exp \\
 \hline
 & \bagpipes &\\
dblplaw $\tau$ & (0, 15) & flat \\
dblplaw $\alpha$ & (0.01, 1000) & log10 \\
dblplaw $\beta$ & (0.01, 1000) & log10 \\
log M$_*$ & (6, 13) & flat \\
Z & (0.1, 2.5) & flat \\
z & (0.01, 20) & flat$^\dagger$ \\
A$_V$ & (0, 4) & flat \\
\enddata
\begin{tablenotes}
    \item Priors and parameter choices for the SED fitting codes we use to infer physical properties from our optimal photometry. Both codes use a Chabrier IMF, an ionization parameter $log(U) = -2$, and a Calzetti dust attenuation law. 
    \item $^\dagger$: only used where EAzY fits have failed.
\end{tablenotes}
\end{deluxetable}

\subsection{BAGPIPES}
In \texttt{BAGPIPES} \citep{Carnall2018} we assume a parametric star formation history in the form of a double power law allowing a rising and falling component of the history, a \citet{Calzetti2000} dust attenuation law and \citet{Chabrier2003} initial mass function (IMF). The stellar models used are those of \citet{Bruzual2003}. Nebular emission is modeled using CLOUDY \citep{Ferland2017}. The code uses nested sampling with either the MultiNest \citet{Buchner2014} or nautilus \citet{Lange2023} sampling algorithms. Specific prior settings are reported in Table \ref{tab:sedparam}.

\subsection{DENSE BASIS}
Using \db\, we perform SED fitting using a non-parametric star formation history as described in \citet{Iyer2019}. Briefly, the method describes an SFH using a tuple (M$_*$, SFR, \{t$_X$\}) where the \{t$_X$\} are a set of lookback times describing when the galaxy formed equally spaced quantiles of its total mass, using 3 \{t$_X$\} parameters ($t_{25}$, $t_{50}$, $t_{75}$) in the current analysis. In addition to the SFH, (and similar to \bagpipes\,) we adopt a \citet{Calzetti2000} dust attenuation law and \citet{Chabrier2003} initial mass function (IMF), using the MILES $+$ MIST isochrones and stellar tracks and CLOUDY photoionization models implemented in Flexible Stellar Population Synthesis (FSPS) \citep{Conroy2009, Conroy2010} to model composite stellar and nebular emission. 
Specific prior settings are reported in Table \ref{tab:sedparam}. To amortize the SED computation while fitting large catalogs, an atlas containing $\mathcal{O}(10^6)$ SEDs corresponding to random draws of parameters from the prior distributions is generated for each field prior to fitting.

\begin{figure*}[t]
\plotone{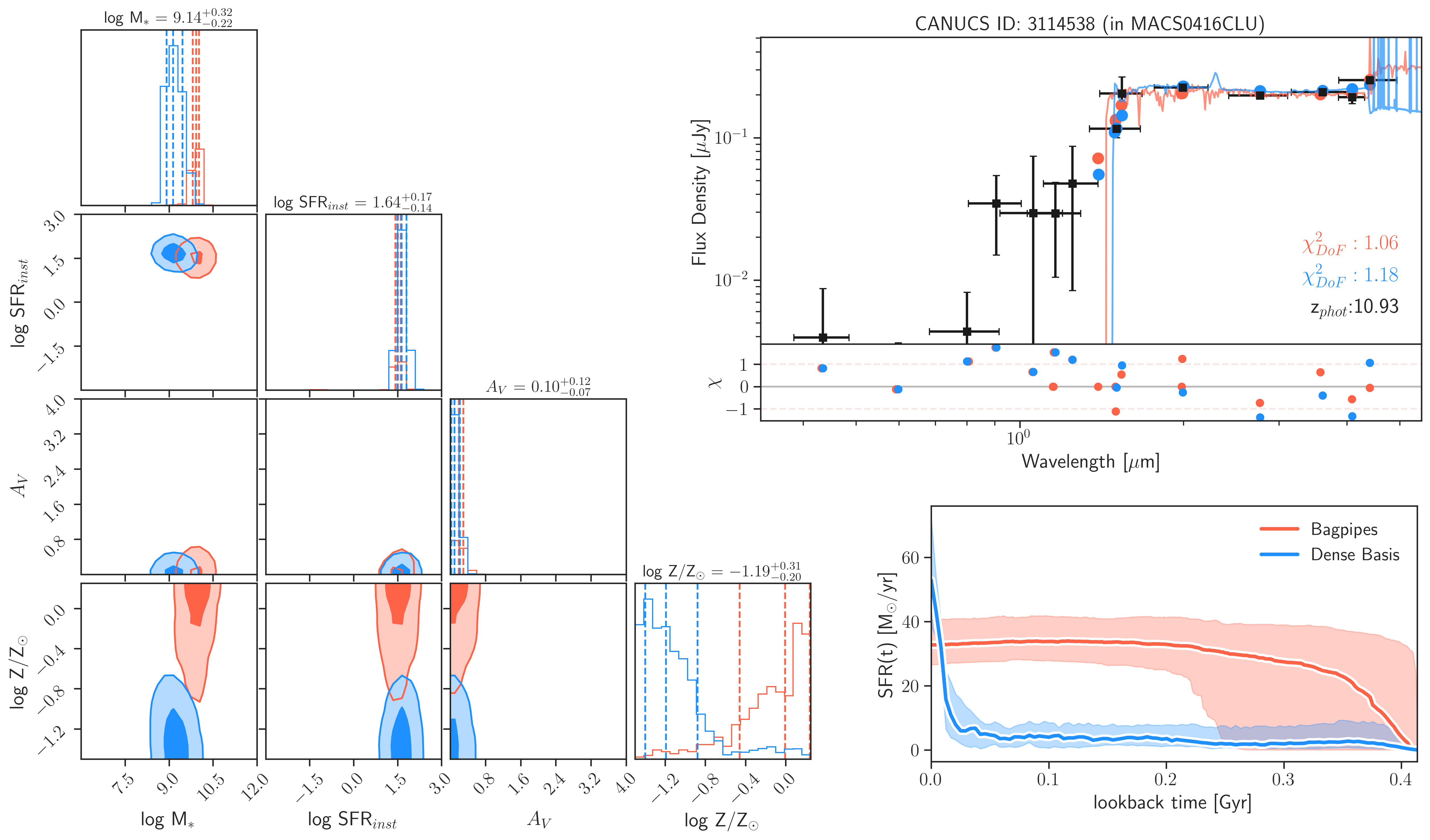}
\plotone{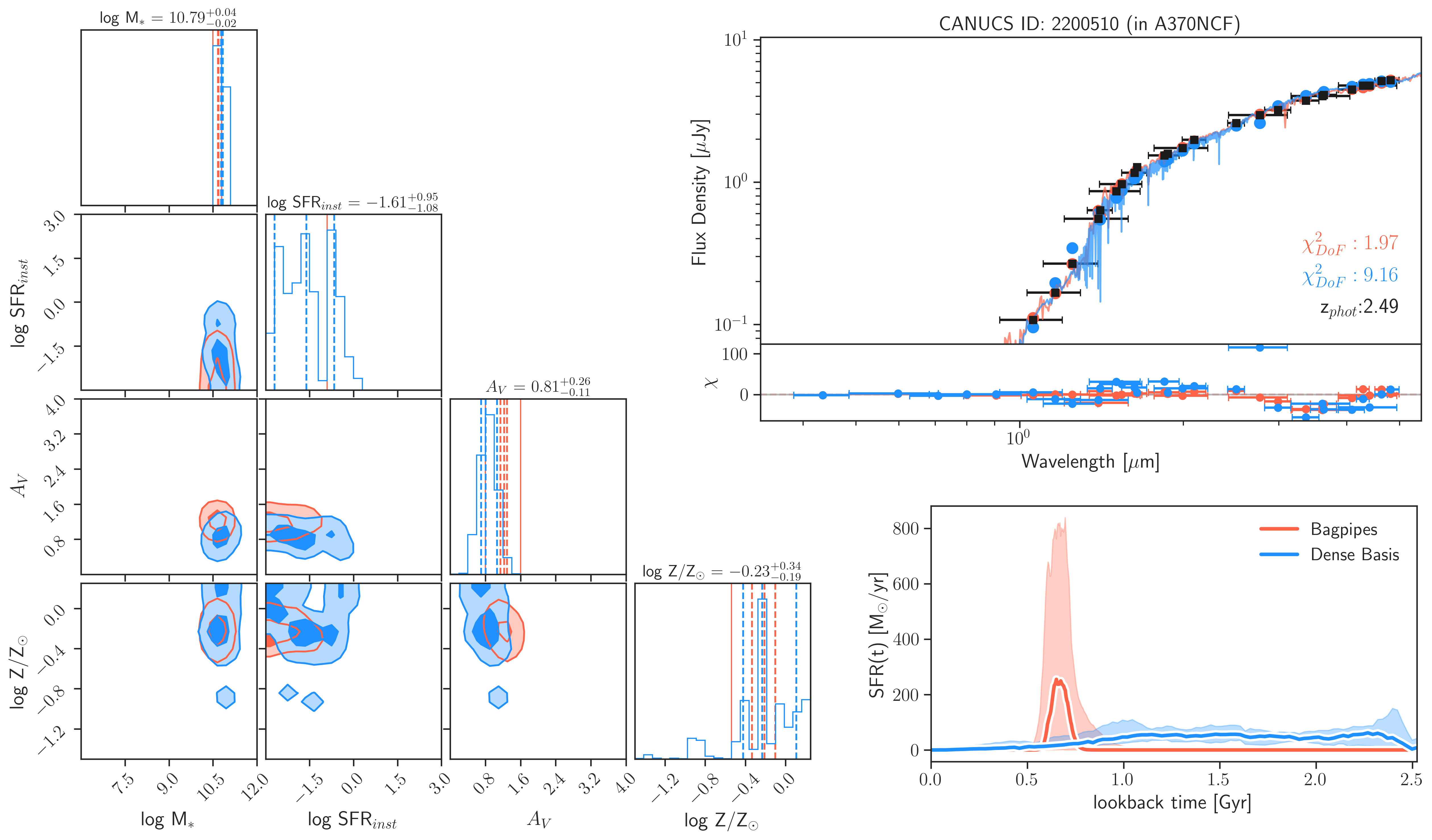}

\caption{SED fitting results for the same selection of galaxies in Figure \ref{fig:SEDs}. For each galaxy, we show the posterior distributions for stellar mass, SFR, A$_V$, and metallicity, the best fit model spectrum, and the star formation history. The reduced $\chi^2$ and photometric redshift are indicated in each panel. Results from \texttt{DENSEBASIS} are shown in blue and \texttt{BAGPIPES} in orange.} \label{fig:example_fits}
\end{figure*}

\subsection{Results}

Here we present basic derived quantities from the SED fits using both codes to validate the fitting procedure. Figure \ref{fig:mass_z} shows the stellar mass versus redshift relation for the full CANUCS sample divided by CLU and NCF. Basic quality cuts on signal to noise ratio, number of bands included in the fitting, and photometry contamination are applied. Recommended cuts for general usage are included with the released catalogs in the form of example notebooks. The cluster galaxy populations are visible, particularly in the CLU panel, as vertical overdensities at the cluster redshifts ($z\sim 0.4$). Figure \ref{fig:sfr10_main_sequence} similarly shows the star formation versus stellar mass relation subdivided by CLU and NCF. A full analysis of the star-forming main sequence in CANUCS will be presented in Mérida et al. (in prep.). Here we only show that we recover the expected correlation between star formation rate and stellar mass with broad consistency to the star-forming main sequence presented by \citet{Speagle2014}.

Some examples of individual SED fits for both codes are shown in Figure \ref{fig:example_fits} and Figures \ref{apx:example_fits_1}-\ref{apx:example_fits_3} in Appendix \ref{apx:db_v_bagpipes}. We show results for the same selection of galaxies in Figure \ref{fig:SEDs}. For each galaxy, we show the posterior distributions for stellar mass, SFR, A$_V$, and metallicity, the best-fit model spectrum, and the star formation history. Results from \texttt{DENSEBASIS} are shown in blue and \texttt{BAGPIPES} in orange. We draw attention that even when both codes produce satisfactory fits to the observed photometry, derived physical properties can differ due to different assumptions and priors, particularly the recipe for the star formation history. 

\begin{figure*}[t]
\hspace{-0.9cm}\begin{tabular}{cc}
CLU & NCF \\
\includegraphics[width=0.5\textwidth]{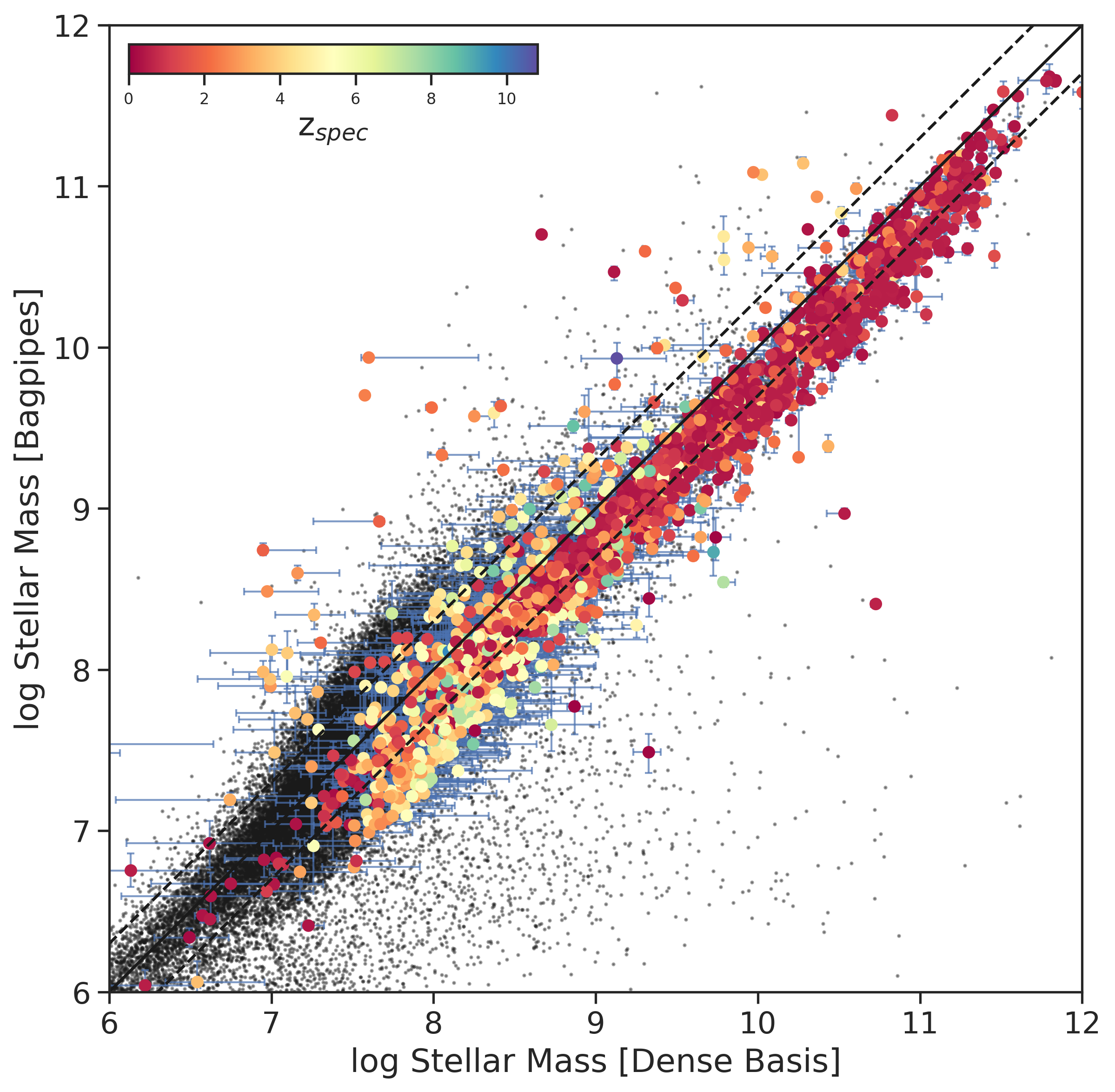}&
\includegraphics[width=0.5\textwidth]{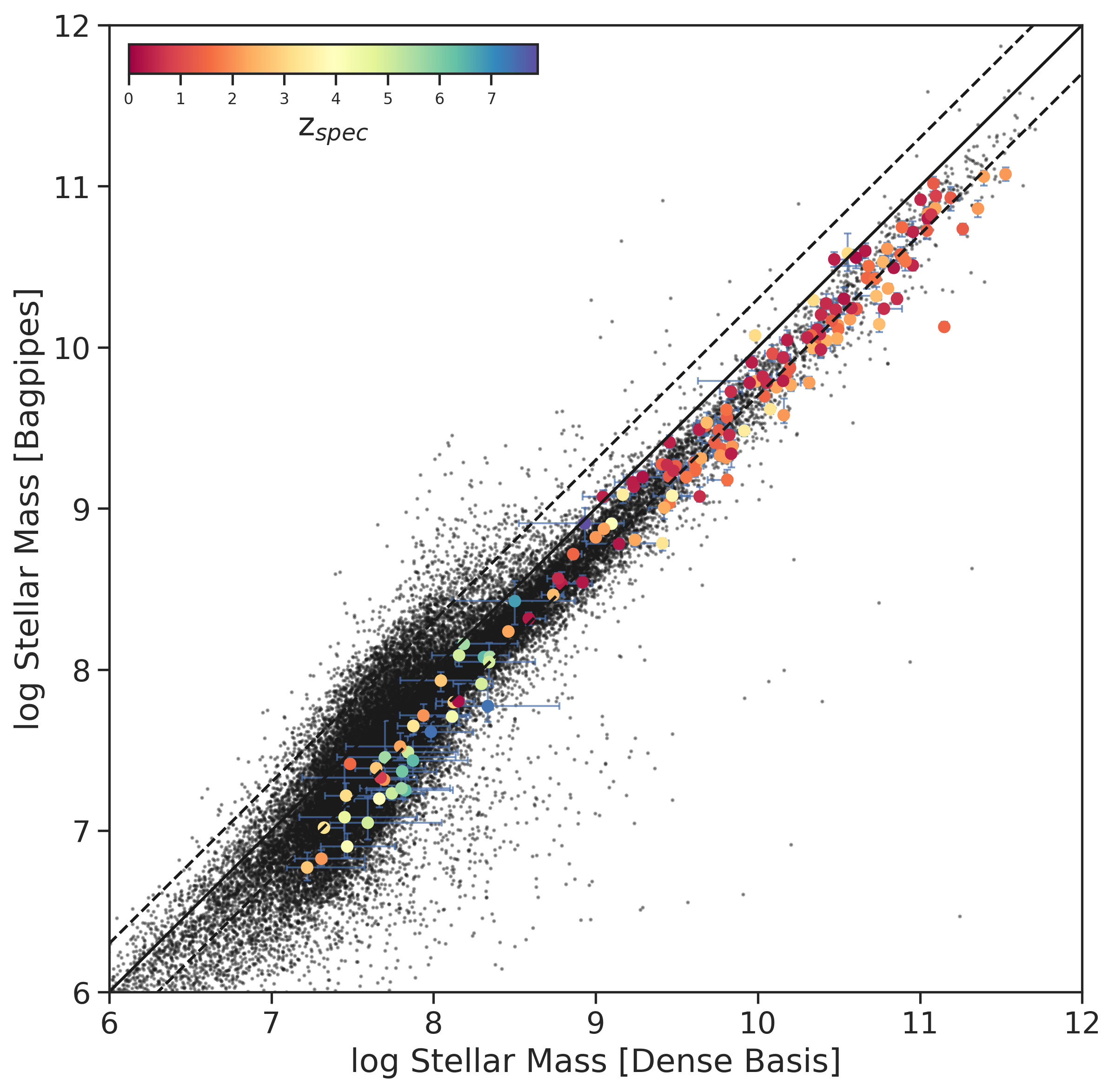}\\
\includegraphics[width=0.5\textwidth]{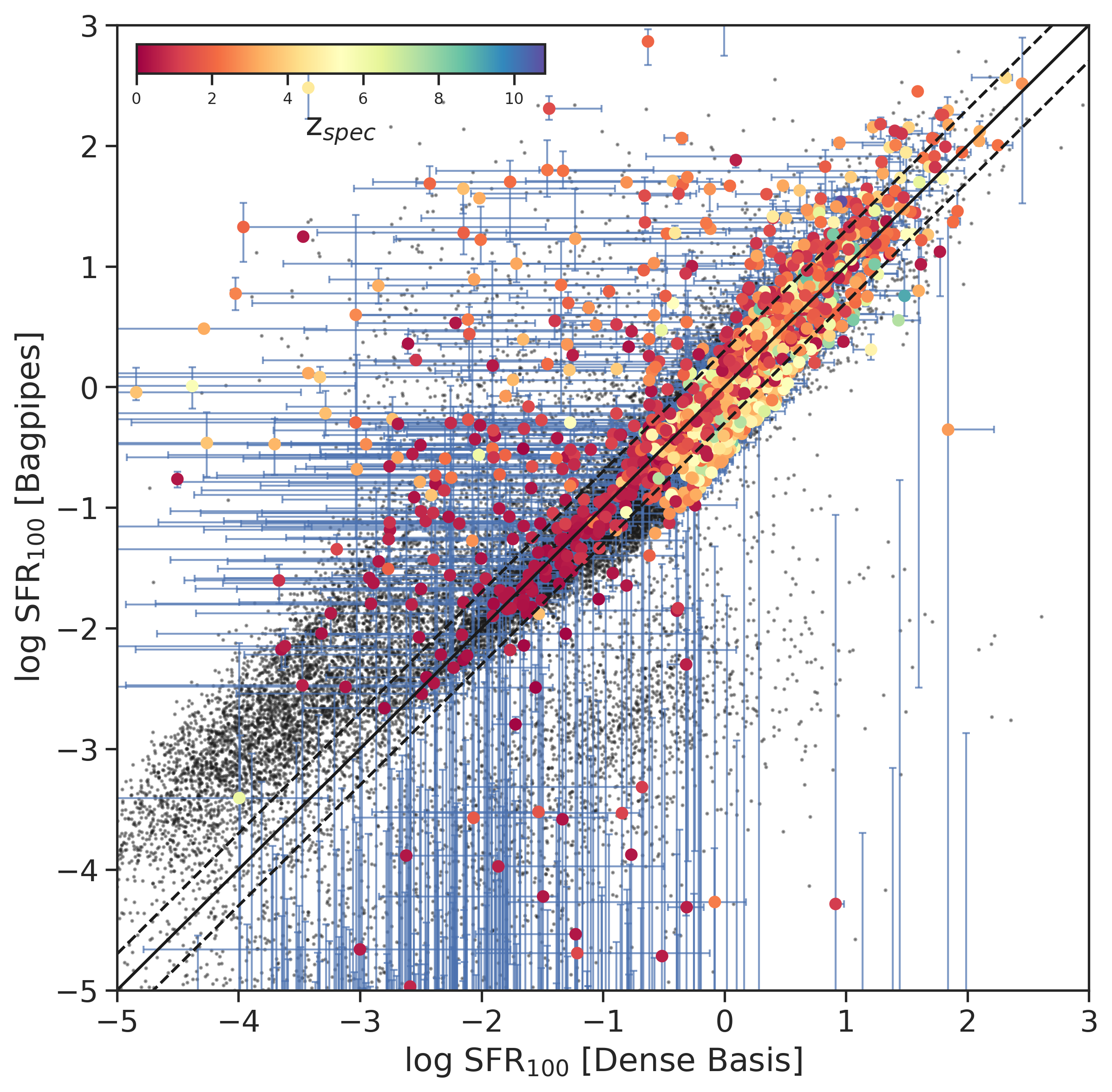}&
\includegraphics[width=0.5\textwidth]{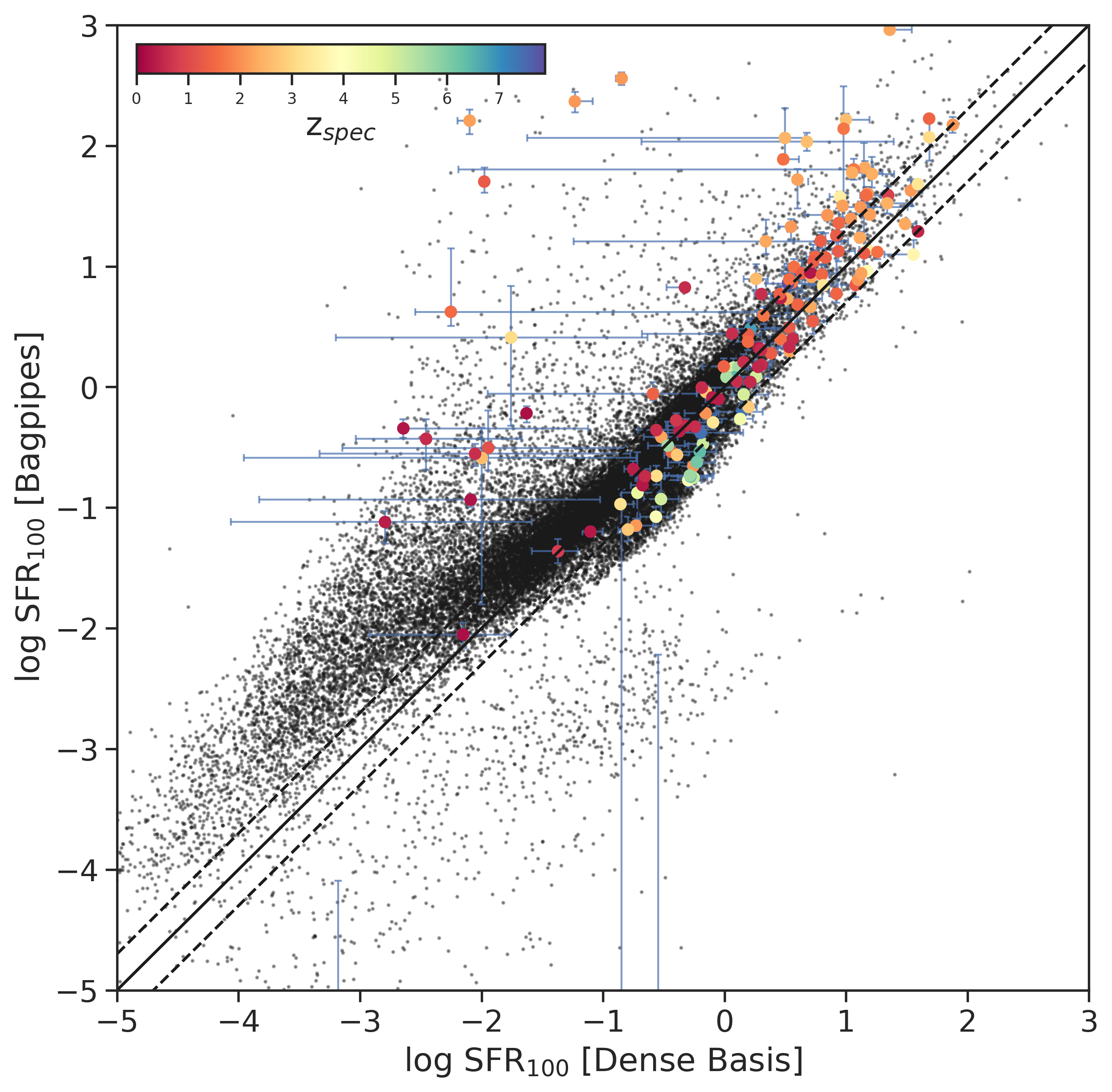}\\
\end{tabular}
\caption{Comparisons of the inferred stellar mass (top) and star formation rate averaged over 100 Myr (bottom) from \texttt{BAGPIPES} and \texttt{DENSEBASIS} SED fits. The left and right columns show galaxies in the CLU and NCF fields respectively. The solid black line shows the 1:1 line and dashed lines show a range of $0.3$ dex. Black points show all galaxies while colored points show the subsample with spectroscopically confirmed redshifts, with corresponding 1$\sigma$ uncertainties from fitting. Masses from the DB are generally higher due to the non-parametric SFHs, which allow additional mass in the form of older stellar populations that can be outshined by younger ones.}
\label{fig:mass_sfr_comparison}
\end{figure*}

Figure \ref{fig:mass_sfr_comparison} directly compares stellar masses (top) and star formation rates (bottom) measured by \texttt{DENSEBASIS} and \texttt{BAGPIPES} in the CLU (left) and NCF (right) fields. Galaxies with spectroscopic redshifts are denoted by colored points. As established by previous works, non-parametric star formation histories generally lead to higher inferred stellar masses \citep[e.g.,][]{Iyer2017, Leja2017, Markov2023, Pacifici2023}. This tendency is partly caused by the more extended non-parametric star formation history allowing the presence of older stellar populations that are outshined by younger ones. We also note a systematic difference in SFR in the  very-low-SFR regime (log(SFR)$\lesssim -2$), such that \texttt{BAGPIPES} finds SFRs $\sim 1$ dex higher. With the released stellar population property catalogs, we provide a flag to alert the user to cases in which the two codes show a significant discrepancy (\texttt{SED\_DISAGREE\_FLAG}). The value of this flag is calculated by comparing the agreement between posteriors from the two codes for five physical properties (stellar mass, current SFR, dust, metallicity and mass-weighted age / $t_{50}$). Since modeling assumptions (e.g. SPS templates, parametric vs non-parametric SFHs etc.) and priors can drive systematic differences in the posteriors, we correct for this effect using the properties of the (N=30) nearest neighbors in physical property space while estimating the posterior agreement. This selects galaxies where the posteriors differ strongly even accounting for differences in modeling assumptions, and should be treated with caution for most science cases. In addition to this, both SED catalogs also contain $\chi^2$ columns from the fitting that can be used to exclude galaxies where the photometry is not well modeled by the different codes, as well as flags (\texttt{DENSEBASIS\_UNCONSTRAINED\_FLAG} and \texttt{BAGPIPES\_UNCONSTRAINED\_FLAG}) that are set when more than one physical property can not be constrained by the SED fits. 

Additionally, we note that users should carefully consider which aperture fits (COLOR03 or COLOR07) are most appropriate for their science. For sources smaller than a diameter of 0\farcs7, the larger aperture adds additional noise to the flux measurements, and tends to drive the fit posteriors to be prior-dominated. On the other hand, for large, extended galaxies, the smaller aperture may not accurately reflect galaxy-wide colors leading to systematics in integrated properties.


\subsection{Rest-frame Colors \& UVJ/ugi Diagrams}


\begin{figure}[t!]
\plotone{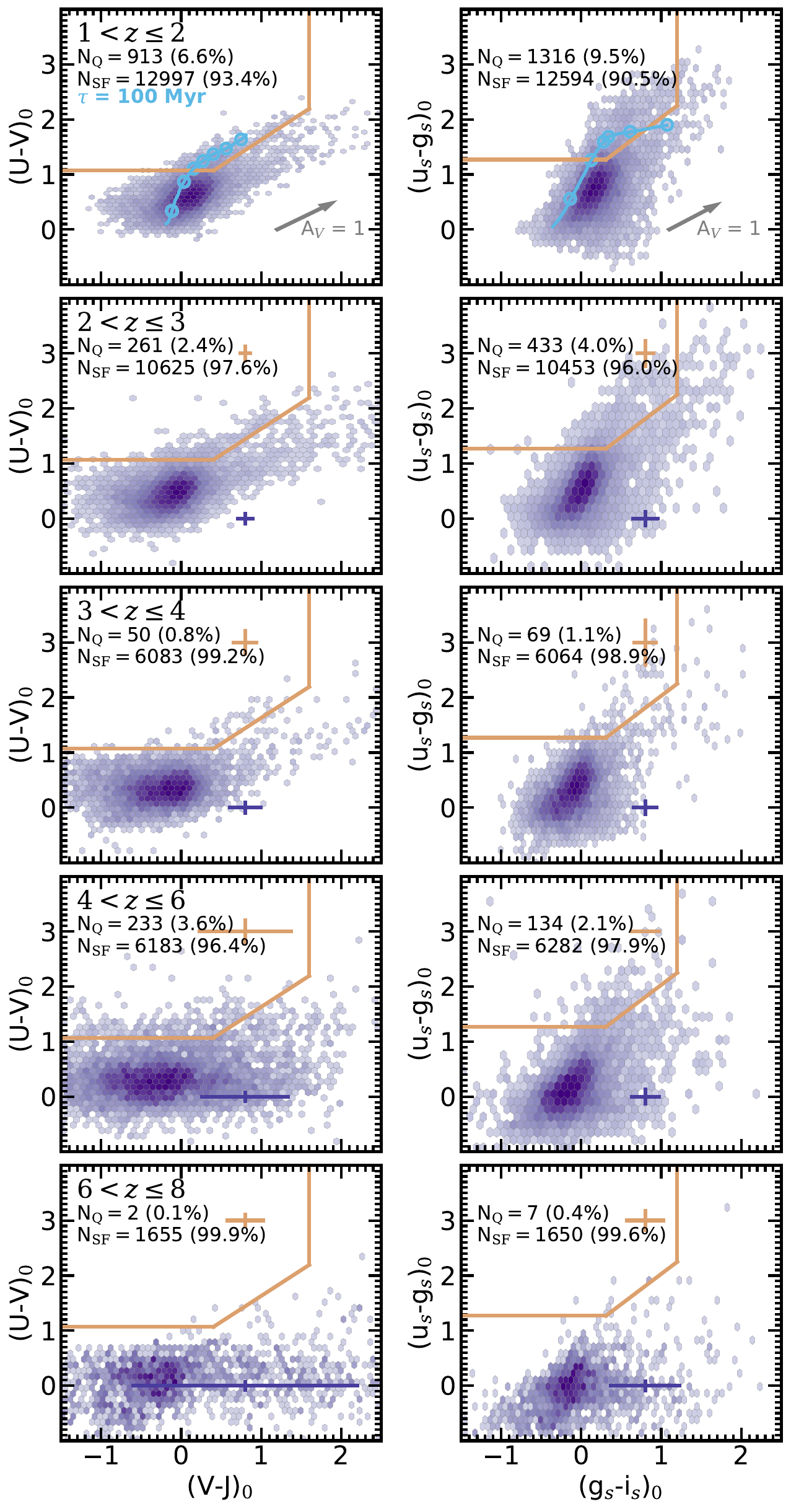}
\caption{UVJ (left) and ugi (right) 2D histograms showing source density in each respective rest-frame color space. Crosses show the 1-$\sigma$ uncertainty on rest-frame colors for quiescent (upper left, inside of the brown wedge) and star-forming galaxies (outside of wedge), respectively. Age tracks shown in blue are from \citet{Antwi-Danso25}.}
\label{fig:UVG_ugi}
\end{figure}


Figure \ref{fig:UVG_ugi} shows UVJ (left) and ugi (right) diagrams for 0.5 $\leq$ z $\leq$ 6, for the all 10 CANUCS fields. 2D histograms show the density distribution, and age tracks the same as \citet{Antwi-Danso25}. Light brown and purple crosses show the 1-$\sigma$ uncertainty on rest-frame colors as a function of redshift for quiescent (inside the upper left wedge) and star forming galaxies (outside of wedge), respectively. 

Following the prescription of \citet{Antwi-Danso2023}, color lines were calibrated for CANUCS on a broad redshift range by iteratively updating the slope and y-intercept of the diagonal line until a clear bimodality emerges between the red and blue sequences.

The UVJ selection criteria are given by:

\begin{align}
    (U-V) > 1.07 \\ (V-J) < 1.6  \\ (U-V) > (V-J) \times 0.92 + 0.69 
\end{align}

The synthetic ugi selection criteria are:

\begin{align}
    (u_s-g_s) > 1.27 \\ (g_s - i_s) < 1.2 \\ (u_s - g_s) > (g_s - i_s) \times 1.1 + 0.93
\end{align}


%
\section{Strong lens modeling}\label{sec:lensing}
We developed strong lensing models of the five galaxy clusters in the CANUCS survey, leveraging the imaging and spectroscopic data presented in this paper. The details of each strong lensing model are published in separate papers (see \citealt{Gledhill24} for Abell 370, \citealt{Rihtarsic25} for MACS0416, Desprez et al., in prep. for MACS0417, Rihtar\v{s}i\v{c} et al., in prep. for MACS1149, and Desprez et al., in prep. for MACS1423). Although the methods varied slightly between clusters, the general procedure is outlined below. 
Each cluster field was first inspected for new multiple images of background sources using the CANUCS imaging data. We used the existing strong lensing models, constrained with pre-JWST multiple image catalogs, to predict the positions of potential counter-images to the sources in the photometric catalogs, assuming \texttt{EAzY} photometric redshifts. We visually inspected a large region ($\sim15''$ to avoid biasing the dataset towards old models) around each prediction and identified multiple image candidates based on their color, photometric redshift and morphology similarity. We confirmed many multiple image systems and obtained their spectroscopic redshifts from the CANUCS NIRISS wide-field slitless and NIRSpec prism spectroscopy. We evaluated the reliability of each multiple-image system and used the most secure systems with spectroscopic redshifts (i.e. the gold catalog) as lens model constraints. 

The lens models were constrained with the parametric lens modeling tool \texttt{Lenstool} \citep{Kneib96,Jullo07,Jullo09}, which models the total (dark matter and baryonic) mass distribution as a superposition of a smooth component modeled with several large-scale halos, and more compact cluster galaxy subhalos. The large-scale smooth halos were described with a dual pseudo isothermal elliptical (dPIE) density profile \citep{2005MNRAS.356..309L,2007arXiv0710.5636E} with eight free parameters. Cluster galaxy subhalos were modeled with smaller dPIE halos, following scaling relations with free normalization \citep{1996MNRAS.280..167J, 1997MNRAS.287..833N}. Some galaxies were modeled separately from the scaling relations (e.g. in the cases of galaxy-galaxy lensing). In MACS 0416 and MACS 1149 we also included the cluster gas component, fitted to X-ray observations \citep[e.g.][]{2018ApJ...864...98B,2017ApJ...842..132B}. 
In most clusters, we tested several different parameterizations and selected the best one using criteria such as $\Delta_{\rm rms}$, Bayesian information criterion (BIC; \citealt{Schwarz78}) and $\chi^2$ per degree of freedom (DoF). We reran the models after rescaling the multiple image position uncertainty so that $\chi^2/\mathrm{DoF}=1$ to get more conservative error estimates and take into account the systematic inaccuracies. 

We used the new lens models to derive the magnification $\mu$ values of the sources in the photometric catalogs. To this end, we use the best-fit lens models and 100 randomly drawn Bayesian samples from the end of the Markov chain Monte Carlo sampling phase when the $\chi^2$ value becomes stable. We used \texttt{EAzY} photometric redshifts or spectroscopic redshifts if available (keyword \texttt{Z\_SPEC} in the photometric catalog). For each source with a valid redshift, we provide magnification values \texttt{MU}, derived from the best-fit lensing model and the 16-th, 50-th and 84-th percentiles (\texttt{MU\_16}, \texttt{MU\_50} and \texttt{MU\_84}, respectively), derived from the Bayesian sample of lens models. Similarly, we provide the best-fit values and percentiles for the eigenvalues of the magnification tensor  \texttt{MU\_TAN} and \texttt{MU\_RAD}, measuring the magnification in the tangential and radial direction, respectively. Cluster members included in the lens models were assigned magnifications of 1 to avoid spurious magnification values derived from their photometric redshifts (for selection of the cluster members refer to individual lens modelling papers. The cluster member catalogs in \texttt{Lenstool} format will be provided with the data release). Apart from magnifications, we also provide convergence, shear, shear component and deflection angle maps (see Sect.~\ref{sec:product_summary}) so a user can calculate magnification and other lensing properties for a given position and redshift of any source.

We note that while magnifications in the NCF fields are much smaller than in the CLU field, they are not negligible and are provided in the catalogs. The upper limit on best-fit magnifications \texttt{MU} in all NCF fields apart from MACS 0417 is 1.2 and the maximum magnification uncertainty ($\mu_{84}-\mu_{16}$) is 0.01. In MACS 0417 NCF field, the upper limit on magnification is 1.6 and the maximal uncertainty is 0.2 (the uncertainty scales with magnification - the upper limit of uncertainty can be estimated as $0.4( \mu -1)$ in MACS 0417 NCF field).



\begin{deluxetable*}{lll}[t]
    \label{tab:data_prod_info}
    \tablecaption{Data Products Included in DR1
  	}
    
    \tablecolumns{3}
    \tablewidth{0pt}
    \tablehead{
    \colhead{Product Type} & \colhead{Filename} & \colhead{Comment} 
    }
    \startdata
        Imaging & \texttt{sci} & Science image \\
        Imaging & \texttt{rms} & RMS uncertainty image \\
        Imaging & \texttt{bgsub-sci} & Background- \& bCG-subtracted science image \\
        Imaging & \texttt{bcgmodel-sci} & Image of the bCG models (CLU only) \\
        Imaging & \texttt{bgsub-psfconv-sci} & Common psf-convolved background- \& bCG-subtracted science image \\
        Imaging & \texttt{psf} & PSF image \\
        Imaging & \texttt{convkernel} & Image of kernel used in PSF convolution \\
        Imaging & \texttt{detection} & Detection image used for source detection \\
        Imaging & \texttt{segmentation} & Segmentation image of detected sources \\
        \hline
        Photometry & \texttt{photometry\_v1\_cat} & PSF-matched photometric catalog \\ 
        SPS & \texttt{bagpipes-cat} & \texttt{bagpipes} SPS catalogs \\  
        SPS & \texttt{densebasis-cat} & \texttt{dense basis} SPS catalogs \\  
        \hline
        Lensing & \texttt{kappa\_dlsds1\_best\_XXmas} & Best-fit convergence map at 50, 100, \& 300mas resolution\\  
        Lensing & \texttt{gamma\_dlsds1\_best\_XXmas} & Best-fit shear map at 50, 100, \& 300mas resolution\\  
        Lensing & \texttt{gamma1/2\_dlsds1\_best\_XXmas} & Best-fit shear component map at 50, 100, \& 300mas resolution\\  
        Lensing & \texttt{dpl1/2\_dlsds1\_best\_XXmas} & Best-fit deflection component map in arcsec at 50, 100, \& 300mas resolution\\  
        Lensing & \texttt{kappa\_dlsds1\_samples\_XXmas} & Convergence map samples at 200, \& 600mas resolution\\  
        Lensing & \texttt{gamma\_dlsds1\_samples\_XXmas} & Shear map samples at 200, \& 600mas resolution\\  
        Lensing & \texttt{gamma1/2\_dlsds1\_samples\_XXmas} & Shear component map samples at 200, \& 600mas resolution\\  
        Lensing & \texttt{dpl1/2\_dlsds1\_samples\_XXmas} & Deflection component map samples at 200, \& 600mas resolution\\  
        Lensing & \texttt{lenstool\_*.txt} & Lens model inputs for reproducibility and best-fit parameter files \\  
        \hline
        Spectra & \texttt{*\_prism-clear\_1208\_*.spec.fits} & 1D and 2D NIRSpec prism spectra \\
    \enddata
\end{deluxetable*}

\section{Summary of Specific Data Products} \label{sec:product_summary}

Here we discuss specific details of products included in this first data release. Filepath naming conventions for every type of data product are included below in Table \ref{tab:data_prod_info}. Data products are grouped by Imaging, Photometry \& SPS fitting, Lensing and Spectra.

\subsection{Imaging Products}

In each field, we release for each available filter the original \texttt{sci} images, background and bCG models (bCG models only available in CLU), bCG- and background-subtracted images (bCG-subtraction only in CLU), PSFs, PSF-matched images, and kernels used for the convolution, all at the 40mas pixel scale. For NIRCam imaging in the short-wave channel (i.e. all filters bluer than F200W/F210M) we also release imaging data products at the 20mas pixel scale. A list of imaging and photometry products and filenames are listed in Table \ref{tab:data_prod_info}.

Note that all files begin with a common naming convention form for the observatory, instrument, target, field, and filter, e.g. $\texttt{hlsp\_canucs\_jwst\_nircam\_macs0416-clu-}$ $\texttt{-\_f150w\_v1}$ for JWST NIRCam F150W in the cluster field of MACS0416.

\subsection{Photometry \& SPS Catalogs}

Photometric catalogs measured on the PSF-matched images at 40mas are provided for each field, as well as SPS catalogs prepared using \bagpipes  \& \texttt{DENSE\_BASIS}. Filename conventions for all catalogs is given in Table \ref{tab:data_prod_info}, and a detailed table listing photometric catalog column names and definitions is given in Appendix \ref{apx:catalog_entries}.

\subsection{Lensing Products}

We provide the following lensing maps for each CLU field: convergence maps (kappa), shear maps (gamma) with both components (gamma1, gamma2), and deflection maps (x and y components) given in arcsec (dpl1, dpl2). The maps, derived from the best-fit lens models are given at 50mas, 100mas, \& 300mas pixel scales. We also provide maps for 100 Bayesian realizations of the lens model that sample the lens model uncertainties. The sample maps are given as multi-extension fits files with 100 extensions. They are provided at 200 and 600mas pixel scales. All maps are given for Dds/Ds=1. For each field we also provide the cluster member catalog in \texttt{Lenstool} format. For published models (MACS0416 and Abell 370), we additionally provide the multiple image catalogs, \texttt{Lenstool} input parameter file and the output parameter file with best-fit parameter values for lens modelling reproducibility. Filename details for all lensing data products are given in Table \ref{tab:data_prod_info}. The updated versions (v2) of the MACS0417, MACS1423 and MACS1149 magnification catalogs, lensing maps and the \texttt{Lenstool} input files will be available after the publication of the corresponding lens modelling papers.

\subsection{Spectra}

NIRSpec prism spectra are made available in a queryable table accessed at https://niriss.github.io/data.html. This table contains all spectra obtained in program 1208, including those in the ``HST-only" regions not covered by the DR1 catalogs and those that did not yield useful redshifts. The four columns SOURCE, Z\_SPEC, ID\_SPECZ and Z\_Q\_REF from the photometric catalogs are included in the table for cross-matching purposes, along with other information pertaining to the MSA observations and independent redshift fits.


\section{Summary} \label{sec:summary}

In this paper we have presented the CANUCS and JWST in Technicolor Cycle 1 \& 2 surveys and the release of first-generation data products for NIRCam and NIRISS imaging and NIRSpec MOS spectroscopy. The CLU fields, centered on the central region of each target galaxy cluster, use both NIRISS and NIRCam to provide NIRCam imaging in 8 filters from 0.9-4.4$\mu$m and continuous NIRISS coverage in F115W, F150W, and F200W (Technicolor adds F090W in the 3 Frontier Fields clusters) for up to 12 JWST filters. The NCF fields offer a suite of 5 wide and 9 medium band filters, and the addition of 8 filters with JWST in Technicolor completes the full set of wide and medium bands, as well as two narrow bands, for up to 22 NIRCam filters spanning 0.7-4.8$\mu$m. All 5 CLU fields and MACS0417 NCF were followed up with low-resolution NIRSpec prism spectroscopy, yielding 747 spectroscopic redshifts within the NIRCam-imaged regions. 

The CANUCS + Technicolor fields constitute some of the most unique and varied data sets taken with JWST to date. With photometric catalogs reaching 3-$\sigma$ depths of \tilda29.5-30mag, in addition to NIRSpec spectra and NIRISS WFSS to come in future, the CLU strong lensing fields and NCF medium band fields add to JWST's growing legacy.


%


%
\clearpage
\begin{acknowledgments}
This work is based on observations made with the NASA/ESA/CSA JWST. The data were obtained from the Mikulski Archive for Space Telescopes at the Space Telescope Science Institute, which is operated by the Association of Universities for Research in Astronomy, Inc., under NASA contract NAS5-03127 for JWST. This research was supported by grants 18JWST-GTO1, 23JWGO2A13, and 23JWGO2B15 from the Canadian Space Agency (CSA), and funding from the Natural Sciences and Engineering Research Council of Canada (NSERC). Support for program JWST-GO-03362 was provided through a grant from the STScI under NASA contract NAS5-03127. This research used the Canadian Advanced Network For Astronomy Research (CANFAR) operated in partnership by the Canadian Astronomy Data Centre and The Digital Research Alliance of Canada with support from the National Research Council of Canada, the Canadian Space Agency, CANARIE and the Canadian Foundation for Innovation. 
YA is supported by JSPS KAKENHI Grant Number 23H00131.
GR, MB, NM, AH, VM, GF, JJ, RT acknowledge support from the ERC Grant FIRSTLIGHT, Slovenian national research agency ARIS through grants N1-0238 and P1-0188, and the program HST-GO-16667, provided through a grant from the STScI under NASA contract NAS5-26555. GW gratefully acknowledges support from the National Science Foundation through grant AST-2205189. The Cosmic Dawn Center is funded by the Danish National Research Foundation (DNRF) under grant \#140. This work is based on data and catalog products from HFF-DeepSpace, funded by the National Science Foundation and Space Telescope Science Institute (operated by the Association of Universities for Research in Astronomy, Inc., under NASA contract NAS5-26555). 
\end{acknowledgments}


\vspace{5mm}
\facilities{\emph{HST}/ACS, \emph{HST}/WFC3\_UVIS, \emph{HST}/WFC3\_IR, \emph{JWST}/NIRCam, \emph{JWST}/NIRISS, \emph{JWST}/NIRSpec}


\software{grizli \citep{Brammer2019},
          EAZY \citep{Brammer2008},
          Dense Basis \citep{Iyer2019},
          SEP \citep{Barbary2016},
          photutils \citep{larry_bradley_2023_7946442},
          astropy \citep{astropy:2022}
          Source Extractor \citep{BERTINE.ARNOUTS1996}
          }


\section*{Data Availability}

Data products presented in this paper are available online through the CANUCS website\footnote{https://niriss.github.io/data.html}, with some products also as MAST High-Level Science Products\footnote{https://archive.stsci.edu/hlsp/canucs}. The CANUCS DOI is doi:10.17909/ph4n-6n76. The JWST in Technicolor DOI is doi:10.17909/cyh7-mm53. The data release HLSP DOI is doi:10.17909/18nv-np70.

\newpage

\appendix

\section{CLU Field RGB Images}\label{apx:clu_RGBs}

\begin{figure*}[h]
\includegraphics[width=1.0\textwidth]{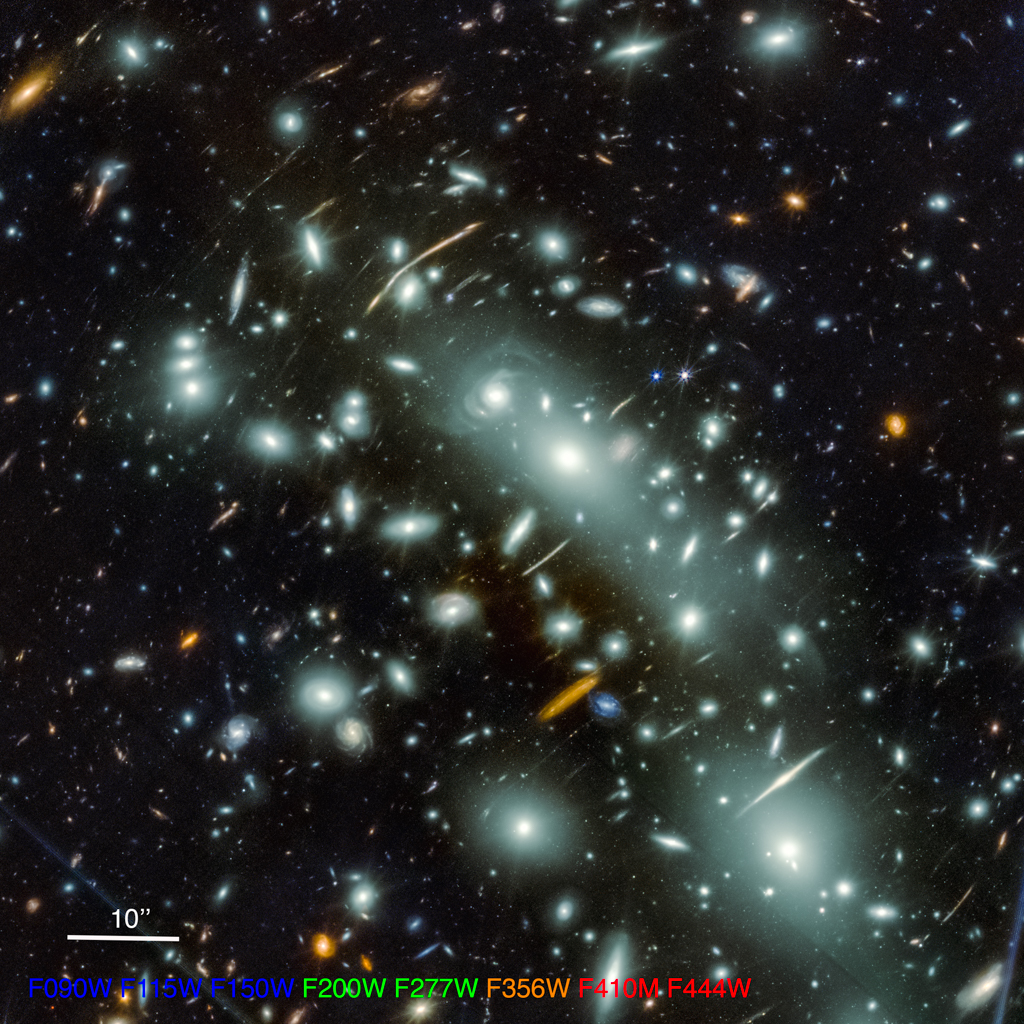}
\caption{RGB image of the center of MACS0416 CLU North is up and East is toward the left.}
\label{apx:macs0416RGB}
\end{figure*}

\begin{figure*}[h]
\includegraphics[width=1.0\textwidth]{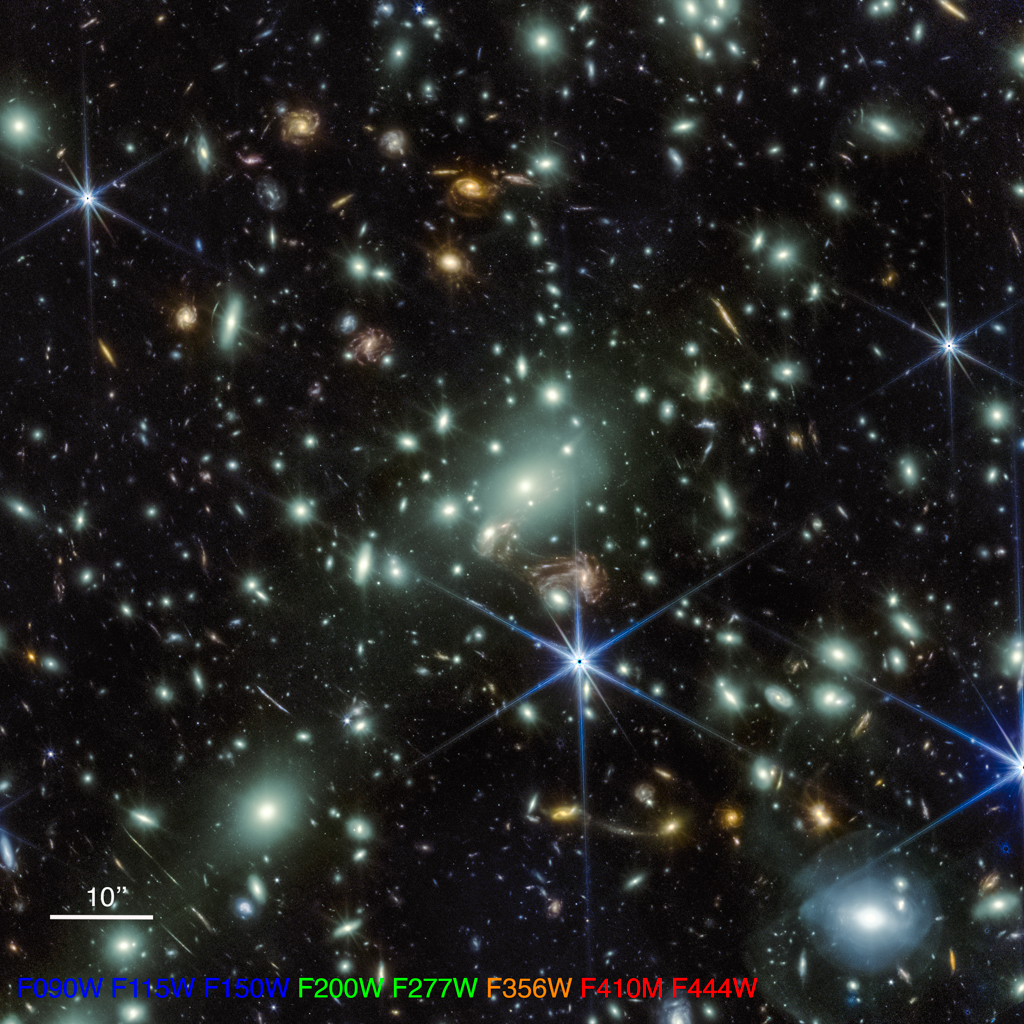}
\caption{RGB image of the center of MACS1149 CLU North is up and East is toward the left.}
\label{apx:macs1149RGB}
\end{figure*}

\begin{figure*}[h]
\includegraphics[width=1.0\textwidth]{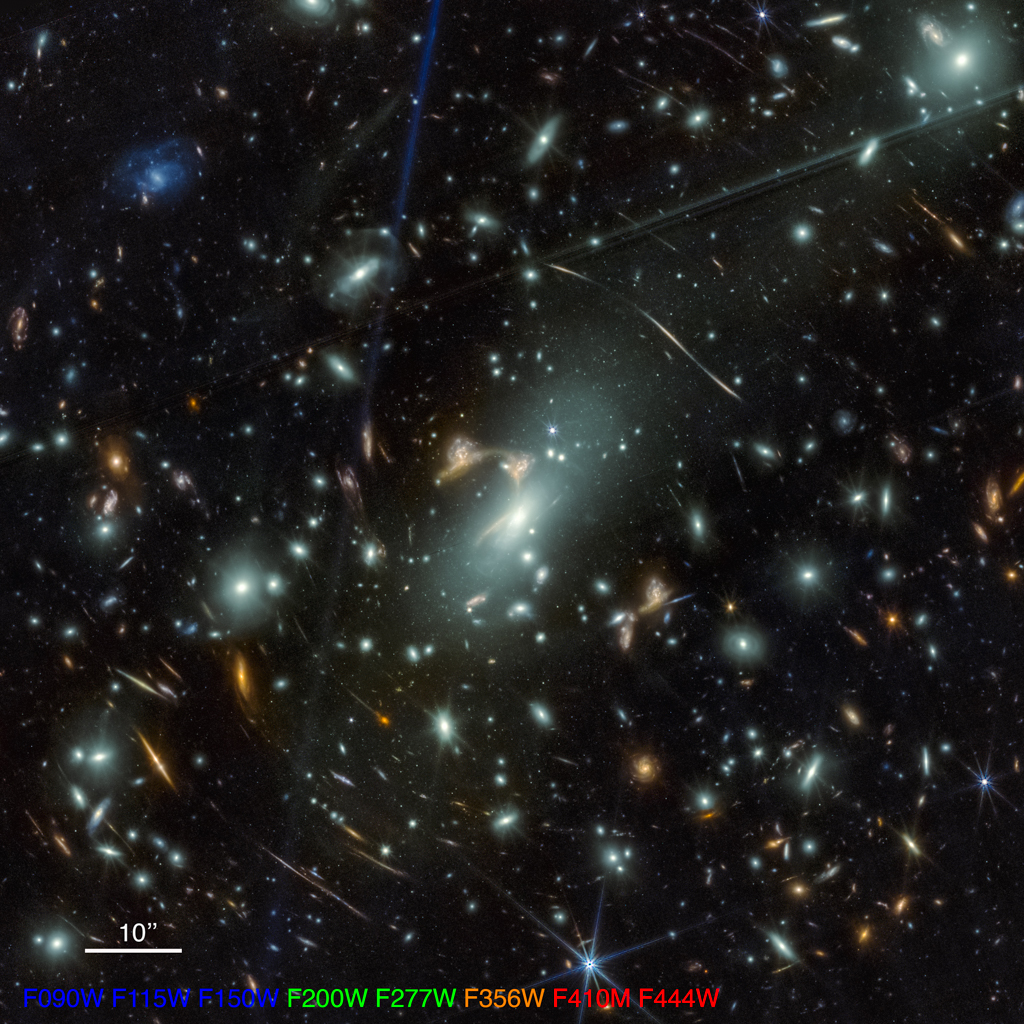}
\caption{RGB image of the center of MACS0417 CLU North is up and East is toward the left.}
\label{apx:macs0417RGB}
\end{figure*}

\begin{figure*}[h]
\includegraphics[width=1.0\textwidth]{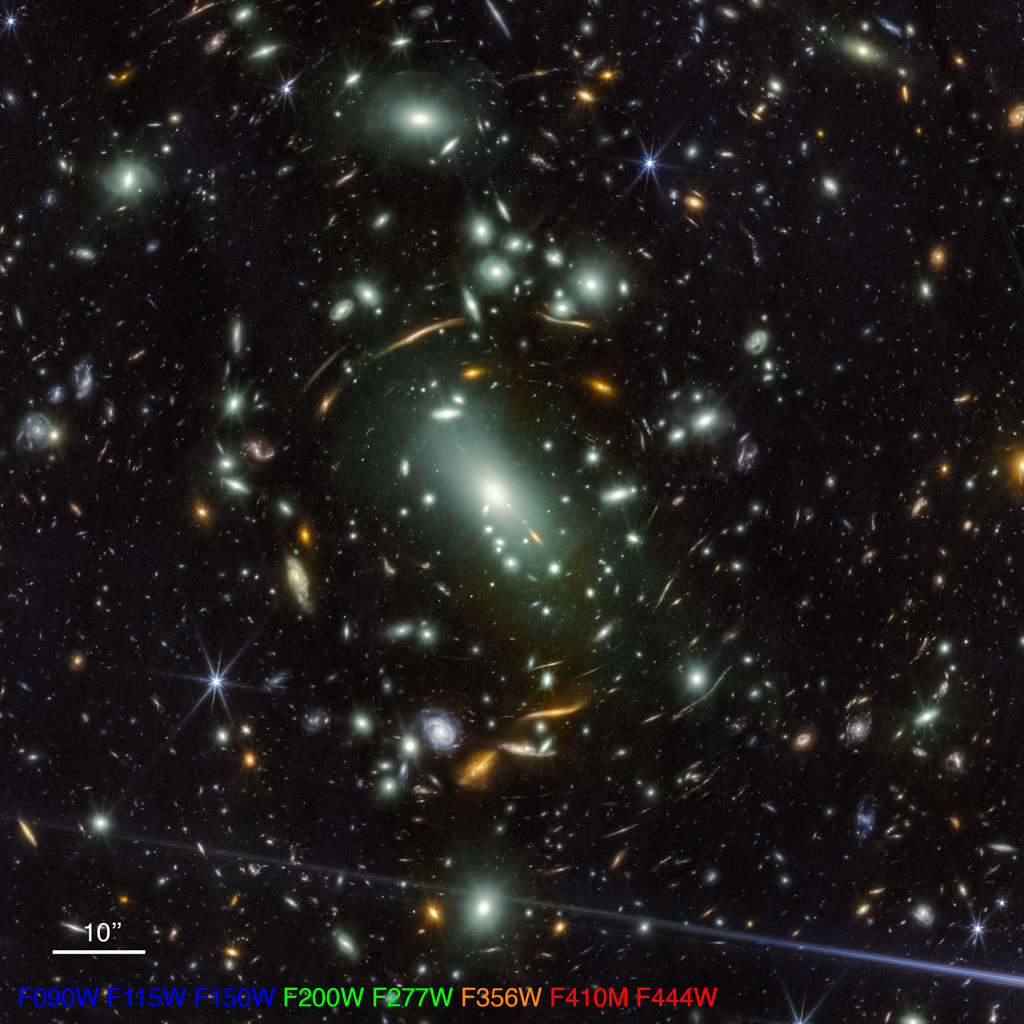}
\caption{RGB image of the center of MACS1423 CLU North is up and East is toward the left.}
\label{apx:macs1423RGB}
\end{figure*}

Figures \ref{apx:macs0416RGB}-\ref{apx:macs1423RGB} show RGB images for the central cluster core of MACS0416, MACS1149, MACS0417, MACS1423, respectively.

\newpage

\newpage

\newpage

\newpage
\newpage
\newpage
\newpage

\section{Catalog entries}\label{apx:catalog_entries}
In this appendix, we give the full details of the table entries in our photometric catalog (\texttt{photometry\_v1\_cat}).
Table \ref{tab:catalog_cols} presents the full entries in the catalog with a short descriptions of each.
We also elaborate more on several entries that are not described in the main text in the following. Readers should refer to the \texttt{README} file as well.

\subsection{The source ID (\texttt{SOURCE} column)}
CANUCS source IDs are 7-digit number. The first digit denotes the lensing cluster where the source is located, and the second digit indicates the pointing in the cluster (CLU or NCF). The last remaining five digits are the unique source ID of the source. For the first digit, 1$=$MACS0417, 2$=$Abell370, 3$=$MACS0416, 4$=$MACS1423, and 5$=$MACS1149. For the second digit, 1$=$CLU and 2$=$NCF.
For example, a source ID of 2100281 in the catalog means ID=281 source in the Abell370 CLU field.

\subsection{Two master flags for problematic photometry (\texttt{FLAG\_FOV\_MASTER\_XX} and \texttt{FLAG\_BPR\_MASTER\_XX})}
There are two master flag columns for each filter (\texttt{FLAG\_FOV\_MASTER\_XX} and \texttt{FLAG\_BPR\_MASTER\_XX}, where \texttt{XX} is the filter name).
These master flags is to identify sources with problematic photometry in the filter, for each aperture available in the catalog separately.
These flags are 8-digit numbers, and each digit corresponds to the eight photometry apertures quoted in the catalog -- Kron, COLOR03, COLOR07, $0.\!\!^{\prime\prime}3$-, $0.\!\!^{\prime\prime}5$-, $0.\!\!^{\prime\prime}7$-, $1.\!\!^{\prime\prime}5$-, $3.\!\!^{\prime\prime}0$-diameter circular aperture photometry, respectively.
"1" indicates one or more problematic pixels are included in the aperture, while "0" means free from them.
The flag \texttt{FLAG\_FOV\_MASTER\_XX} is to set for pixels that are not covered by the filter \texttt{XX} footprint, and the flag \texttt{FLAG\_BPR\_MASTER\_XX} is set for bad pixels identified in the weight maps.
For example, if a source has "11100011" for \texttt{FLAG\_BPR\_MASTER\_XX} flag, then the Kron, COLOR03, COLOR07, $1.\!\!^{\prime\prime}5$-, $3.\!\!^{\prime\prime}0$-diameter circular aperture photometry are not reliable since the aperture contains one or more bad pixels in the filter \texttt{XX}, while $0.\!\!^{\prime\prime}3$-, $0.\!\!^{\prime\prime}5$-, and $0.\!\!^{\prime\prime}7$-diameter circular aperture photometry are fine.

This flag is useful when the user wants to build a sample using their favorite reference filter and/or apertures for their science.
For most of quality flags in the catalog, we use F277W for the reference filter and Kron or $0.\!\!^{\prime\prime}3$-diameter aperture.
This may not be desired for some science cases, e.g., F410M - F444W medium-band color excess selection does not require valid F277W photometry in principle, but rather needs to select valid F410M and F444W phtometry in the aperture of interest.
Users can make a sample based on the master flags in F410M and F444W filters in such a case.

\subsection{Flag for HST/ACS observed sources (\texttt{FLAG\_ACS})}
Optical observations by \hst\ are important particularly in $z<6$ galaxy science, since the rest-frame UV emissions including the Lyman break are observed at $\lambda_{\rm obs}<0.8\ \mu$m where NIRCam filters do not cover (except for F070W; F070W observation is available only in the three Technicolor NCF fields).
Each of the 10 fields has two or more \hst\ optical filter observations from \hst/ACS WFC or \hst/WFC3 UVIS instrument (see Table \ref{tbl:filts_and_fields}), and users can select sources that are observed with all optical \hst\ filters available in the field via the \texttt{FLAG\_ACS} flag.
Note that the flag name is the same even in \hst/WFC3 UVIS fields (MACS0417 NCF and MACS1423 NCF).

\subsection{Literature reference for spectroscopic redshifts (\texttt{REF\_SPECZ})}
The reference for spec-$z$ is given (when $z_{\rm spec}$ is available) in the \texttt{REF\_SPECZ} column.
Reference for our CANUCS NIRSpec observations presented in this work is "\texttt{SA25}".

From ancillary $z_{\rm spec}$ measurements, we use the following catalogs in each field: MACS0417 has two catalogs from \citet["\texttt{jauzac19}"]{Jauzac2019MNRAS} and V. Sok et al. (in prep., "\texttt{Viz25}"); Abell370 has five catalogs from GLASS \citep["\texttt{treu15,schmidt14}";][]{Schmidt2014ApJ,Treu2015ApJ}, \citet["\texttt{lagattuta17}"]{Lagattuta2017MNRAS}, \citet["\texttt{shipley18}"]{Shipley2018}, \citet["\texttt{richard20}"]{Richard2021AA}, and V. Sok et al. (in prep., "\texttt{Viz25}"); MACS0416 has five catalogs from CLASH \citep["\texttt{clash}";][Rosati et al. in prep.]{Balestra2016ApJS,Caminha2017AA}, GLASS \citep["\texttt{glass}";][]{Schmidt2014ApJ,Treu2015ApJ,Hoag2016ApJ}, \citet["\texttt{jauzac}"]{Jauzac2014}, \citet["\texttt{shipley18}"]{Shipley2018}, and \citet["\texttt{richard20}"]{Richard2021AA}; MACS1423 has two catalogs from GLASS \citep["\texttt{glass}";][]{Schmidt2014ApJ,Treu2015ApJ} and V. Sok et al. (in prep., "\texttt{Viz25}"); MACS1149 has four catalogs from GLASS \citep["\texttt{glass2}";][]{Schmidt2014ApJ,Treu2015ApJ}, CLASH \citep["\texttt{clash}";][]{Grillo2016ApJ,Treu2016ApJ}, \citet["\texttt{shipley18}"]{Shipley2018}, and V. Sok et al. (in prep., "\texttt{Viz25}").

\begin{deluxetable}{lcl}
    \label{tab:catalog_cols}
    \tablecaption{Catalog entries
  	}
    \tablewidth{0pt}
    \tablehead{
    \colhead{Column name} & \colhead{unit} & \colhead{Description}
    }
    \startdata
		\texttt{SOURCE} &  & Unique identifier of the source \\
        \texttt{RA} & deg & Right ascension in J2000 \\
        \texttt{DEC} & deg & Declination in J2000 \\
        \texttt{X} & pix & X centroid in the image coordinate \\
        \texttt{Y} & pix & Y centroid in the image coordinate \\
        \texttt{X\_MIN} & pix & X lower boundary of the minimum bounding box \\
        \texttt{X\_MAX} & pix & X upper boundary of the minimum bounding box \\
        \texttt{Y\_MIN} & pix & Y lower boundary of the minimum bounding box \\
        \texttt{Y\_MAX} & pix & Y upper boundary of the minimum bounding box \\
        \texttt{FLUX\_KRON\_TOTAL\_XX} & nJy & Total flux in the filter XX, measured in a fixed Kron elliptical aperture in each filter \\
        \texttt{FLUXERR\_KRON\_TOTAL\_XX} & nJy & 1$\sigma$ flux uncertainty on \texttt{FLUX\_KRON\_TOTAL\_XX} \\
        \texttt{FLUX\_COLOR03\_TOTAL\_XX} & nJy & Total flux in each filter XX, measured by scaling $0.^{\prime\prime}3$-diameter aperture photometry to the total \\
        \texttt{FLUXERR\_COLOR03\_TOTAL\_XX} & nJy & 1$\sigma$ flux uncertainty on \texttt{FLUX\_COLOR03\_TOTAL\_XX} \\
        \texttt{FLUX\_COLOR07\_TOTAL\_XX} & nJy & Total flux in each filter XX, measured by scaling $0.^{\prime\prime}7$-diameter aperture photometry to the total\\
        \texttt{FLUXERR\_COLOR07\_TOTAL\_XX} & nJy & 1$\sigma$ flux uncertainty on \texttt{FLUX\_COLOR07\_TOTAL\_XX} \\
        \texttt{KRON\_RADIUS} & & Unscaled first-moment Kron radius \\
        \texttt{PA} & deg & Position angle of the major-axis, measured from the x-axis in counter-clockwise\\
        \texttt{A} & pix & Semi-major axis\\
        \texttt{B} & pix & Semi-minor axis\\
        \texttt{AREA\_ISO} & pix$^2$ & Total area of the source segment in the $\chi$-mean detection image \\
        \texttt{FLUX\_RADIUS} & arcsec & Circular radius enclosing the 50 \% of the total flux in the F150W 20mas PSF-unmatched image \\
        \texttt{FLAG\_POINTSRC} & & True if the source is identified as a point source (see Section \ref{subsubsec:flagging_stars})\\
        \texttt{FLAG\_GAIA} & & True if the source is cross-matched with the Gaia DR3 catalog (see Section \ref{subsubsec:flagging_stars})\\
        \texttt{ID\_GAIA} & & Source ID in the Gaia DR3 catalog\\
        \texttt{FLAG\_DEBLEND} & & True if the source is deblended in the detection\\
        \texttt{DIST\_FOV\_FULL} & arcsec & Distance between the source centroid and the edge of the detection map\\
        \texttt{DIST\_FOV\_NIRCAM} & arcsec & Distance between the source centroid and the edge of the NIRCam image\\
        \texttt{FLAG\_EXCLUDE} & & True if the Kron aperture of the source contains the masked region defined by \texttt{mask\_exclude}\\
        \texttt{DIST\_EXCLUDE} & arcsec & Distance between the source centroid and the masked region defined by \texttt{mask\_exclude}\\
        \texttt{MEDIAN\_BCGMODEL\_F444W} & nJy & Median value of the BCG model fluxes within the segment area in F444W\\
        \texttt{LOCAL\_BACKGROUND\_XX} & nJy/pix$^2$ & Local background level in the filter XX; this is already corrected in all photometry\\
        \texttt{SNR\_DETIMG} & & S/N in the $\chi$-mean detection image \\
        \texttt{FLAG\_ACS} & & True if the source has valid Kron flux measurement in all optical wavelength filters on \hst \\
        \texttt{FLAG\_NIRCAM} & & True if the source has valid Kron flux measurement in all NIRCam filters\\
        \texttt{FLAG\_HIGH\_BCGCONTAMI} & & True if the BCG model flux is larger than the science flux in F444W (see Section \ref{subsubsec:flag_highbcg})\\
        \texttt{USE\_PHOT} & & True if all criteria described in Section \ref{subsubsec:use_flag} are met\\
        \texttt{USE\_PHOT\_APER03} & & Similar to \texttt{USE\_PHOT} but using $0.^{\prime\prime}3$-diameter aperture photometry instead \\
        \texttt{FLAG\_BCG} & & True if the source is a BCG that is removed from the final image (see Section \ref{subsec:bcgs_icl} and \ref{subsec:bcg_photometry})\\
        \texttt{FLUX\_APER03\_XX} & nJy & Fixed circular aperture flux in the filter XX with $0.^{\prime\prime}3$ diameter \\
        \texttt{FLUXERR\_APER03\_XX} & nJy & 1$\sigma$ flux uncertainty on \texttt{FLUX\_APER03\_XX} \\
        \texttt{FLUX\_APER05\_XX} & nJy & Fixed circular aperture flux in the filter XX with $0.^{\prime\prime}5$ diameter \\
        \texttt{FLUXERR\_APER05\_XX} & nJy & 1$\sigma$ flux uncertainty on \texttt{FLUX\_APER05\_XX} \\
        \texttt{FLUX\_APER07\_XX} & nJy & Fixed circular aperture flux in the filter XX with $0.^{\prime\prime}7$ diameter \\
        \texttt{FLUXERR\_APER07\_XX} & nJy & 1$\sigma$ flux uncertainty on \texttt{FLUX\_APER07\_XX} \\
        \texttt{FLUX\_APER15\_XX} & nJy & Fixed circular aperture flux in the filter XX with $1.^{\prime\prime}5$ diameter \\
        \texttt{FLUXERR\_APER15\_XX} & nJy & 1$\sigma$ flux uncertainty on \texttt{FLUX\_APER15\_XX} \\
        \texttt{FLUX\_APER30\_XX} & nJy & Fixed circular aperture flux in the filter XX with $3.^{\prime\prime}0$ diameter \\
        \texttt{FLUXERR\_APER30\_XX} & nJy & 1$\sigma$ flux uncertainty on \texttt{FLUX\_APER30\_XX} \\
    \enddata
\end{deluxetable}

\subsection{Spectroscopic redshift quality (\texttt{Z\_Q})}
Each spec-$z$ literature has each own $z_{\rm spec}$ quality flag definition.
Referring to these quality flags in literature, we have a standardized $z_{\rm spec}$ quality flag (\texttt{Z\_Q}) in this catalog,, where we define as 1$=$low, 2$=$medium, and 3$=$high. The original $z_{\rm spec}$ quality flag in the literature is also stored in the column \texttt{Z\_Q\_REF}, but we recommend users primarily refer to \texttt{Z\_Q} column.

\begin{deluxetable}{lcl}\tablenum{7}
    \tablecaption{ \textit{Continued}
  	}
    \tablewidth{0pt}
    \tablehead{
    \colhead{Column name} & \colhead{unit} & \colhead{Description}
    }
    \startdata
        \texttt{FLAG\_FOV\_MASTER\_XX} &  & Master flag to find sources whose apertures contain pixels outside the FoV in the filter XX (see text)\\
        \texttt{FLAG\_BPR\_MASTER\_XX} &  & Master flag to find sources whose apertures contain bad pixels in the filter XX (see text)\\
        \texttt{Z\_ML} & & Best photometric redshift estimation (see Section \ref{subsec:photz} for photometric redshifts) \\
        \texttt{Z\_ML\_CHI2} & & $\chi^2$ value of the best templates at \texttt{Z\_ML} \\
        \texttt{NUSEFILT} & & The number of filters used in the \texttt{EAzY} fitting\\
        \texttt{Z025}/\texttt{Z160}/\texttt{Z500}/\texttt{Z840}/\texttt{Z975} & & Photometric redshift posterior percentiles, where 025 stands for 2.5\%, etc. \\
        \texttt{RESTNUV} & nJy & Rest-frame $NUV$ band flux in the fixed $0.^{\prime\prime}3$-diameter aperture photometry \\
        \texttt{RESTNUV\_ERR} & nJy & 1$\sigma$ flux uncertainty on \texttt{RESTNUV} \\
        \texttt{RESTU} & nJy & Rest-frame $U$ band flux in the fixed $0.^{\prime\prime}3$-diameter aperture photometry \\
        \texttt{RESTU\_ERR} & nJy & 1$\sigma$ flux uncertainty on \texttt{RESTU} \\
        \texttt{RESTB} & nJy & Rest-frame $B$ band flux in the fixed $0.^{\prime\prime}3$-diameter aperture photometry \\
        \texttt{RESTB\_ERR} & nJy & 1$\sigma$ flux uncertainty on \texttt{RESTB} \\
        \texttt{RESTV} & nJy & Rest-frame $V$ band flux in the fixed $0.^{\prime\prime}3$-diameter aperture photometry \\
        \texttt{RESTV\_ERR} & nJy & 1$\sigma$ flux uncertainty on \texttt{RESTV} \\
        \texttt{RESTJ} & nJy & Rest-frame $J$ band flux in the fixed $0.^{\prime\prime}3$-diameter aperture photometry \\
        \texttt{RESTJ\_ERR} & nJy & 1$\sigma$ flux uncertainty on \texttt{RESTJ} \\
        \texttt{RESTUS} & nJy & Rest-frame synthetic $u$ band flux in the fixed $0.^{\prime\prime}3$-diameter aperture photometry \\
        \texttt{RESTUS\_ERR} & nJy & 1$\sigma$ flux uncertainty on \texttt{RESTUS} \\
        \texttt{RESTGS} & nJy & Rest-frame synthetic $g$ band flux in the fixed $0.^{\prime\prime}3$-diameter aperture photometry \\
        \texttt{RESTGS\_ERR} & nJy & 1$\sigma$ flux uncertainty on \texttt{RESTGS} \\
        \texttt{RESTIS} & nJy & Rest-frame synthetic $i$ band flux in the fixed $0.^{\prime\prime}3$-diameter aperture photometry \\
        \texttt{RESTIS\_ERR} & nJy & 1$\sigma$ flux uncertainty on \texttt{RESTIS} \\
		\texttt{ID\_SPECZ} &  & Source ID in the ancillary spec-$z$ catalog (when available) \\
        \texttt{RA\_SPECZ} & deg & Right Ascension in the spec-$z$ literature \\
        \texttt{DEC\_SPECZ} & deg & Declination in the spec-$z$ literature \\
        \texttt{RA\_SPECZ\_CORR} & deg & Right Ascension in the spec-$z$ literature corrected for the astrometry difference \\
        \texttt{DEC\_SPECZ\_CORR} & deg & Declination in the spec-$z$ literature corrected for the astrometry difference \\
        \texttt{Z\_SPEC} &  & Spectroscopic redshift from the literature or CANUCS; -99 when no spec-$z$ is available \\
        \texttt{REF\_SPEC} &  & Literature reference for the spectroscopic redshift \\
        \texttt{Z\_Q\_REF} & & Spectroscopic redshift quality in the literature \\
        \texttt{Z\_Q} & & Standardized spectroscopic redshift quality; 1=low, 2=medium, 3=high \\
        \texttt{MU} & & Magnification factor due to the gravitational lensing, derived from the best lens model \\
        \texttt{MU\_16}/\texttt{MU\_50}/\texttt{MU\_84} & & Lens magnification posterior percentiles \\
        \texttt{MU\_RAD} & & Magnification tensor eigenvalue measured along the radial direction from the best lens model \\
        \texttt{MU\_RAD\_16}/\texttt{MU\_RAD\_50}/\texttt{MU\_RAD\_84} & & Posterior percentiles for \texttt{MU\_RAD}\\
        \texttt{MU\_TAN} & & Magnification tensor eigenvalue measured along the tangential direction from the best lens model \\
        \texttt{MU\_TAN\_16}/\texttt{MU\_TAN\_50}/\texttt{MU\_TAN\_84} & & Posterior percentiles for \texttt{MU\_TAN} \\
    \enddata
    \begin{tablenotes}
      \small
      \item {\bf Notes}
      \item \ 1. XX is a filter name. 
      \item \ 2. Source ID (``\texttt{NUMBER}'') is a 7-digit number. The first digit number indicates the cluster where the source is located, and the second digit number indicates the field in the cluster. The last 5 digits are the unique source ID of the source.
      \item \ 3. Synthetic $ugi$ filters are defined by \citet{Antwi-Danso_2023}
    \end{tablenotes}
\end{deluxetable}

For our CANUCS NIRSpec observations, $z_{\rm spec}$ measurements based on multiple emission lines or clear continuum breaks are assigned \texttt{Z\_Q}$=$3, while those based on single emission lines with the aid of photometric data are assigned \texttt{Z\_Q}$=$2.
On the other hand, for ancillary $z_{\rm spec}$, we only include most secure redshifts based on multiple emission lines in most cases, so all ancillary $z_{\rm spec}$ are assigned \texttt{Z\_Q}$=$3.

\section{DENSE BASIS vs BAGPIPES Comparison}\label{apx:db_v_bagpipes}

\begin{figure*}[h]
\plotone{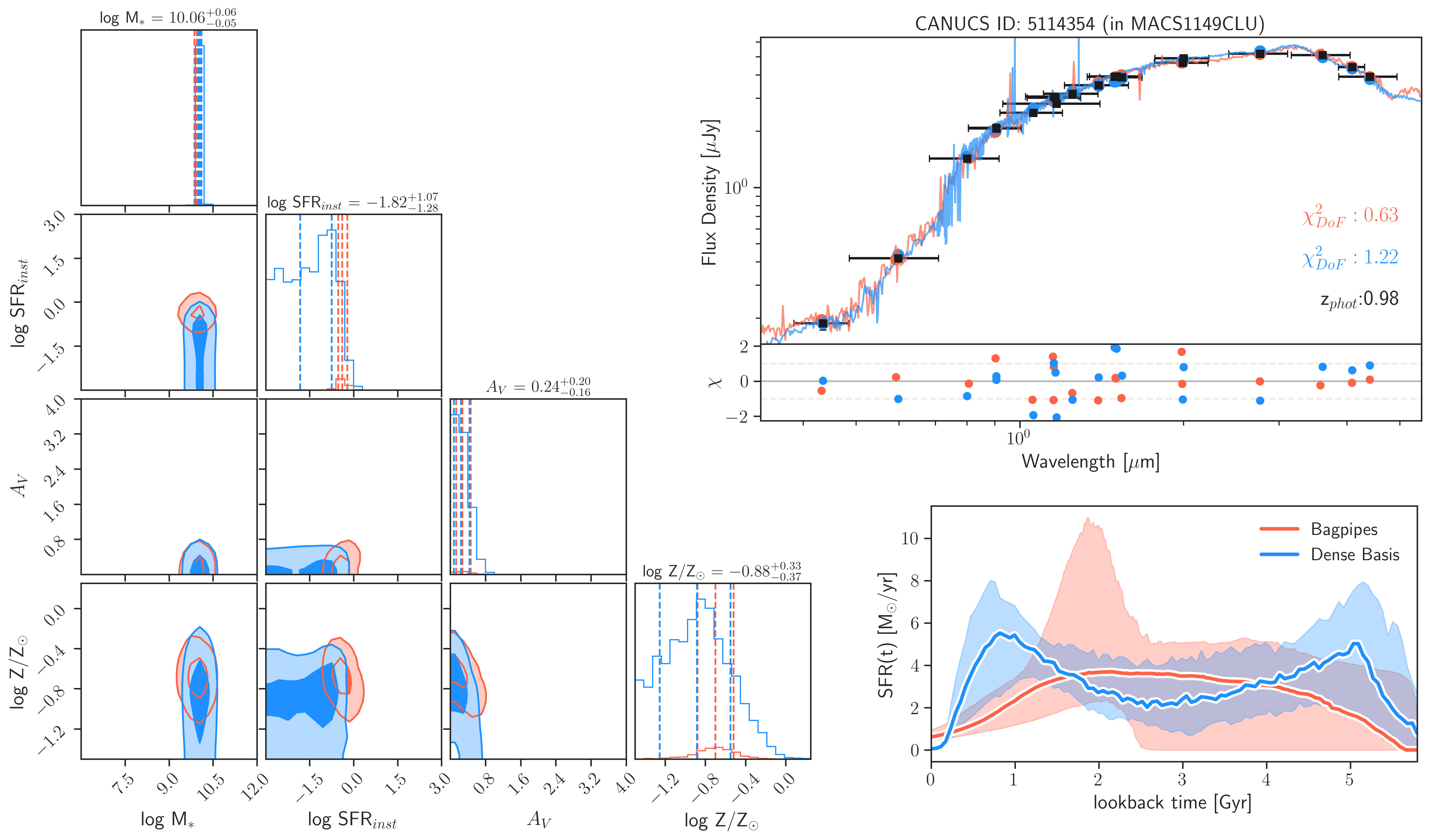}
\plotone{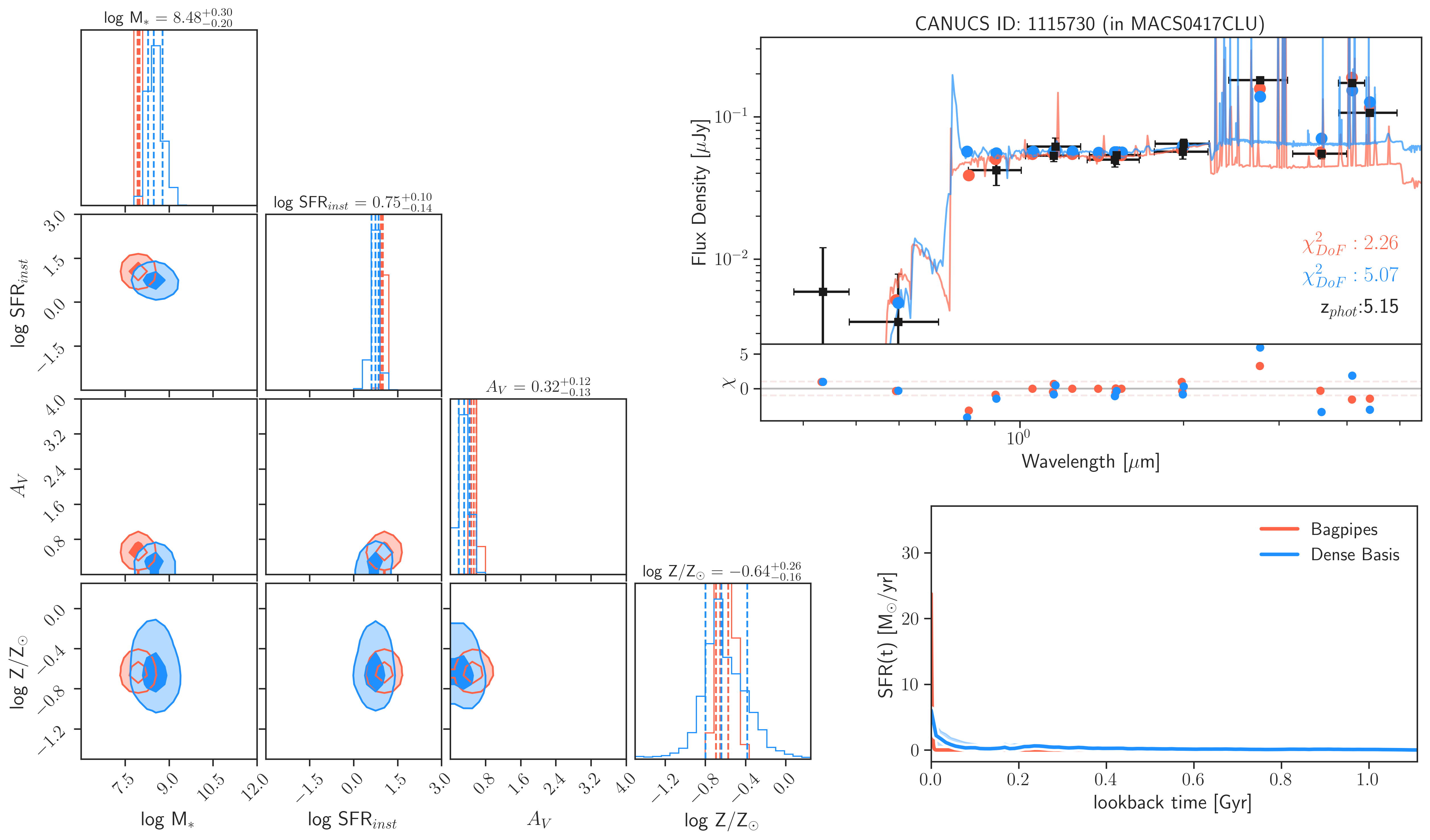}
\caption{Same as Figure \ref{fig:example_fits}, for galaxies 5114354 \& 1115730 from Figure \ref{fig:SEDs}.}
\label{apx:example_fits_1}
\end{figure*}

\begin{figure*}[]
\plotone{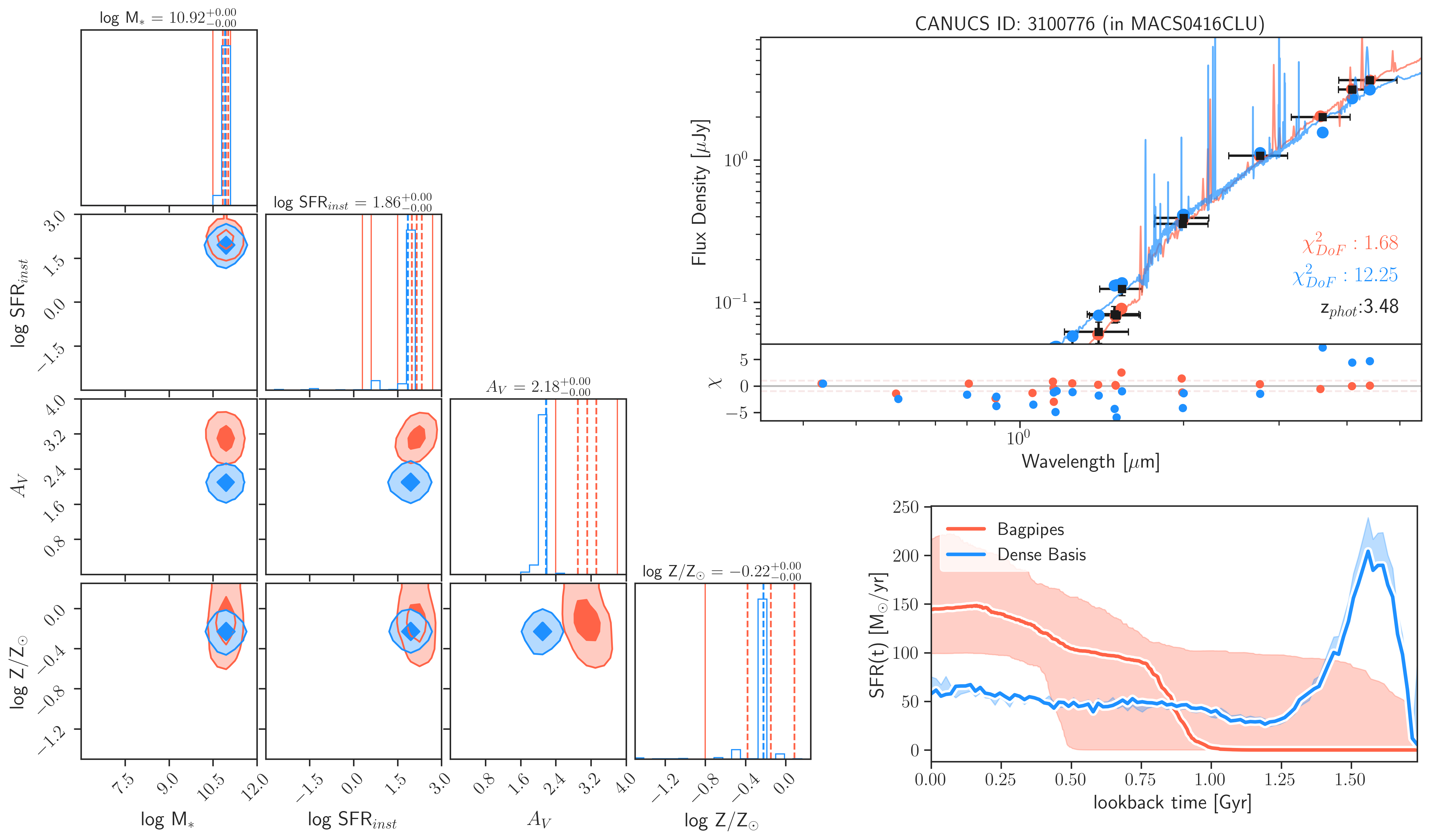}
\plotone{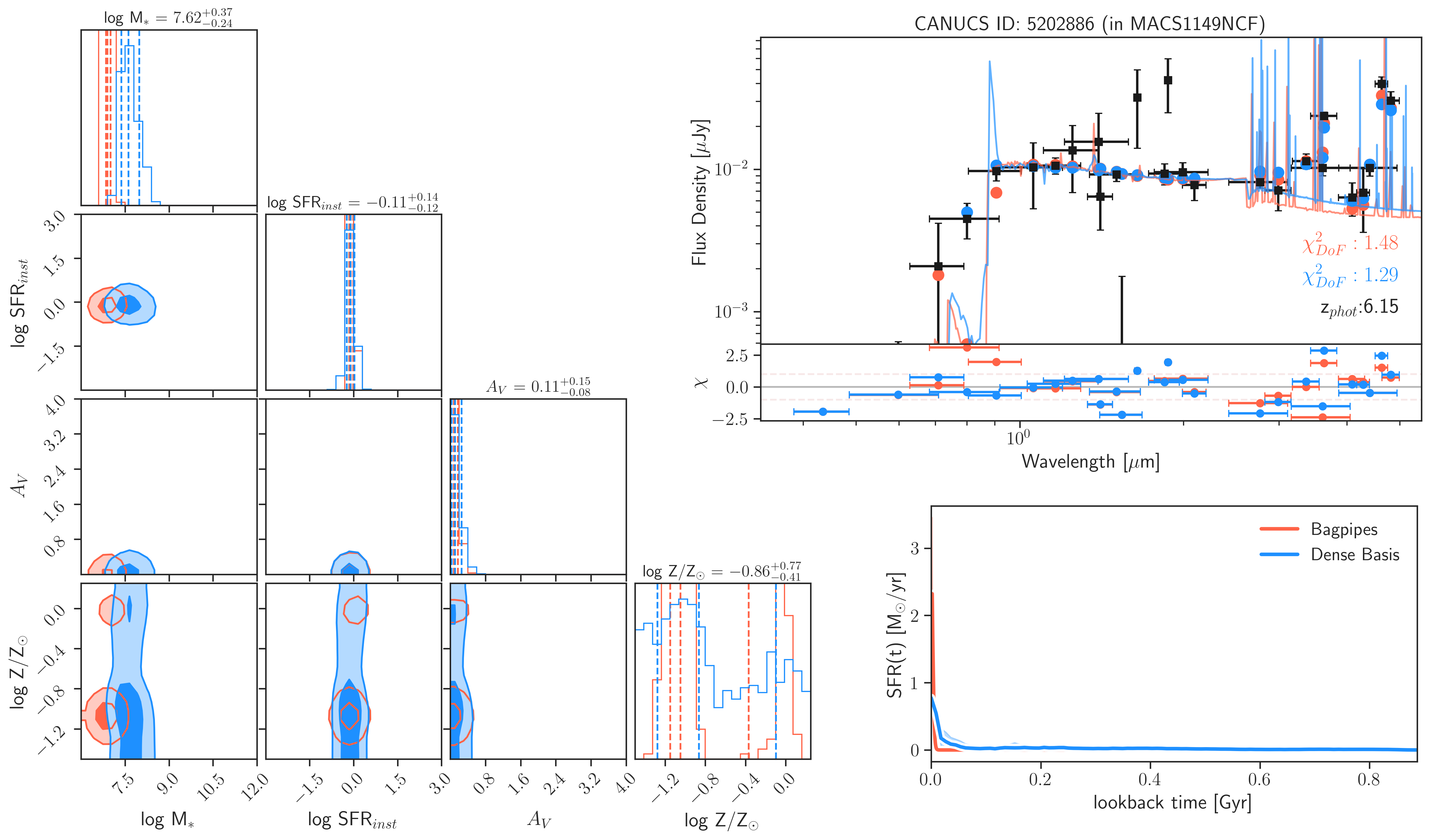}
\caption{Same as Figure \ref{fig:example_fits}, for galaxies 3100776 \& 5202886 from Figure \ref{fig:SEDs}.}
\label{apx:example_fits_2}
\end{figure*}

\begin{figure*}[]
\plotone{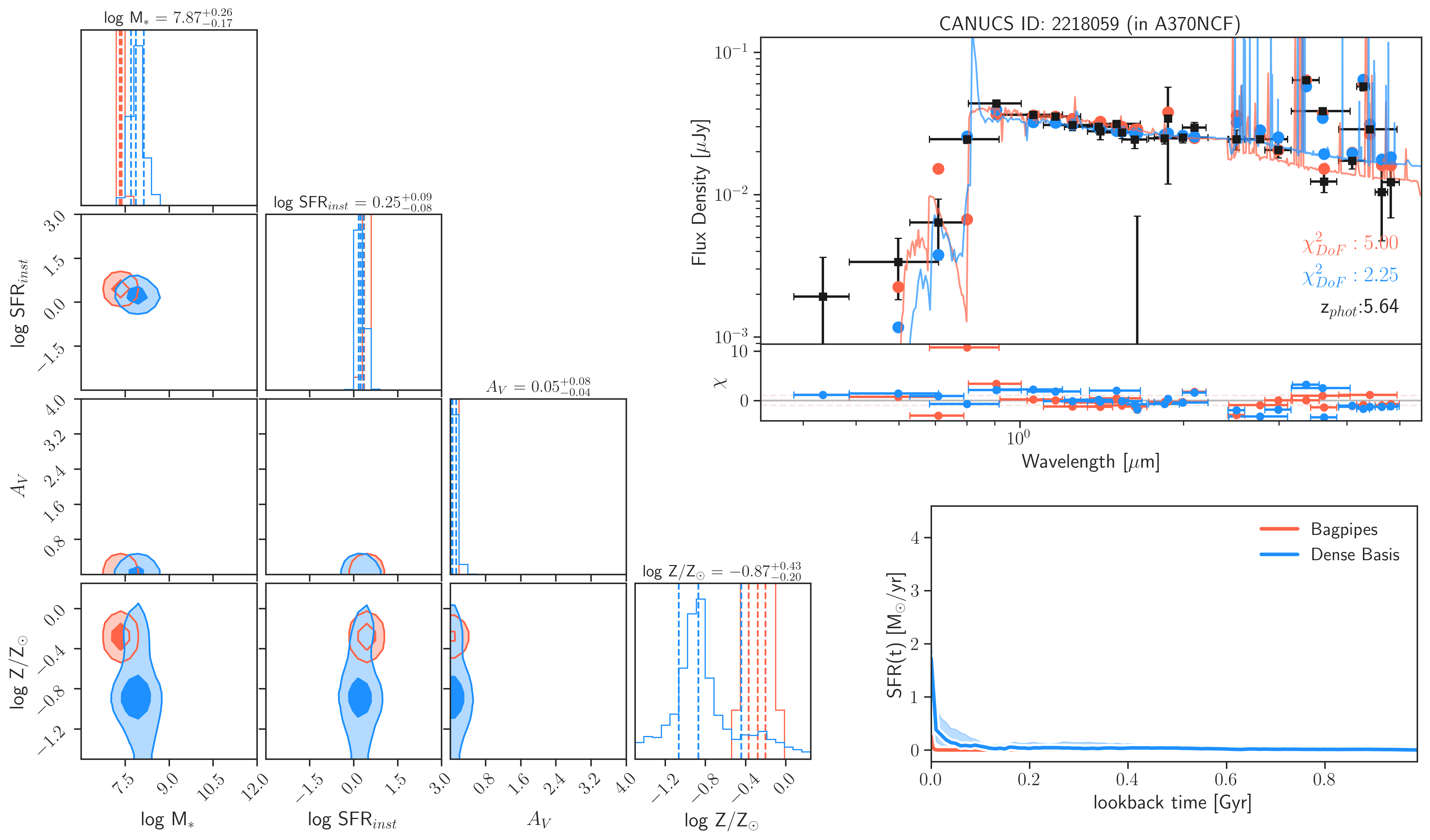}
\plotone{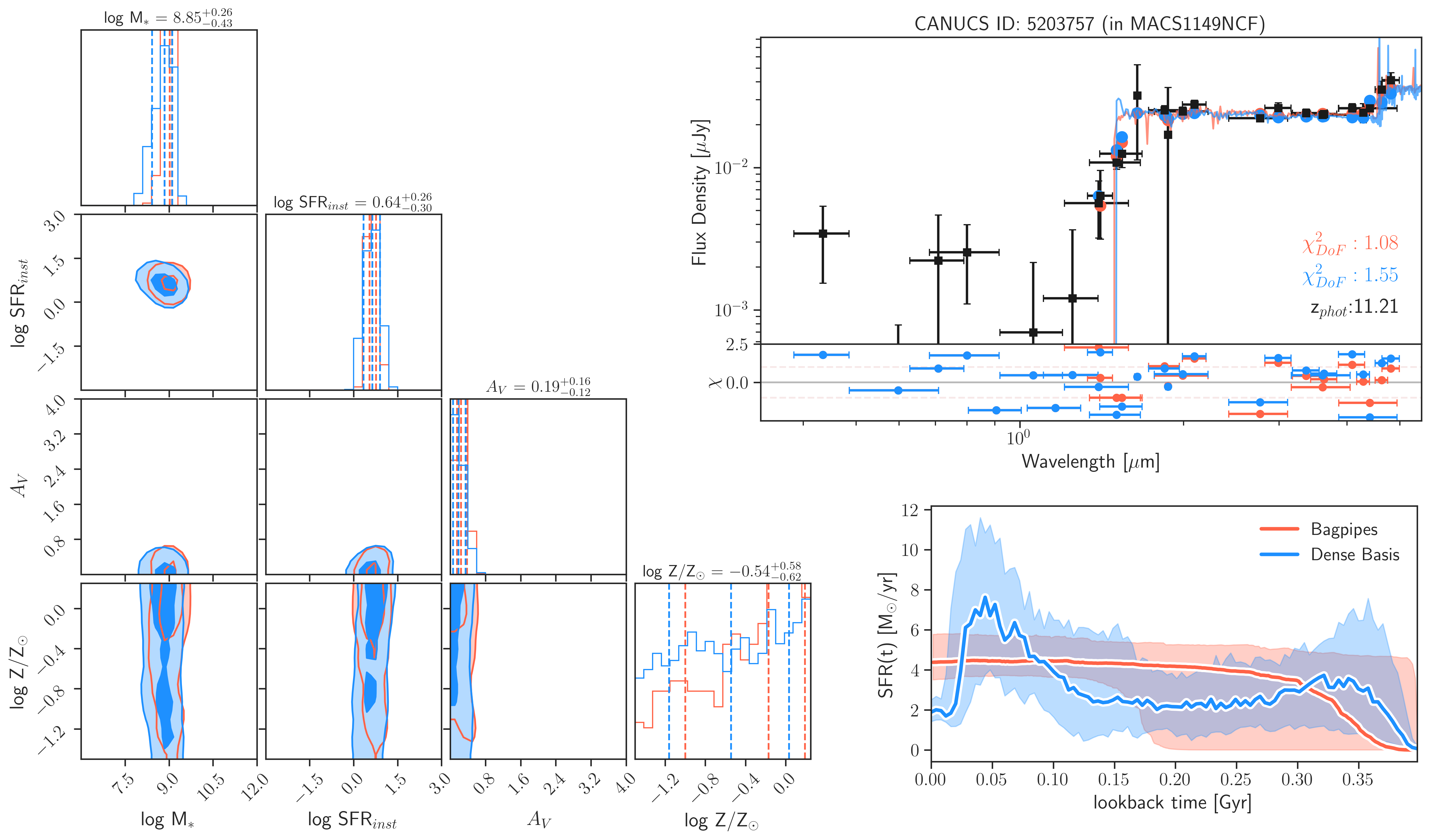}
\caption{Same as Figure \ref{fig:example_fits}, for galaxies 2218059 \& 5203757 from Figure \ref{fig:SEDs}.}
\label{apx:example_fits_3}
\end{figure*}

Some additional examples of SED fitting using both the \texttt{DENSE BASIS} \& \bagpipes codes are shown in Figures \ref{apx:example_fits_1}-\ref{apx:example_fits_3}.

\newpage

\newpage

\bibliography{canucs_dr1}{}
\bibliographystyle{aasjournal}

\end{document}